\documentclass[acmsmall,review=false,nonacm=true]{acmart}

\def\BibTeX{{\rm B\kern-.05em{\sc i\kern-.025em b}\kern-.08emT\kern-.1667em\lower.7ex\hbox{E}\kern-.125emX}}

\usepackage[colorinlistoftodos]{todonotes}
\usepackage[utf8]{inputenc}
\usepackage{hyperref}
\usepackage{graphicx}
\usepackage{xspace}
\usepackage{mdframed}
\usepackage{adjustbox}
\usepackage{tablefootnote}
\usepackage{longtable}
\usepackage{subcaption}

\usepackage{alltt}

\begin{document}

\title{Learning Software Configuration Spaces:\\ A Systematic Literature Review}

\author{Juliana Alves Pereira}
\affiliation{
  \institution{Univ Rennes, Inria, CNRS, IRISA}
  \city{Rennes}
  \country{France}}
\email{juliana.alves-pereira@irisa.fr}

\author{Hugo Martin}
\affiliation{
  \institution{Univ Rennes, Inria, CNRS, IRISA}
  \city{Rennes}
  \country{France}}
\email{hugo.martin@irisa.fr}

\author{Mathieu Acher}
\affiliation{
  \institution{Univ Rennes, Inria, CNRS, IRISA}
  \city{Rennes}
  \country{France}}
\email{mathieu.acher@irisa.fr}

\author{Jean-Marc Jézéquel}
\affiliation{
  \institution{Univ Rennes, Inria, CNRS, IRISA}
  \city{Rennes}
  \country{France}}
\email{Jean-Marc.Jezequel@irisa.fr}

\author{Goetz Botterweck}
\affiliation{
  \institution{University of Limerick, Lero–The Irish Software Research Centre}
  \city{Limerick}
  \country{Ireland}}
\email{goetz.botterweck@lero.ie}

\author{Anthony Ventresque}
\affiliation{
  \institution{University College Dublin}
  \city{Dublin}
  \country{Ireland}}
\email{anthony.ventresque@ucd.ie}

\begin{abstract}
Most modern software systems (operating systems like Linux or Android, Web browsers like Firefox or Chrome, video encoders like ffmpeg, x264 or VLC, mobile and cloud applications, etc.) are highly-configurable. Hundreds of configuration options, features, or plugins can be combined, each potentially with distinct functionality and effects on execution time, security, energy consumption, etc. Due to the combinatorial explosion and the cost of executing software, it is quickly impossible to exhaustively explore the whole configuration space. 
Hence, numerous works have investigated the idea of learning it from a small sample of configurations' measurements. The pattern "sampling, measuring, learning" has emerged in the literature, with several practical interests for both software developers and end-users of configurable systems. 
In this survey, we report on the different application objectives (\textit{e.g.}, performance prediction, configuration optimization, constraint mining), use-cases, targeted software systems and application domains.
We review the various strategies employed to gather a representative and cost-effective sample. We describe automated software techniques used to measure functional and non-functional properties of configurations. We classify machine learning algorithms and how they relate to the pursued application.
Finally, we also describe how researchers evaluate the quality of the learning process.
 The findings from this systematic review show that the potential application objective is important; there are a vast number of case studies reported in the literature from the basis of several domains and software systems. 
Yet, the huge variant space of configurable systems is still challenging and calls to further investigate the synergies between artificial intelligence and software engineering. 
\end{abstract}

%

\keywords{Systematic Literature Review, Software Product Lines, Machine Learning, Configurable Systems}

\maketitle 

\section{Introduction} \label{introduction}



End-users, system administrators, software engineers, and scientists have at their disposal thousands of options (a.k.a. features or parameters) to configure various kinds of software systems in order to fit their functional and non-functional needs (execution time, output quality, security, energy consumption, etc).
It is now ubiquitous that software comes in many variants and is highly configurable through conditional compilations, command-line options, runtime parameters, configuration files, or plugins. Software product lines (SPLs), software generators, dynamic, self-adaptive systems, variability-intensive systems are well studied in the literature and enter in this class of configurable software systems~\cite{svahnberg2005, pohl2005, apel2013book, sayagh2018, benavides2010, cohenASE2018, DBLP:journals/computer/HallsteinsenHPS08, DBLP:journals/computer/MorinBJFS09,temple2017a}.  

From an abstract point of view, a software configuration is simply a combination of options' values. Though customization is highly desirable, it introduces an enormous complexity due to the combinatorial explosion of possible variants.
For example, the Linux kernel has 15,000+ options and most of them can have 3 values: "yes", "no", or "module". Without considering the presence of constraints to avoid some combinations of options, there may be $3^{15,000}$ possible variants of Linux -- the estimated number of atoms in the universe is $10^{80}$ and is already reached with 300 Boolean options. Though Linux is an extreme case, many software systems or projects exhibit a very large configuration space; it has several consequences. 
 
On the one hand, developers struggle to maintain, understand, and test configuration spaces since they can hardly analyze or execute all variants in every possible settings. According to several studies~\cite{halin:hal-01829928, sayagh2018}, the flexibility brought by variability is expensive as configuration failures represent one of the most common types of software failures. On the other hand, end-users fear software variability and stick to default configurations~\cite{DBLP:conf/sigsoft/XuJFZPT15,zheng2007} that may be sub-optimal (\textit{e.g.}, the software system will run very slowly) or simply inadequate (\textit{e.g.}, the quality of the output will be terrible). 
 
\begin{figure}
	\centering
	\includegraphics[width=1.0\textwidth]{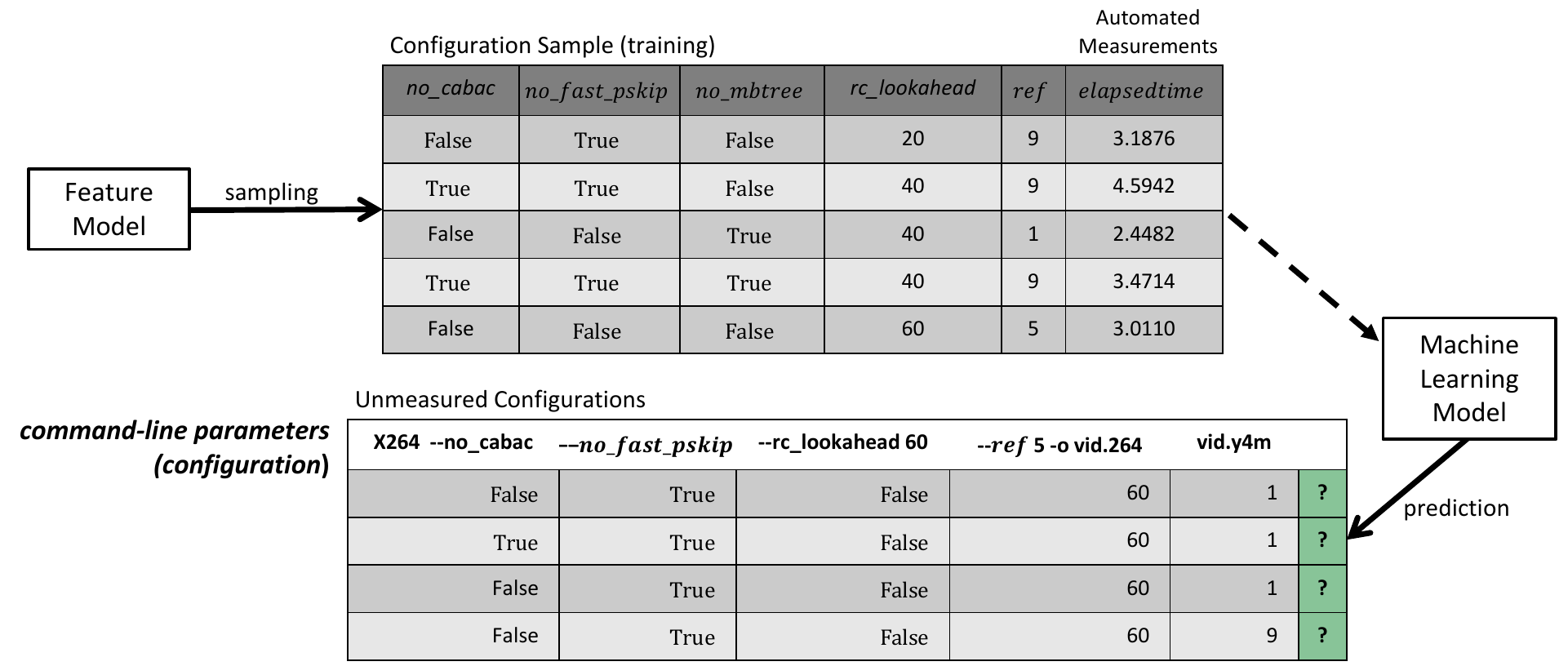}
	\caption{Features, configurations, sample, measurements, and learning.}
    \label{fig:introduction}
\end{figure}

Since it is hardy possible to fully explore all software configurations, the use of machine learning techniques is a quite natural and appealing approach. The basic idea is to learn out of a \emph{sample} of configurations' observations and hopefully generalize to the whole configuration space. There are several applications ranging from performance prediction, configuration optimization, software understanding to constraint mining -- we will give a more exhaustive list in this survey. 
For instance, end-users of x264 (a configurable video encoder) can estimate in advance the execution time of the command-line at the center of Fig.~\ref{fig:introduction},
since a machine learning model has been crafted to predict the performance of configurations. End-users may want to use the fastest configuration, or know all configurations that meet an objective (\textit{e.g.}, encoding time should be less than 10 seconds). Developers of x264 can be interested in understanding the effects of some options and how options interact. 

For all these use-cases, a pattern has emerged in the scientific literature: \emph{"sampling, measuring, learning"}. 
The basic principle is that a procedure is able to learn out of a sample of configurations' measurements (see Fig.~\ref{fig:introduction}).
Specifically, many software configuration problems can actually be framed as statistical machine learning problems under the condition a sample of configurations' observations is available. For example, the prediction of the performance of individual configurations can be formulated as a regression problem; appropriate learning algorithms (\textit{e.g.}, CART) can then be used to predict performance of untested, new configurations. 
In this respect, it is worth noticing the dual use of \emph{feature} in the software or machine learning fields: features either refer to software features (\textit{a.k.a.} configuration options) or to variables a regressor aims to relate. 
A way to reconcile and visualize both is to consider a configuration matrix as depicted in Fig.~\ref{fig:introduction}.
Each row is a configuration together with observations. In the example of Fig.~\ref{fig:introduction}, the first configuration has \textsf{no\_cabac} set to False value and \textsf{ref} set to 9 value while the encoding time is 3.1876 seconds. We can use a sample of configurations to train a machine learning model (a regressor) with predictive variables being command-line parameters of x264. Unmeasured configurations could then be predicted.

Learning software configuration spaces is, however, not a pure machine learning problem and there are a number of specific challenges to address at the intersection of software engineering and artificial intelligence. For instance, the sampling phase involves a number of difficult activities: \textit{(1)} picking configurations that are valid and conform to constraints among options -- one needs to resolve a satisfiability problem; \textit{(2)} instrumenting the executions and observations of software for a variety of configurations -- it can have an important computational cost and is hard to engineer especially when measuring non-functional aspects of software; \textit{(3)} meanwhile, we expect that the sample is representative to the whole population of valid configurations otherwise the learning algorithm may hardly generalize to the whole configuration space. The general problem is to find the right strategy to decrease the cost of labelling software configurations while minimizing the prediction errors. From an empirical perspective, one can also wonder to what extent learning approaches are effective for real-world software systems present in numerous domains. 

While several studies have covered different aspects of configurable systems over the last years, there has been no secondary study (such as systematic literature reviews) that identifies and catalogs individual contributions for machine learning configuration spaces.
Thus, there is no clear consensus on what techniques are used to support the process, including which quantitative and qualitative properties are considered and how they can be measured and evaluated, as well as how to select a significant sample of configurations and what is an ideal sample size.
This stresses the need for a secondary study to build knowledge from combining findings from different approaches and present a complete overview of the progress made in this field. 
To achieve this aim, we conduct a \textit{Systematic Literature Review} (SLR)~\cite{kitchenham2007}
 to identify, analyze and interpret all available important research in this domain. 
 We systematically review research papers in which the process of sampling, measuring, and learning configuration spaces occurs -- more details about our research methodology are given in Section~\ref{methodology}.
Specifically, we aim of synthesizing evidence to answer the following six research questions:

\begin{itemize}
	\item \textit{RQ1. What are the concrete applications of learning software configuration spaces?}
    
	\item \textit{RQ2. Which sampling methods are adopted when learning software configuration spaces?} 
    
    \item \textit{RQ3. Which techniques are used to gather measurements of functional and non-functional properties of configurations?}
    
	\item \textit{RQ4. Which learning techniques are used?}
    
    \item \textit{RQ5. How are learning-based techniques validated?}
    
    \item \textit{RQ6. What are the limitations faced by the current techniques and open challenges that need attention in the future?}
\end{itemize}

To address \textit{RQ1}, we analyze the application objective of the study (\textit{i.e.}, why they apply learning-based techniques). 
It would allow us to assess whether the proposed approaches are applicable. 
With respect to \textit{RQ2}, we investigate which sampling methods are used in the literature.
With respect to \textit{RQ3}, we give an in-depth view of how each study measure a sample of configurations.
Next, \textit{RQ4} systematically identifies learning techniques used in the literature for exploring the SPL configuration space. 
In addition, \textit{RQ5} follows identifying 
which sampling design and evaluation metrics are used for evaluation. 
Finally, analyzing existing techniques allows identifying evidence about their maturity and limitations, such as which domains are missing or have not been considered, addressed by \textit{RQ6}.

By answering these questions, we make the following five contributions:

\begin{enumerate}
	\item  We identified six main different application areas: \textit{pure prediction}, \textit{interpretability}, \textit{optimization}, \textit{dynamic configuration}, \textit{evolution}, and \textit{mining constraints}. 
    \item We provide a framework classification of four main stages used for learning: \textit{Sampling}, \textit{Measuring}, \textit{Learning}, and \textit{Validation}. 
    \item We describe 23 high-level sampling methods, 5 measurement strategies, 51 learning techniques, and 50 evaluation metrics used in the literature. 
	As case study, we identify 71 real-world configurable systems targeting several domains, and functional and non-functional properties. 
	We relate and discuss the learning and validation techniques with regards to their application objective.
	\item We identify a set of open challenges faced by the current approaches, in order to guide researchers and practitioners to use and build appropriate solutions.
	\item We build a Web repository to make our SLR results publicly available for the purpose of reproducibility and extension.
\end{enumerate}

Overall, the findings of this SLR reveal that there is a significant body of work specialized in learning software configurable systems with an important application in terms of software technologies, application domains, or goals. 
There is a wide variety in the considered sampling or learning algorithms as well as in the evaluation process, mainly due to the considered subject systems and application objectives. 
Practitioners and researchers can benefit from the findings reported in this SLR as a reference when they select a learning technique for their own settings.
To this end, this review provides a classification and catalog of specialized techniques in this field.

The rest of the paper is structured as follows. 
In Section~\ref{methodology}, we describe the research protocol used to conduct the SLR.
In Section~\ref{preliminaries}, we categorize a sequence of key learning stages used by the ML state-of-the-art literature to explore highly configurable systems. 
In Section~\ref{results}, we discuss the research questions. 
In Section~\ref{threats_validity}, we discuss the threats to the validity of our SLR.
In Section~\ref{related_work}, we describe similar secondary studies and indicate how our survey differs from them.
Finally, in Section~\ref{conclusion}, we present the conclusions of our work.

\section{The Review Methodology} \label{methodology}

We followed the SLR guidelines by Kitchenham and Charters~\cite{kitchenham2007} to systematically investigate the use of learning techniques for exploring the SPL configuration space. 
In this section, we present the SLR methodology that covers three main phases: \textit{planning the review}, \textit{conducting the review} and \textit{reporting the review}.
We report the details about each phase so that readers can assess their rigor and completeness, and reproduce our findings. 

\subsection{Planning the Review} \label{planning}
The steps involved in \textit{planning the review} are: identification of the need for a review, specification of the research questions, and development of a review protocol.
\paragraph{The need for a systematic review.}
The main goal of this SLR is to systematically investigate and summarize the state-of-the-art of the research concerning learning techniques in the context of software configurable systems. 
The purpose of conducting this SLR has partially been addressed in the introduction and was motivated by the lack of a systematic study carried on this topic. 
 According to \cite{kitchenham2007} the findings of an SLR is expected to provide a valuable overview of the status of the field to the community through the summarization of existing empirical evidence supported by current scientific studies.
The outcomes of such an overview can identify whether, or under what conditions, the proposed learning approaches can support various use-cases around configurable systems and be practically adopted (\textit{e.g.}, for which context a specific learning technique is much suitable). 
 By mean of this outcome, we can detect the limitations in current approaches to properly suggest areas for further investigation. 

\paragraph{The research questions.}
The goal of this SLR is to answer the following main research question:
\textit{What studies have been reported in the literature on learning software configuration spaces since the introduction of Software Product Lines in the early 1990s~\cite{kang1990} to date (2019)?}
However, this question is too broad, so we derived the six sub-questions defined in Section~\ref{introduction}, so as to focus on specific aspects. \textit{RQ1} classifies the papers with regards to their application objective, i.e., for which particular task the approach is suited and useful.
We can group studies into similar categories and then compare them. 
It is also of interest to identify the practical motivations behind learning approaches. We verified whether the authors indicated a specific application for their approach; otherwise, we classified the approach as pure prediction.
\textit{RQ2--RQ5} seek to understand key steps of the learning process. 
\textit{RQ2} reviews the set of sampling methods used in the literature.
\textit{RQ3} describes which subject software systems, application domains, and functional and non-functional properties of configurations are measured and how the measurement process is conducted.
\textit{RQ4} classifies the set of learning-based techniques used in the literature. 
\textit{RQ5} aims to characterize the evaluation process used by researcher, including the sample design and supported evaluation metric(s). 
 Finally, addressing these questions will allow us to answer \textit{RQ6}.
\textit{RQ6} identifies trends and challenges in the current state-of-the-art approaches, as well as analysis their maturity to summarize our findings and propose future works.

\paragraph{The review protocol.}
We searched for all relevant papers published up to May 31st 2019. 
The search process involved the use of 5 scientific digital libraries\footnote{We decided not to use Google Scholar due to search engine limitations, such as the very strict size of the search string.}: 
IEEE Xplore Digital Library\footnote{http://ieee.org/ieeexplore}, ACM Digital Library\footnote{http://dl.acm.org}, Science Direct\footnote{http://www.sciencedirect.com}, Springer-link\footnote{http://link.springer.com}, and Scopus\footnote{http://www.scopus.com}.  These search engines were selected because they are known as the top five preferred on-line databases in the software engineering literature~\cite{hoda2017}. 
 We restricted the search to publication titles and abstracts.
However, the library Springer-link only enables a full-text search.
Therefore, we first used the full-text option to generate an initial set of papers (the results were stored in a .bib file).
Then, we created a script to perform an expert search in the title and abstract over these results.

Each author of this paper had specific roles when performing this SLR.
Pereira applied the search string to the scientific databases and exported the results (\textit{i.e.}, detailed information about the candidate papers) into a spreadsheet.
After this stage, papers were selected based on careful reading of the titles and abstracts (and if necessary checking the introduction and conclusion).
Each identified candidate paper in accordance with the selection criteria defined in Section~\ref{conducting} were identified as potentially relevant.
When Pereira decided that a paper was not relevant, she provided a short rationale why the paper should not be included in the study.
In addition, another researcher checked each excluded paper at this stage.
To minimize potential biases introduced into the selection process, any disagreement between researchers were put up for discussion between all authors until a consensus agreement was obtained. 
This step was done in order to check that all relevant papers were selected.

The search in such databases is known to be challenging due to different search limitations, \textit{e.g.} different ways of constructing the search string. 
Thus, apart from the automatic search, we also consider the use of snowballing~\cite{wohlin2014} as a complementary approach.
Through snowballing, we searched for additional relevant primary studies by following the references from all preliminary selected studies (plus excluded secondary studies). As we published some works related to the topic of the survey, we used our knowledge and expertise to complement the pool of relevant papers. 

During the data extraction stage, each paper was assigned to one researcher.
Pereira coordinated the allocation of researchers to tasks based on the availability of each researcher. 
The researcher responsible for extracting the data of a specific selected paper applied the snowballing technique to the correspondent paper. 
Pereira applied the snowballing technique for excluded secondary studies.
Each primary study was then assigned to another researcher for review.
Once all the primary studies were reviewed, the extracted data was compared. Whenever there were any discrepancies either about the data reported or about the list of additional selected papers derived from the snowballing process, we again resolved the problem through discussions among all authors. 

\subsection{Conducting the Review} \label{conducting}
The steps involved in \textit{conducting the review} are: definition of the search string, specification of the selection criteria, and specification of the data extraction process.
\paragraph{The search string.}
According to Kitchenham et al.~\cite{kitchenham2007} there is no silver bullet for identifying good search strings since, very often, the terminology used in the field is not standardized. When using broad search terms, a large number of irrelevant papers may be found in the search which makes the screening process very challenging.
The search string used in this review was first initiated by selecting an initial list of relevant publications by using our expertise in the field.

We identified in the title and abstract the major terms to be used for systematic search of the primary studies. Then, we searched for synonyms related to each major term.
Next, we performed several test searches with alternative links between keywords through the different digital libraries.
The results from the test searches were continuously discussed among the authors to refine the search string until we were fully satisfied with the capability of the string to detect as much of the initial set of relevant publications as possible.
Following this iterative strategy and after a series of test executions and reviews, we obtained Table~\ref{tab:keywords} that structures the set of search terms and keywords.

\begin{table}
	\centering
	\begin{tabular}[t]{p{0.16\textwidth}p{0.75\textwidth}}
    	\hline
		 \textbf{Term} & \textbf{Keywords} \\
        \hline
		Product Line & product line, configurable (system, software), software configurations, configuration of a software, feature (selection, configuration)\\
		Learning Techniques & learning techniques, (machine, model, statistical) learning\\
		Performance Prediction & performance (prediction, model, goal), (software, program, system) performance, (prediction of, measure) non-functional properties\\
		Predict & predict, measure, transfer learning, optimal (configuration, variant), adaptation rules, constraints, context\\
		Medicine & gene, medicine, disease, patient, biology, diagnosis, molecular, health, brain, biomedical \\
        \hline
	\end{tabular}
	\caption{Keywords used to build the search strings.}
	\label{tab:keywords}
\end{table}

Specifically, Table~\ref{tab:keywords} shows the \textit{term} we are looking for and related synonyms that we considered as \textit{keywords} in our search.
Keywords associated to \texttt{Product Line} allow us to include studies that focus on configurable systems.
By combining keywords associated to \texttt{Learning Techniques} and \texttt{Performance Prediction}, we can find studies that focus on the use of learning-based techniques for exploring the variability space.
In addition, keywords associated to \texttt{Predict} (most specific term) allow us to focus on the application objective of such works.
We decided to include the keywords associated to \texttt{Predict} so as to identify the context of the study and have a more restricted number of primary studies.
Otherwise, the keywords \texttt{(Product Line AND (Learning Techniques OR Performance Prediction))} return a broad number of studies, \textit{e.g.} studies addressing the use of learning techniques for product line testing or configuration guidance.
In addition, we used keywords from \texttt{Medicine} to exclude studies in this field from our search.
The final result is the following search string: 

\begin{mdframed}[backgroundcolor=gray!10] 
	\begin{center}
		\texttt{(Product Line AND (Learning Techniques OR\\ Performance Prediction) AND Predict) AND NOT Medicine}
	\end{center}
\end{mdframed}

The terms \texttt{Product Line} , \texttt{Learning Techniques}, \texttt{Performance Prediction}, and \texttt{Predict} are represented as a disjunction of the keywords in Table~\ref{tab:keywords}.
The search string format we used was slightly modified to meet the search engine requirements of the different scientific databases.
For example, the scientific library Science Direct limits the size of the search string.
Thus, when searching in this library, we had to split the search string to generate an initial set of papers.
Then, we created a script to perform an expert search of all keywords in the title and abstract over these results, \textit{i.e.} we made every effort to ensure that the search strings used were logically and semantically equivalent to the original string in Table~\ref{tab:keywords}.
The detailed search strings used in each digital search engine are provided in the Web supplementary material~\cite{pereira2019}.

\paragraph{The selection criteria.}
The selection of studies was conducted by applying a set of selection criteria for retrieving a relevant subset of publications.
First, we only selected papers published up to May 31st 2019 that satisfied all of the following three \textit{Inclusion Criteria} (IC):

\begin{itemize}
	\item[\textit{IC1}] The paper is available on-line and in English;
    
    \item[\textit{IC2}] The paper should be about \textit{configurable software systems}.

    \item[\textit{IC3}] The paper deals with techniques to statistically learn data from a sample of configurations (see Section~\ref{RQ1}). 
    When different extensions of a paper were observed, \textit{e.g.}, an algorithm is improved by parameter tuning, we intentionally classified and evaluated them as separate primary studies for a more rigorous analysis.
    
\end{itemize}


Moreover, we excluded papers that satisfied at least one of the following four \textit{Exclusion Criteria} (EC):
\begin{itemize}
	\item[\textit{EC1}] Introductions to special issues, workshops, tutorials, conferences, conference tracks, panels, poster sessions, as well as editorials and books.
    \item[\textit{EC2}] Short papers (less than or equal to 4 pages) and work-in-progress.
    \item[\textit{EC3}] Pure artificial intelligence papers. 
    \item[\textit{EC4}] Secondary studies, such as literature reviews, 
    articles presenting lessons learned, position or philosophical papers, with no technical contribution. However, the references of these studies were read in order to identify other relevant primary studies for inclusion through the snowballing technique (see Section~\ref{planning}). Moreover, we consider secondary studies in the related work section.
\end{itemize}

We find it useful to give some examples of approaches that were \emph{not} included: 
\begin{itemize}
    \item the use of sampling techniques without learning (\textit{e.g.}, the main application is testing or model-checking a software product line). That is, the sample is not used to train a machine learning model but rather for reducing the cost of verifying a family of products. For a review on sampling for product line testing, we refer to~\cite{medeiros2016, lopez2015, machado2014, lee2012, thum2014classification}. We also discuss the complementary between the two lines of work in Section~\ref{RQ6}; 
    \item the use of state-of-the-art recommendations and visualization techniques for configuration guidance (\textit{e.g.}, \cite{pereira2018a} and~\cite{murashkin2013}) and
    optimization methods based on evolutionary algorithms (\textit{e.g.}, \cite{guo2011} and~\cite{sayyad2013}) since a sample of configurations' measurements is not considered.
    \item the use of learning techniques to predict the existence of a software defect or vulnerability based on source code analysis (\textit{e.g.}, in \cite{putri2017} and \cite{stuckman2017}, features do not refer to configurable systems, instead features refer to properties or metrics of the source code). 
\end{itemize}


\paragraph{The data extraction.}
The data extraction process was conducted using a structured extraction form in Google Sheets\footnote{https://www.google.com/sheets/about/} to synthesize all data required for further analyze in such a way that the research questions can be answered.
In addition, Google Sheets allow future contributions to be online updated by shareholders.
First, all candidate papers were analyzed regarding the selection criteria.
The following data were extracted from each retrieved study:

\begin{itemize}
	\item Date of search, scientic database, and search string.
	\item Database, authors, title, venue, publication type (\textit{i.e.}, journal, conference, symposium, workshop, or report), publisher, pages, and publication year.
	\item Inclusion criteria IC1, IC2, and IC3 (yes or no)?
	\item Exclusion criteria EC1, EC2, and EC3 (yes or no)?
	\item Selected (yes or no)? If not selected, justification regarding exclusion.
\end{itemize} 

Once the list of primary studies was decided, each selected publication was then read very carefully and the content data for each selected paper was captured and extracted in a second form.
The data extraction aimed to summarize the data from the selected primary studies for further analysis of the research questions and for increasing confidence regarding their relevance.
All available documentation from studies served as data sources, such as thesis, websites, tool support, as well as the communication with authors (\textit{e.g.}, emails exchanged).
The following data were extracted from each selected paper: 

\begin{itemize} 
    \item \textit{RQ1}: Scope of the approach. We classified the approach according to the following six categories: \textit{pure prediction}, \textit{interpretability of configurable systems}, \textit{optimization}, \textit{dynamic configuration}, \textit{mining constraints}, and \textit{evolution}. 
    \item \textit{RQ2}: Sampling technique(s).
    \item \textit{RQ3}: Information about subject systems (\textit{i.e.}, reference, name, domain, number of features, and valid configurations) and the measurement procedure. We collected data about the measured (non-)functional properties and the adopted strategies of measurement.
	\item \textit{RQ4}: Short summary about the adopted learning techniques. 
    \item \textit{RQ5}: Evaluation metrics and sample designs used by approaches for the purpose of training and validating machine learning models. 
    \item \textit{RQ6}: Description of the main challenges and open issues raised by the authors of each selected study. We captured challenges and open issues from the future work, conclusion, or threat to validity sections of a primary study. In the end, we grouped gaps and open challenges according to the learning phase and the scope they address.
\end{itemize}


\subsection{Reporting the Review} \label{reporting}
\begin{figure}
	\centering
	\includegraphics[width=0.9\textwidth]{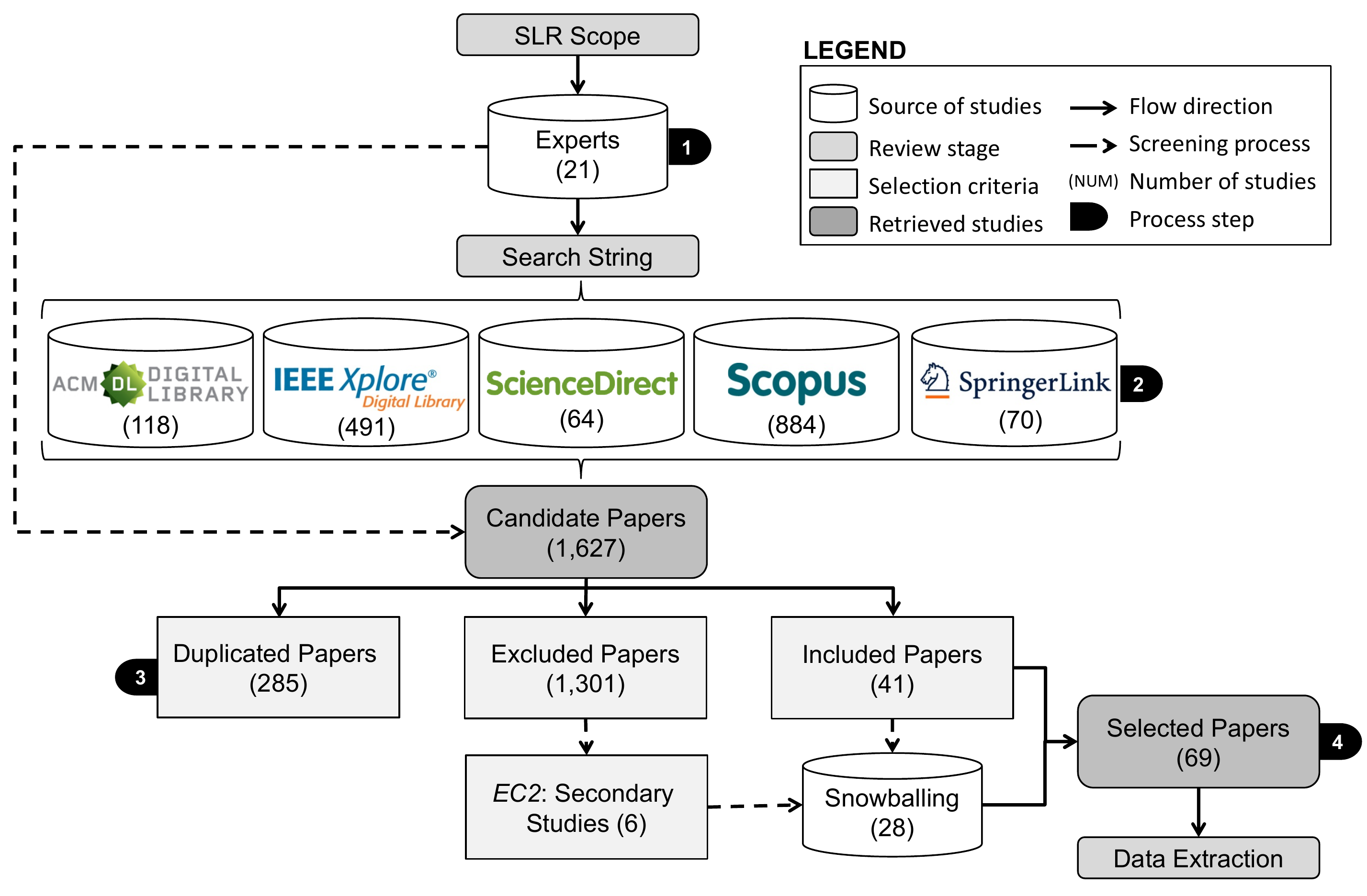}
	\caption{Flow of the paper selection process: papers retrieved from each digital library and details of the selection phases.}
    \label{fig:selection_process}
\end{figure}

The paper selection process is shown in Figure~\ref{fig:selection_process}. 
For identifying the candidate primary studies, we followed the review protocol described in Section~\ref{planning}.
As a first step, we defined our SLR scope.
Then, we used our expertise in the field to obtain an initial pool of relevant studies.
Based on this effort, we defined the search string (see Section~\ref{conducting}). 
As a second step, we applied the search string to the scientific digital libraries.
At the end of step 2, the initial search from all sources resulted in a total of 1,627 candidate papers, which includes the 21 papers from our initial pool of relevant studies.
Figure~\ref{fig:selection_process} shows the number of papers obtained from each digital library.

As a third step, after removing all duplicated papers (285), we carried out the selection process at the content level.
During this step, 1,301 papers were excluded, yielding a total of 41 selected papers for inclusion in the review process.  
A fourth step of the filtering was performed to select additional relevant papers through the snowballing process.
This step considered all included papers, as well as removed secondary studies, which resulted in the inclusion of 28 additional papers.
This resulted to the selection of 69 primary papers for data extraction.
The Web supplementary material~\cite{pereira2019} provides the results of the search procedure from each of these steps.


\section{Survey Pattern: Sampling, Measuring, Learning} \label{preliminaries}

\begin{figure}
	\centering
	\includegraphics[width=1\textwidth]{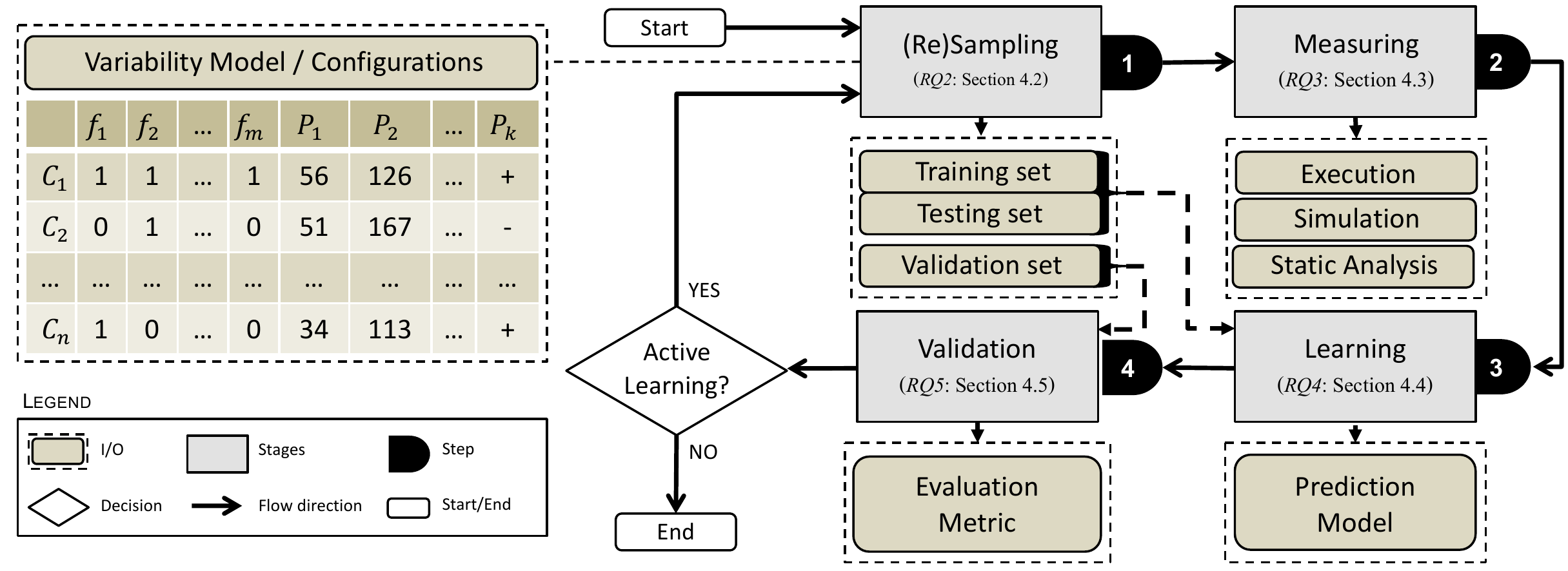}
	\caption{Employed ML stages to explore SPL configuration spaces.}
    \label{fig:ML_overview}
\end{figure}

Understanding how the system behavior
varies across a large number of variants of a configurable system is essential for supporting end-users to choose the desirable product. It is also useful of developers in charge of maintaining such software systems. 
In this context, machine learning-based techniques have been widely considered to predict configurations' behavior and assist stakeholders in making informed decisions.
Throughout our surveying effort, we have observed that such approaches follow a 4-stage process: \textit{(1)} sampling; \textit{(2)} measuring; \textit{(3)} learning; and \textit{(4)} validation. 
The stages are sketched in Fig.~\ref{fig:ML_overview}. 
The dashed-line boxes denote the inputs and outputs of each stage.
The process starts by building and measuring an initial sample of valid configurations.
The set of valid configurations in an SPL is predefined at design time through variability models usually expressed as a feature model \cite{pohl2005}.
Then, these measurements are used to learn a prediction model.
Prediction models help stakeholders to better understand characteristics of complex configurable software systems.
They try to describe the 
behavior of all valid configurations.
Finally, the validation step computes the accuracy of the prediction model.
In addition, some works use active learning~\cite{guo2017, jehooh2017, nair2018a, westermann2012, zuluaga2016, xi2004, grebhahn2017} to improve the sample in each interaction based on previous accuracy results until it reaches a configuration that has a satisfactory accuracy.
Next, we describe in detail each step.

\paragraph{Sampling.}
Decision makers may decide to select or deselect features to customize a system.
Each feature can have an effect on the system non-functional properties.
The quantification of the non-functional properties of each individual feature is not enough in most cases, as unknown feature interactions among configuration options may cause unpredictable measurements.
Interactions occur when combinations among features share a common component or require additional component(s). 
Thus, understanding the correlation between feature selections and system non-functional properties is important for stakeholders to be able to find an appropriated system variant that meets their requirements.
In Fig.~\ref{fig:ML_overview}, let $C=\{C_1, C_2,..., C_n\}$ be the set of $n$ valid configurations, and $C_i=\{f_1, f_2,..., f_m\}$ with $f_j \in \{0,1\}$ a combination of $m$ selected (\textit{i.e.}, 1) and deselected (\textit{i.e.}, 0) features. 
A straightforward way to determine whether a specific variant meets the requirements is to measure its target non-functional property $P$ and repeat the process for all $C$ variants of a system, and then \textit{e.g.} search for the cheapest configuration $C_i$ with $C_i \in C$.
However, it is usually unfeasible to benchmark all possible variants, due to the exponentially growing configuration space. 
ML techniques address this issue making use of a small measured \textit{sample} $S_C = \{s_1,..., s_k\}$ of configurations, where $S_C \subseteq C$, and the number of samples $k$ and the prediction error $\epsilon$ are minimal.
With the promise to balance measurement effort and prediction accuracy, several sample strategies have been reported in the literature (see Table~\ref{tab:RQ2} in Section~\ref{RQ2}).
For example, Siegmund et al.~\cite{siegmund2011, siegmund2012a, siegmund2013, siegmund2015} explore several ways of sampling configurations, in order to find the most accurate prediction model.
Moreover, several authors~\cite{guo2017, jehooh2017, nair2018a, westermann2012, zuluaga2016, xi2004, grebhahn2017} have tried to improve the prediction power of the model by updating an initial sample based on information gained from the previous set of samples through active learning (see Section~\ref{RQ4}).
The sample might be partitioned into training, testing and validation sets which are used to train and validate the prediction model (see Section~\ref{RQ5}).

\paragraph{Measuring.}
This stage measures the set of non-functional properties $\{p_1,..., p_l\}$ of a configuration sample $S_C = \{s_1,..., s_k\}$, where $p_1 = \{p_1(s_1),..., p_1(s_k)\}$.
Non-functional properties are measured either by \textit{execution}, \textit{simulation}, \textit{static analysis}, \textit{user feedback} or \textit{synthetic measurements}.

\textit{Execution} consists of executing the configuration samples and monitoring the measurements of non-functional properties at runtime.
Although execution is much more precise, it may incur in unacceptable measurement costs since it is often not possible to create suddenly potentially important scenarios in the real environment.
To overcome this issue, some approaches have adopted measurement by simulation.
Instead of measuring out of real executions of a system which may result in high costs or risks of failure, \textit{simulation} learns the model using offline environmental conditions that approximate the behavior of the real system faster and cheaper. 
The use of simulators allows stakeholders to understand the system behavior during early development stages and identify alternative solutions in critical cases. 
Moreover, simulators can be programmed offline which eliminates any downtime in online environments.
In addition to execution and simulation, \textit{static analysis} infers measurements only by examining the code, model, or documentation. 
For example, the non-functional property cost can be measured as the required effort to add a feature to a system under construction by analyzing the system cycle evolution, such as the number of lines of code, the development time, or other functional size metrics.
Although static analysis may not be always accurate, it is much faster than collecting data dynamically by execution and simulation. 
Moreover, partial configurations can be also measured.
Finally, instead of measuring configurations statically or dynamically, some authors also make use of either \textit{user feedback} (UF) or \textit{synthetic measurements} (SM).
In contrast to static and dynamic measurements, both approaches do not rely on systems artifacts.
\textit{User feedback} relies only on domain experts knowledge to label configurations (e.g., whether the configuration is acceptable or not).
 \textit{Synthetic measurements} are based on the use of learning techniques to generate artificial (non-)functional values to configurable systems~\cite{siegmund2017}. Researchers can use the THOR generator to mimic and experiment with properties of real-world configurable systems (e.g., performance distributions, feature interactions). 
 

\paragraph{Learning.}
In this stage, we learn a prediction model based on a given sample of measured configurations $P(S_C)$ to infer the behavior of non-measured configurations $P(C-S_C)$.
The sampling set $S_C$ is divided into a \textit{training set} $S_T$ and a \textit{validation set} $S_V$, where $S_C = S_T + S_V$.
The training set is used as input to learn a prediction model, \textit{i.e.} describe how configuration options and their interactions influence the behavior of a system. 
For parameter tuning, interactive sampling, and active learning, the training set is also partitioned into training and testing sets (see Section~\ref{RQ4}).

Some authors \cite{aken2017, jamshidi2017a, jamshidi2017b, jamshidi2018, valov2017} applied transfer learning techniques to accelerate the learning process.
Instead of building the prediction model from scratch, transfer learning reuses the knowledge gathered from samples of other relevant related sources to a target source. 
It uses a regression model that automatically captures the correlation between target and related systems.
Correlation means the common knowledge that is shared implicitly between the systems.
This correlation is an indicator that there is a potential to learn across the systems.
If the correlation is strong, the transfer learning method can lead to a much more accurate and reliable prediction model more quickly by reusing measurements from other sources. 

\paragraph{Validating.}
The validation stage quantitatively analysis the quality of the sample $S_T$ for prediction using an evaluation metric on the validation set $S_V$. 
To be practically useful, an ideal sample $S_T$ should result in a \textit{(i)} low prediction error; 
\textit{(ii)} small model size; 
\textit{(iii)} reasonable measurement effort.
The aim is to find as few samples as possible 
to yield an understandable and accurate ML model in short computation time. 
In Section~\ref{RQ5}, we detail the evaluation metrics used by the selected primary studies to compute accuracy.\\[-5pt]

Overall, 
exploring the configuration space based on a small sample of configurations is a critical step since in practice, the sample may not contain important feature interactions nor reflect the system real behavior accurately. 
To overcome this issue, numerous learning approaches have been proposed in the last years.
Next, we analyze the existing literature by investigating the more fine-grained characteristics of these four learning stages. 

\section{Results and Discussion of the Research Questions} \label{results}
In this section, we discuss the answers to our research questions defined in Section~\ref{introduction}.
In Section~\ref{RQ1}, we identify the main goal of the learning process. 
Next, in Sections~\ref{RQ2}, \ref{RQ3}, \ref{RQ4} and \ref{RQ5} we deeply analyze how each study address each learning stage defined in Section~\ref{preliminaries} to accomplish the goals described in Section~\ref{RQ1}.
Finally, Section~\ref{RQ6} presents a set of limitations and open challenges in order to point out possible future directions in this domain.

\subsection{RQ1: What are the concrete applications of learning software configuration spaces?}
\label{RQ1}
\begin{figure}
    \centering
    \begin{subfigure}[b]{0.3\textwidth}
        \includegraphics[width=\textwidth]{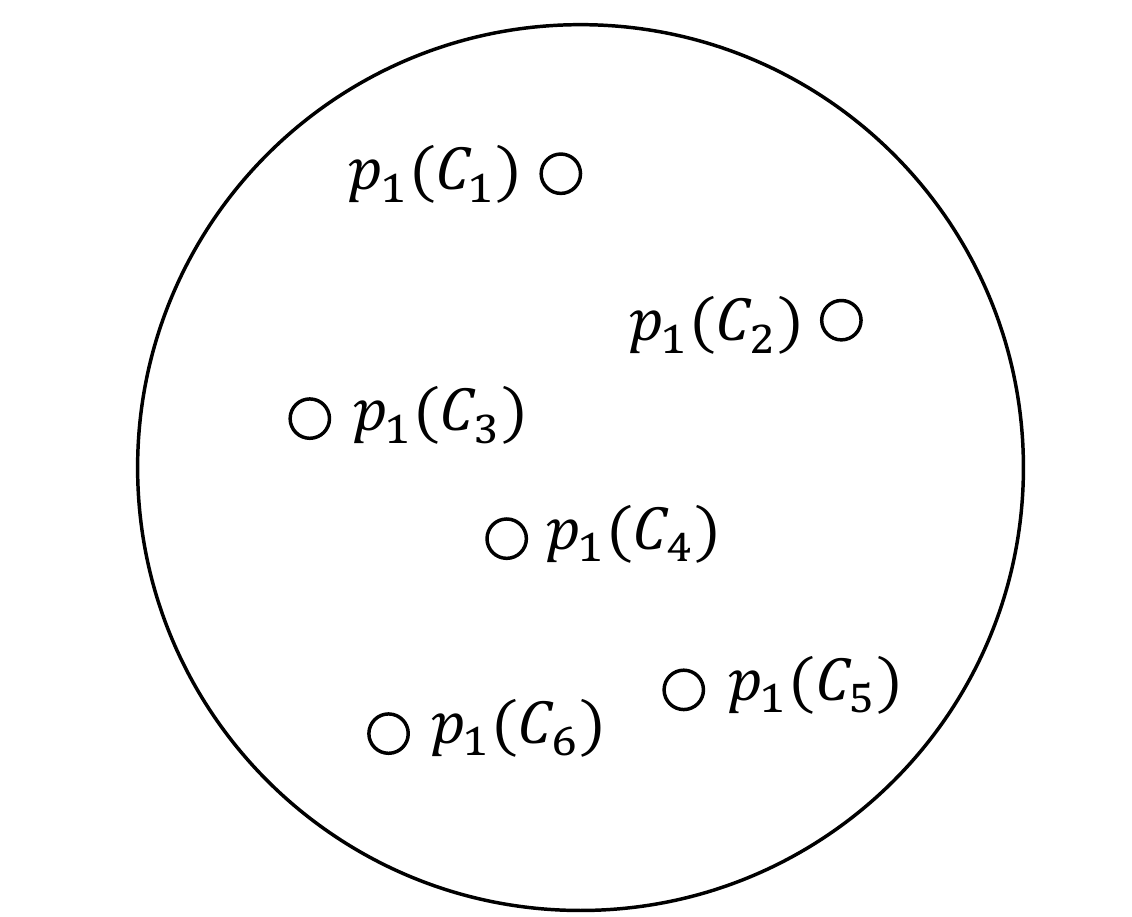}
        \caption{\textit{A1}: Pure Prediction.}
        \label{fig:A1}
    \end{subfigure}
    ~ 
    \begin{subfigure}[b]{0.3\textwidth}
        \includegraphics[width=\textwidth]{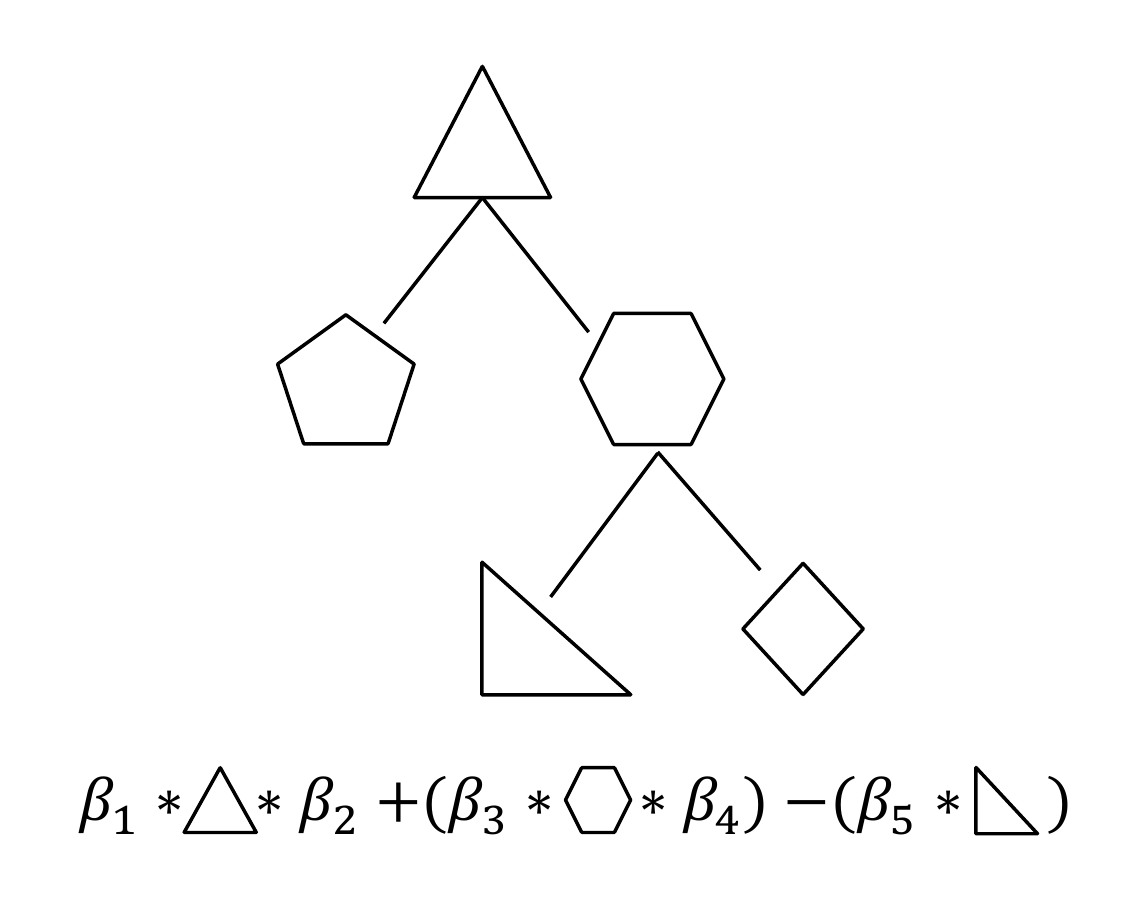}
        \caption{\textit{A2}: Interpretability.}
        \label{fig:A2}
    \end{subfigure}
    \begin{subfigure}[b]{0.3\textwidth}
        \includegraphics[width=\textwidth]{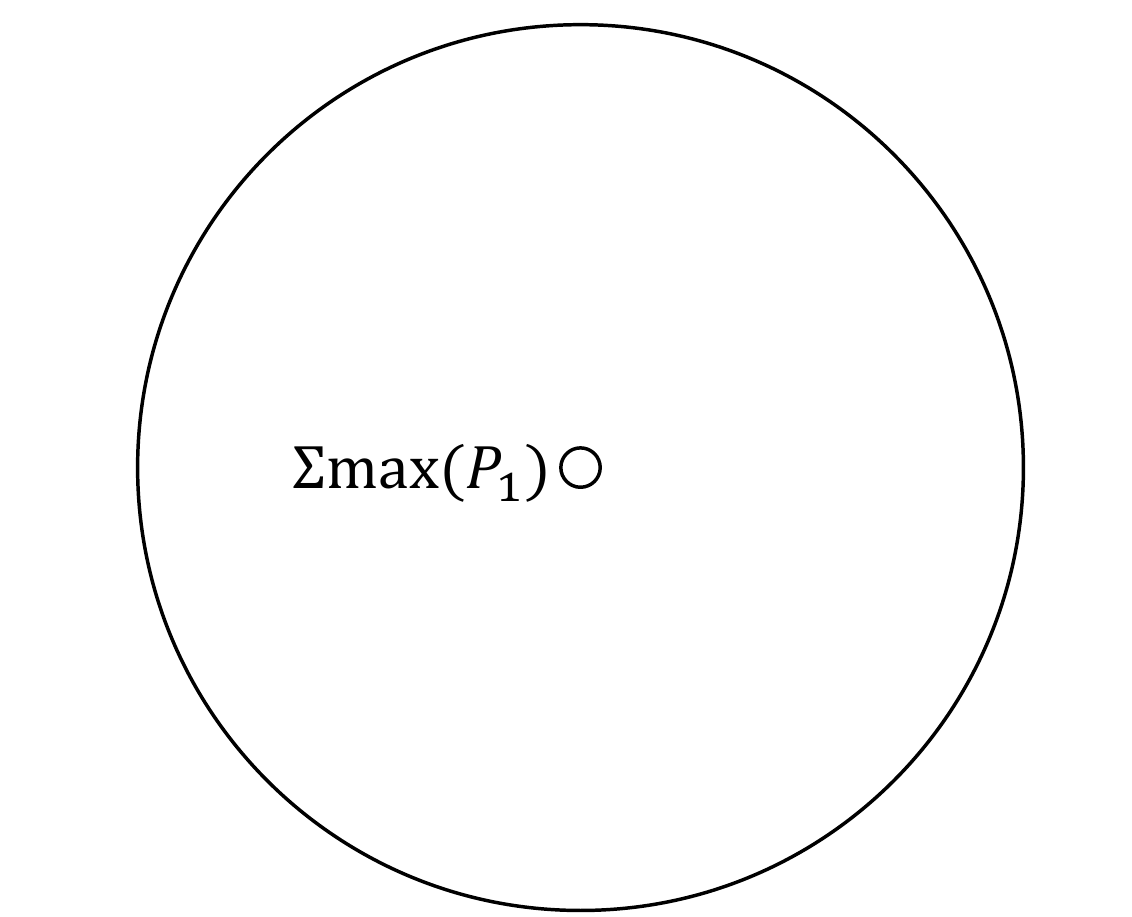}
        \caption{\textit{A3}: Optimization.}
        \label{fig:A3}
    \end{subfigure}

    \begin{subfigure}[b]{0.3\textwidth}
        \includegraphics[width=\textwidth]{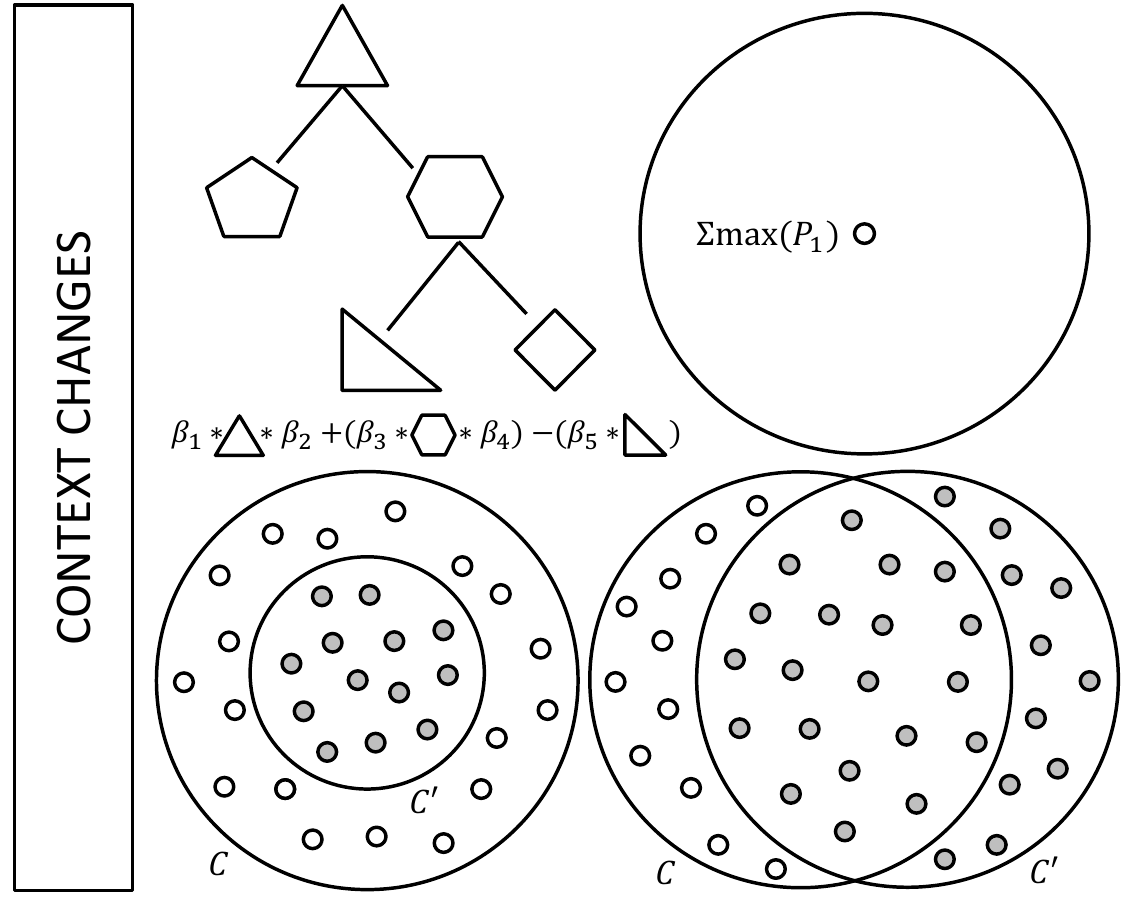}
        \caption{\textit{A4}: Dynamic Configuration.}
        \label{fig:A4}
    \end{subfigure}
    \begin{subfigure}[b]{0.3\textwidth}
        \includegraphics[width=\textwidth]{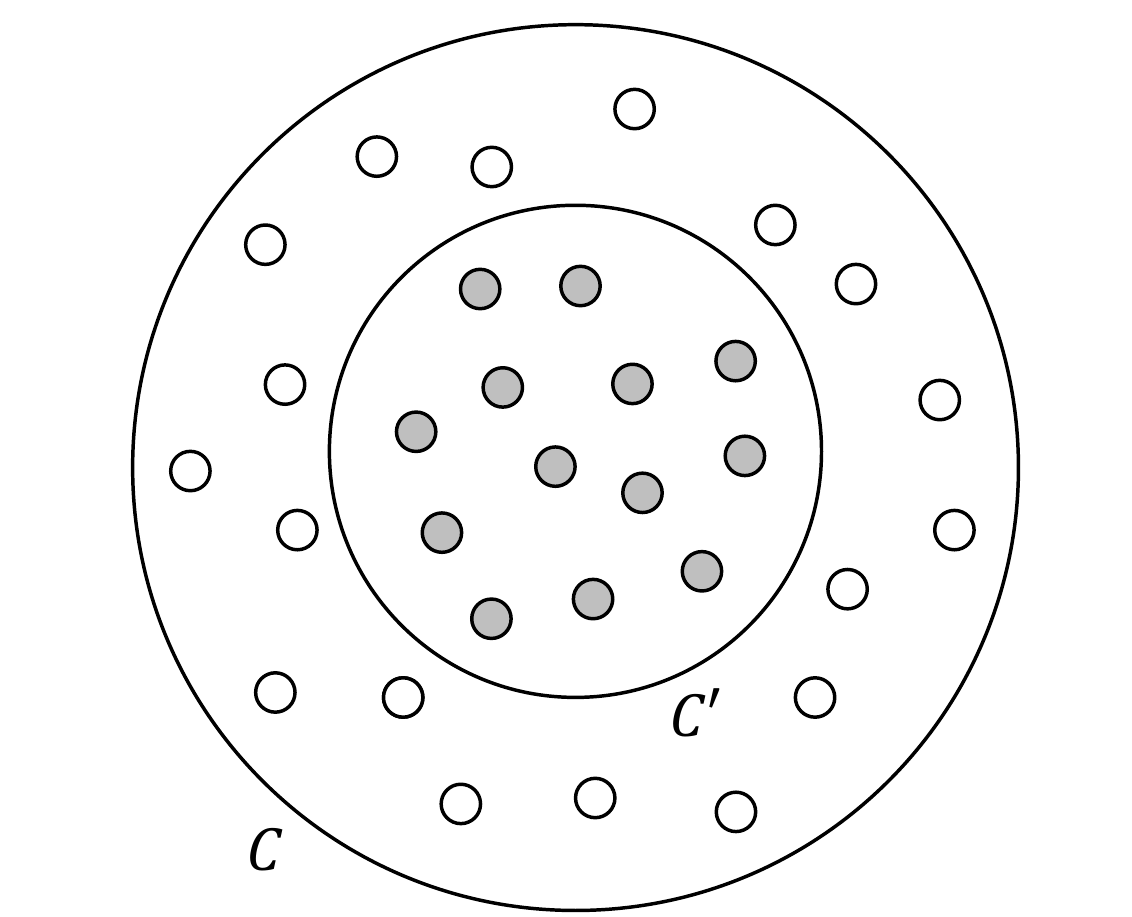}
        \caption{\textit{A5}: Mining Constraints}
        \label{fig:A5}
    \end{subfigure}
    \begin{subfigure}[b]{0.3\textwidth}
        \includegraphics[width=\textwidth]{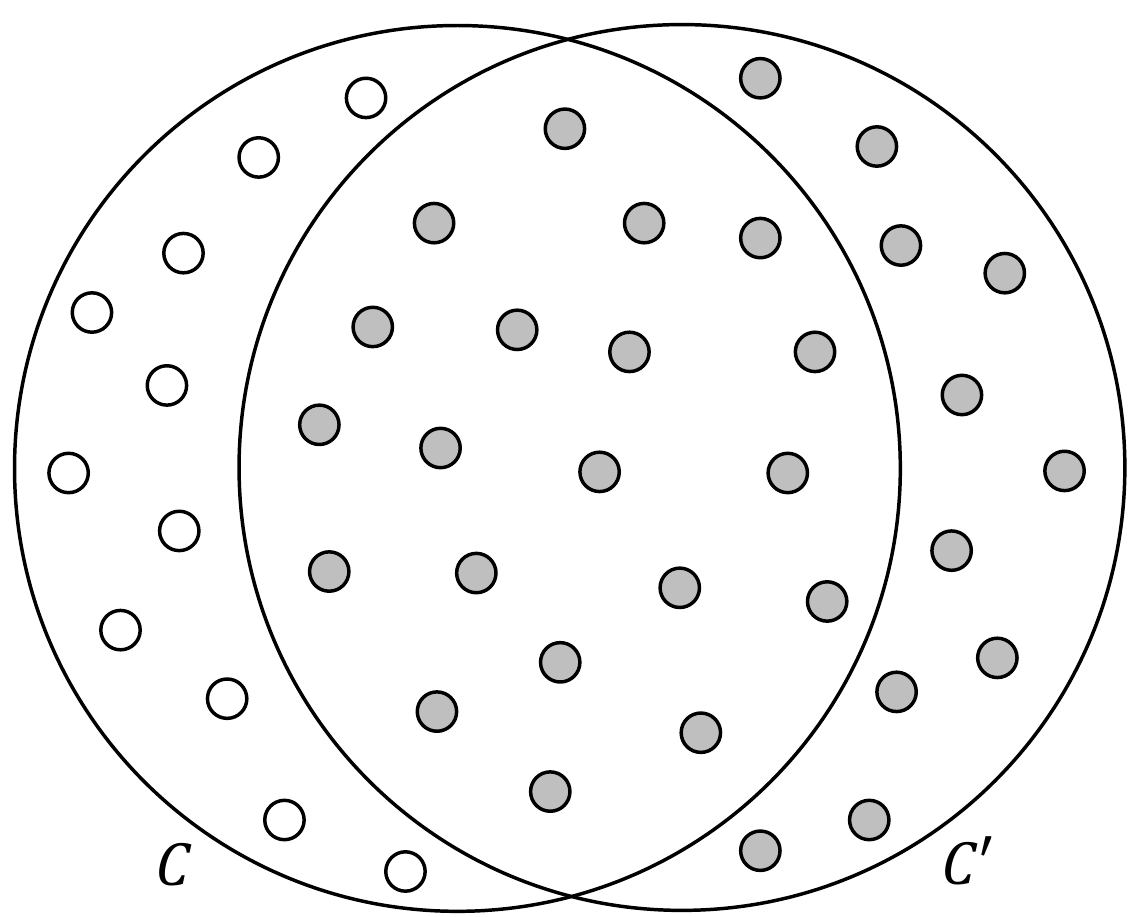}
        \caption{\textit{A6}: Evolution.}
        \label{fig:A6}
    \end{subfigure}
    \caption{Study application objective.}\label{fig:RQ1_applicability}
\end{figure}

In this section, we analyze the application objective of the selected studies since learning may have different practical interests and motivations. 
 Learning techniques have been used in the literature to target six different scenarios (see Figure~\ref{fig:RQ1_applicability} and Table~\ref{tab:RQ1}). We describe each individual application and give prominent examples of papers entering in this category. 
 


\begin{itemize}
    \item[\textit{A1}] \textit{Pure Prediction.}
    The aim is to accurately predict labels of unmeasured configurations. Labels can be qualitative (\textit{e.g.}, whether the software configuration has a defect) or quantitative (\textit{e.g.}, the execution time in seconds of a software configuration). The outcome is to associate through prediction some properties to all configurations of the space (see Figure~\ref{fig:A1}). 
    Guo et al.~\cite{guo2013} is a seminal paper with respect to the use of statistical learning for predicting performances.
    In this scenario, other factors such as the model comprehension and the computation time are less important.
    In some engineering contexts, the sole prediction of a property of a configuration has limited practical interest \emph{per se} and is sometimes used as a basis for targeting other applications (\textit{e.g.}, configuration optimization). 
    
    \item[\textit{A2}] \textit{Interpretability of configurable systems.}  
    Understanding the correlation between configuration options and system quality is important for a wide variety of tasks, such as optimization, program comprehension and debugging. 
    To this end, these studies aim at learning an accurate model that is fast to compute and simple to interpret (\textit{i.e.}, easy for developers to get an understandable overview of how configuration options interact, see Figure~\ref{fig:A2}). 
    For example, Kolesnikov et al.~\cite{kolesnikov2018} explore how so-called performance-influence models quantify options' influences and can be used to explain the performance behavior of a configurable system as a whole. 
    
    \item[\textit{A3}] \textit{Optimization.} 
    Instead of labeling all configurations,
    optimization approaches aim at finding a (near-)optimal valid configuration to best satisfy requirements (see Figure~\ref{fig:A3}). 
    According to Ochoa et al.~\cite{ochoa2018}, there are three types of stakeholders' requirements: resource constraints (threshold such as \texttt{response time} $<$ \texttt{1 hour}), stakeholders' preferences (\textit{e.g.}, \texttt{security} is extremely more preferable and relevant than \texttt{response time}) and optimization objectives (\textit{e.g.}, minimization of response time).
    Although the specification of requirements may reduce the configuration space~\citep{acher2013,bkak2016,ochoa2015,eichelberger2016,frantz2012}, searching for the most appropriate configuration is still an overwhelming task due to the combinatorial explosion. 
    Learning approaches exposed in \textit{e.g.}~\cite{jehooh2017, nair2017, nair2018a} propose a recursive search method to interactively add new samples to train the prediction model until it reaches a sample with an acceptable accuracy.
    \item[\textit{A4}] \textit{Dynamic Configuration.} 
    There are many dynamic systems (\textit{e.g.}, robotic systems) that aim to manage run-time adaptations of software to react to (uncertain) environmental changes. Without self-adaptation, requirements would be violated. 
    There are several works
    that explicitly define a set of adaptation rules in the variability model during design-time. 
    However, anticipating all contextual changes and defining appropriate adaptation rules earlier is often hard to domain engineers due to the huge variability space and the uncertainty of how the context may change at run-time.
    Therefore, approaches classified in this group use learning techniques to constantly monitor the environment to detect contextual changes that require the system to adapt (see Figure~\ref{fig:A4}).
    It learns the influences of contexts online in a feedback loop under time and resource constraints to reach one of the previous defined application objective.
    For example, learning techniques are used in the dynamic scenario to support the synthesis of contextual-variability models, including logical constraints~\cite{temple2017a, krismayer2017}. 
    Other approaches~\cite{jamshidi2017a, weckesser2018, samreen2016, porter2007, jamshidi2019} use learning techniques with the goal of finding a configuration that is optimal and consistent with the current, dynamic context.

    \item[\textit{A5}] \textit{Mining Constraints.}
    In a configurable system, not all combinations of options' values are possible (\textit{e.g.}, some options are mutually exclusive). Variability models are used to precisely define the space of valid configurations, typically through the specification of logical constraints among options.
    However, the identification of constraints is a difficult task and it is easy to forget a constraint leading to 
    configurations that do not compile, crash at run-time, or do not meet a particular resource constraint or optimization goal. 
    To overcome this issue, learning techniques can be used to discover additional constraints that would exclude unacceptable configurations. These studies seek an accurate and complete set of constraints to restrict the space of possible configurations. 
    Finally, it creates a new variability model by adding the identified constraints to the original model (see Figure~\ref{fig:A5}).
    Therefore, the aim is to accurately remove invalid configurations that were never derived and tested before.
    Mining constraints approaches work with qualitative properties, such as \textit{video quality}~\cite{temple2016, temple2018} and \textit{defects}~\cite{yilmaz2006, krismayer2017, gargantini2017, amand2019}. 
    
    \item[\textit{A6}] \textit{Evolution.} 
    In the evolution scenario, a configurable system will inevitably need to adapt 
    to satisfy real-world changes in external conditions, such as 
    changes in requirements, 
    design and performance improvements, and 
    changes in the source code.
    Thus, new configuration options become available and valid, while existing configurations may become obsolete and invalid (see Figure~\ref{fig:A6}).
    Consequently, the new variability model structure may influence certain non-functional properties.
    In this scenario, it is important to make sure that the learning stage is informed by \textit{evolution} about changes and the set of sample configurations is readjusted
    accordingly with the new variability model by excluding invalid configurations and 
    considering the parts of the configuration space not yet covered. 
    In this case, the measurements for each configuration that includes a new feature or a new interaction is updated. 
\end{itemize}

\begin{mdframed}[backgroundcolor=gray!10] 
	 The idea of "sampling, measuring, learning" has a wide application objective and can be used for supporting developers or end-users in six main different tasks: \textit{Pure Prediction}, \textit{Interpretability of Configurable Systems}, \textit{Optimization}, \textit{Dynamic Configuration}, \textit{Mining Constraints}, and \textit{SPL Evolution}. It is also possible to combine different tasks (\textit{e.g.}, mining constraints for supporting dynamic configuration~\cite{temple2017a, krismayer2017}).
\end{mdframed}

\begin{table} 
	\centering
    \begin{tabular}{p{0.06\textwidth} p{0.7\textwidth}}
    	\hline
		    \textbf{App.} & \textbf{Reference} \\ 
		\hline
            \textit{A1}	&	\cite{chen2005, guo2013, guo2017, sarkar2015, siegmund2011, siegmund2012a, siegmund2013, siegmund2015, valov2015, zhang2015, yilmaz2014, kolesnikov2017, couto2017, kaltenecker2019, jamshidi2018, siegmund2017, lillacka2013, siegmund2013b, queiroz2016, zhang2016, song2013}	\\
            \textit{A2}	&	\cite{jamshidi2017b, kolesnikov2018, sincero2010, valov2017, siegmund2017, duarte2018, etxeberria2014}	\\
            \textit{A3}	&	\cite{jehooh2017, murwantara2014, nair2017, nair2018a, siegmund2012b, westermann2012, martinez2018, nair2018c, aken2017, jamshidi2016, zuluaga2016, xi2004, alipourfard2017, zheng2007, siegmund2017, siegmund2008,ghamizi2019,grebhahn2017,saleem2015,bao2018,svogor2019,elafia2018,ding2015,thornton2013,xu2008,hutter2011, osogami2007}	\\
            \textit{A4}	&	\cite{jamshidi2017a, weckesser2018, samreen2016, porter2007, jamshidi2019, temple2017a, krismayer2017, sharifloo2016, siegmund2017, duarte2018, chen2009}	\\
            \textit{A5}	&	\cite{temple2016, acher2018, temple2018, yilmaz2006, gargantini2017, amand2019, siegmund2017, temple2017a, krismayer2017, safdar2017}	\\
            \textit{A6}	&	\cite{sharifloo2016, zheng2007, siegmund2017}	\\ 
        \hline
	\end{tabular}
	\caption{Applicability of selected primary studies. \textit{A1}: Pure Prediction; \textit{A2}: Interpretability of Configurable Systems; \textit{A3}: Optimization; \textit{A4}: Dynamic Configuration; \textit{A5}: Mining Constraints; \textit{A6}: SPL Evolution.}
	\label{tab:RQ1}
\end{table}


\subsection{RQ2: Which sampling methods are adopted when learning software configuration spaces?}
\label{RQ2}
In this section, we analyze the set of sampling methods used by learning-based techniques in the literature.
Table~\ref{tab:RQ2} shows the sample methods adopted by each study.
The first column is about the method 
and the second column identifies the study reference(s).
There are 23 high-level sample methods documented in the literature.
Next, we describe the particularities of the most used sample methods.

 \begin{table} 
 	\centering
        \begin{tabular}{p{0.38\textwidth} p{0.55\textwidth}}
     	\hline
 		    \textbf{Sample Method} & \textbf{Reference} \\ 
 		\hline
            Random	&	\cite{guo2013, guo2017, jamshidi2017a, jamshidi2017b, jehooh2017, nair2018a, temple2016, temple2017a, valov2015, weckesser2018, zhang2015, acher2018, jamshidi2019, temple2018, siegmund2015, valov2017, nair2017, nair2018c, kaltenecker2019, siegmund2017, alipourfard2017, siegmund2008, zhang2016, thornton2013, svogor2019, grebhahn2017, chen2009, osogami2007}	\\
            Knowledge-wise heuristic	&	\cite{siegmund2011, siegmund2013, siegmund2017, siegmund2008, xu2008}	\\
            Feature-coverage heuristic	&	\cite{kaltenecker2019, sarkar2015, siegmund2012a, siegmund2011, siegmund2013, siegmund2015, yilmaz2014, yilmaz2006, lillacka2013, grebhahn2017}	\\
            Feature-frequency heuristic	&	\cite{sarkar2015, siegmund2012a, siegmund2015, siegmund2011, siegmund2013, siegmund2012b, lillacka2013, grebhahn2017, etxeberria2014}	\\ 
            Family-based simulation	&	\cite{siegmund2013b}	\\
            Multi-start local search	&	\cite{hutter2011}	\\
            Plackett-Burman design	&	\cite{siegmund2015, grebhahn2017}	\\
            central composite design	&	\cite{grebhahn2017}	\\
            D-optimal design	&	\cite{grebhahn2017}	\\
            Breakdown	&	\cite{westermann2012}	\\
            Sequence type trees	&	\cite{krismayer2017}	\\
            East-west sampling	&	\cite{nair2018c}	\\
            Exemplar sampling	&	\cite{nair2018c}	\\
            Constrained-driven sampling	&	\cite{gargantini2017}	\\
            Diameter uncertainty strategy	&	\cite{zuluaga2016}	\\
            Historical dataset of configurations	&	\cite{aken2017, saleem2015}	\\
            Latin hypercube sampling 	&	\cite{xi2004, jamshidi2016, bao2018}	\\
            Neighborhood sampling	&	\cite{porter2007, jamshidi2018}	\\
            Input-based clustering	&	\cite{ding2015}	\\
            Distance-based sampling	&	\cite{kaltenecker2019, ghamizi2019}	\\
            Genetic sampling	&	\cite{martinez2018, jamshidi2016, safdar2017}	\\
            Interaction tree discovery	&	\cite{song2013}	\\
            Arbitrarily chosen	&	\cite{chen2005, murwantara2014, sincero2010, samreen2016, amand2019, queiroz2016, duarte2018, chen2009}	\\
         \hline
     \end{tabular}
 	\caption{Sample methods reported in the literature.}
	\label{tab:RQ2}
 \end{table}

\textbf{Random sampling.} Several studies have used random sampling~\cite{guo2013, guo2017, jamshidi2017a, jamshidi2017b, jehooh2017, nair2018a, temple2016, temple2017a, valov2015, weckesser2018, zhang2015, acher2018, jamshidi2019, temple2018, siegmund2015, valov2017, nair2017, nair2018c, kaltenecker2019} with different notions of randomness.

Guo et al.~\cite{guo2013} consider four sizes of random samples for training: $N$, $2N$, $3N$, and $M$, where $N$ is the number of features of a system, and $M$ is the number of minimal valid configurations covering each pair of features.
They choose size $N$, $2N$, and $3N$, because measuring a sample whose size is linear in the number of features is likely feasible and reasonable in practice, given the high cost of measurements by execution (see Section~\ref{RQ3}).
Valov et al. and Guo et al.~\cite{guo2017, valov2015, valov2017} use a random sample to train, but also to cross-validate their machine learning model.
Several works~\cite{nair2018a, nair2017, jehooh2017, zhang2015} seek to determine the number of samples in an adaptive, progressive sampling manner and a random strategy is usually employed. 
Nair et al.~\cite{nair2018a, nair2017} and Jehooh et al.~\cite{jehooh2017} aim at optimizing a configuration.
At each iteration, they randomly add an arbitrary number of configurations to learn a prediction model until they reach a model that exhibits a desired satisfactory \textit{accuracy} (see Section~\ref{RQ4}). They consider several sizes of samples from tens to thousands of configurations.
To focus on a reduced part of the configuration space, Nair et al.~\cite{nair2018a} and Jehooh et al.~\cite{jehooh2017} determine statistically significant parts of the configuration space that contribute to good performance through active learning. 
In order to have a more representative sample, Valov et al.~\cite{valov2017} adopted stratified random sampling.
This sampling strategy exhaustively divides a sampled population into mutually exclusive subsets of observations before performing actual sampling. 

Prior works~\cite{guo2013, guo2017, valov2017} relied on the random selection of features to create a configuration, followed by a filter to eliminate invalid configurations (\textit{a.k.a}, pseudo-random sampling).
Walker's alias sampling~\cite{valov2017} is an example of pseudo-random sampling.
Quasi-random sampling (\textit{e.g.}, sobol sampling) is similar
to pseudo-random sampling, however they are specifically designed to cover a sampled population more uniformly~\cite{alipourfard2017, valov2017}.
However, pseudo-random sampling may result in too many invalid configurations, which makes this strategy inefficient.
To overcome this issue, several works~\cite{nair2018a, jehooh2017, acher2018, zhang2015, temple2016, temple2017a, temple2018, weckesser2018, jamshidi2017a, jamshidi2017b, jamshidi2019} use solver-based sampling techniques (\textit{a.k.a.}, true random sampling).

\textbf{Sampling and heuristics.} Instead of randomly choosing configurations as part of the sample, several heuristics have been developed. The general motivation is to better cover features and features' interactions as part of the sample. The hope is to \emph{better capture the essence of the configuration space with a lower sampling size}. We describe some heuristics hereafter. 

\paragraph{Knowledge-wise heuristic.} 
This heuristic selects a sample of configurations based on its influence on the target non-functional properties.
Siegmund et al.~\cite{siegmund2011, siegmund2013} sampling method measures each feature in the feature model plus all known feature interactions defined by a domain expert. 
Experts detect feature interactions by analyzing the specification of features, implementation assets, and source code, which require substantial domain knowledge and exhaustive analysis. 
SPLCoqueror\footnote{http://fosd.de/SPLConqueror.} provides to stakeholders an environment in which they can document and incorporate known feature interactions. 
For each defined feature interaction, a single configuration is added to the set of sample for measurement.
THOR~\cite{siegmund2017} is a generator for synthesizing synthetic yet realistic variability models where users (researchers) can specify the number of interactions and the degree of interactions. 

\paragraph{Feature-coverage heuristic.} 
To automatically detect all first order feature interactions, Siegmund et al.~\cite{siegmund2011, siegmund2012a, siegmund2013, siegmund2015} use a pair-wise measurement heuristic.
This heuristic assumes the existence of a feature interaction between each pair of features in an SPL.
It includes a minimal valid configuration for each pair of features being selected. 
Pair-wise requires a number of measurements that is quadratic in the number of optional features.
Some authors~\cite{sarkar2015, kaltenecker2019, lillacka2013} also use a 3-wise feature coverage heuristic to discover interactions among 3 features. 
Siegmund et al.~\cite{siegmund2012a} propose a \textit{3rd-order coverage heuristic} that considers each minimal valid configuration where three features interact pair-wise among them (adopted by~\cite{lillacka2013}).
They also propose the idea that there are hot-spot features that represent a performance-critical functionality within a system. These hot-spot features are identified by counting the number of interactions per features from the feature-coverage and higher-order interaction heuristics.
Yilmaz et al.~\cite{yilmaz2014} adopted even a 4-wise feature coverage heuristic, and Yilmaz et al.~\cite{yilmaz2006} a 5 and 6-wise heuristic.
As there are n-th order feature coverage heuristics, the sample set might be likely unnecessarily large which increases measurement effort substantially.
However, not every generated sample contains features that interacts with each other. 
Thus, the main problem of this strategy is that it requires prior knowledge to select a proper coverage criterion.
To overcome this issue, state-of-the-art approaches might use the interaction-wise heuristic to fix the size of the initial sample to the number of features or potential feature interactions of a system~\cite{siegmund2017}.


\paragraph{Feature-frequency heuristic.} 
The feature-frequency heuristic considers a set of valid configurations in which each feature is selected and deselected, at least, once.
Sarkar et al.~\cite{sarkar2015} heuristic counts the number of times a feature has been selected and deselected. Sampling stops when the counts of features selected and deselected is, at least, at a predefined threshold.
Nair et al.~\cite{nair2017} analysis the number of samples required by using the previous heuristic~\cite{sarkar2015} against a rank-based random heuristic. 
Siegmund et al.~\cite{siegmund2011, siegmund2012a, siegmund2012b, siegmund2013, siegmund2015} quantify the influence of an individual feature by computing the delta of two minimal configurations with and without the feature. 
They then relate to each feature a minimum valid configuration that contains the current feature, which requires the measurement of a configuration per feature.
Hence, each feature can exploit the previously defined configuration to compute its delta over a performance value of interest.
In addition, to maximize the number of possible interactions, Siegmund et al.~\cite{siegmund2015} also relate to each feature a maximal valid configuration that contains the current feature.\\[-5pt]

There are several others sampling heuristics, such as \textit{Plackett-Burman design}~\cite{siegmund2015, grebhahn2017} for reasoning with numerical options; \textit{Breakdown}~\cite{westermann2012} (random breakdown, adaptive random breakdown, adaptive equidistant breakdown) for breaking down (in different sectors) the parameter space; \textit{Constrained-driven sampling}~\cite{gargantini2017} (constrained CIT, CIT of constraint validity, constraints violating CIT, combinatorial union, unconstrained CIT) to verify the validity of combinatorial interaction testing (CIT) models; and many others (see Table~\ref{tab:RQ2}).

\textbf{Sampling and transfer learning.} Jamshidi et al.~\cite{jamshidi2017a, jamshidi2017b, jamshidi2018} aim at applying transfer learning techniques to learn a prediction model (see Section~\ref{RQ4}).
Jamshidi et al.~\cite{jamshidi2017a} consider a combination of random samples from \textit{target} and \textit{source} systems for training: \{0\%, 10\%, ..., 100\%\} from the total number of valid configurations of a source system, and \{1\%, 2\%,..., 10\%\} from the total number of valid configurations of a target system. 
In a similar scenario, Jamshidi et al.~\cite{jamshidi2017b} randomly select an arbitrary number of valid configurations from a system before and after environmental changes (\textit{e.g.}, using different hardware, different workloads, and different versions of the system).
In another scenario, Jamshidi et al.~\cite{jamshidi2018} use transfer learning to sample.
Their sampling strategy, called L2S, exploits common similarities between source and target systems.
L2S progressively learns the interesting 
regions of the target configuration space, based on transferable knowledge from the source.

\textbf{Arbitrarily chosen sampling.} 
Chen et al.~\cite{chen2005} and Murwantara et al.~\cite{murwantara2014} have arbitrarily chosen a set of configurations as their sample is based on their current available resources. 
Sincero et al.~\cite{sincero2010} use a subset of valid configurations from a preliminary set of (de)selected features. 
This sample is arbitrarily chosen by domain experts based on the use of features which will probably have a high influence on the properties of interest. 
In the context of investigating temporal variations, Samreen et al.~\cite{samreen2016} consider on-demand instances at different times of the day over a period of seven days with a delay of ten minutes between each pair of runs. 
In a similar context, Duarte et al.~\cite{duarte2018} and Chen et al.~\cite{chen2009} also sample configurations under different workloads (e.g., active servers and requests per second) at different times of the day.
An important insight is that there are engineering contexts in which the sampling strategy is imposed and can hardly be controlled. 

\textbf{All configurations (no sampling).}
Sampling is not applicable for four of the selected primary studies~\cite{kolesnikov2018, kolesnikov2017, couto2017, zheng2007}, mainly for experimental reasons.
For example, Kolesnikov et al.~\cite{kolesnikov2018, kolesnikov2017} consider all valid configurations in their experiments and use an established learning technique to study and analyze the trade-offs among prediction error, model size, and computation time of performance-prediction models.
For the purpose of their study, they were specifically interested to explore the evolution of the model properties to see the maximum possible extent of the corresponding trade-offs after each iteration of the learning algorithm. 
So, they performed a whole-population exploration of the largest possible learning set (\textit{i.e.}, all valid configurations).
In a similar scenario, Kolesnikov et al.~\cite{kolesnikov2017} explored the use of control-flow feature interactions to identify potentially interacting features based on detected interactions from performance prediction techniques using performance measurements).
Therefore, they also performed a whole-population exploration of all valid configurations.

\textbf{Reasoning about configurations validity.} Sampling is realized either out of an enumerated set of configurations (e.g., the whole ground truth) or a variability model (e.g., a feature model). The former usually assumes that configurations of the set are logically valid. The latter is more challenging, since picking a configuration boils down to resolve a satisfiability or constraint problem. 

Acher et al.~\cite{acher2018} and Temple et al.~\cite{temple2016, temple2017a, temple2018} encoded variability models as \textit{Constraint Programming} (CSP) by using the Choco solver, while Weckesser et al.~\cite{weckesser2018} and Siegmund et al.~\cite{siegmund2017} employed SAT solvers. 
Constraint solver may produce clustered configurations with similar features due to the way solvers enumerate solutions (\textit{i.e.}, often the sample set consists of the closest $k$ valid configurations).
Therefore, these strategies do not guarantee true randomness as in pseudo-random sampling. 
Moreover, using CSP and SAT solvers to enumerate all valid configurations are often impractical~\cite{mendonca2008, pohl2011}.  
Thus, Jehooh et al.~\cite{jehooh2017} encoded variability models as \textit{Binary Decision Diagrams} (BDDs)~\cite{akers1978}, for which counting the number of valid configurations is straightforward.
Given the number of valid configurations $n$, they randomly and uniformly select the $k^{th}$ configuration, where $k \in \{1...n\}$, \textit{a.k.a.} randomized true-random sampling. 
Kaltenecker et al.~\cite{kaltenecker2019} perform a comparison among pseudo-random sampling, true-random sampling, and randomized true-random sampling.

\emph{Reasoning with numerical options.} 
Ghamizi et al.~\cite{ghamizi2019} transform numeric and enumerated attributes into alternative Boolean features to be handle as binary option.
Temple et al.~\cite{temple2016, temple2017a, temple2018} adopted random sampling of numeric options, \textit{i.e.} real and integer values.
First, their approach randomly selects a value for each feature within the boundaries of its domain. 
Then, it propagates the values to other features with a solver to avoid invalid configurations. 
In a similar scenario, Siegmund et al.~\cite{siegmund2015} and Grebhahn et al.~\cite{grebhahn2017} adopted pseudo-random sampling of numeric options.
Siegmund et al.~\cite{siegmund2015} claim that it is very unusual that numeric options have value ranges with undefined or invalid holes and that constraints among numeric options appear rarely in configurable systems. 
Grebhahn et al.~\cite{grebhahn2017} adopted different reasoning techniques over binary and numeric options, then they compute the Cartesian product of the two sets to create single configurations used as input for learning. 
In the scenario of reasoning with numerical options, Amand et al.~\cite{amand2019} arbitrarily select equidistant parameters values.\\

\begin{mdframed}[backgroundcolor=gray!10] 
    Though random sampling is a widely used baseline, numerous other sampling algorithms and heuristics have been devised and described. There are different trade-offs to find when sampling configurations \textit{(1)} minimization of invalid configurations due to constraints' violations among options; \textit{(2)} minimization of the cost (\textit{e.g.}, size) of the sample; \textit{(3)} generalization of the sample to the whole configuration space. The question of a one-size-fits-all sampling strategy remains open and several factors are to be considered (targeted application, subject systems, functional and non-functional properties, presence of domain knowledge, etc.).
\end{mdframed}


\subsection{RQ3: Which techniques are used to gather measurements of functional and non-functional properties of configurations?}
\label{RQ3}
The measuring step takes as input a sample of configurations and measures, for each configuration, their functional properties or \textit{non-functional properties} (NFPs). 
In this section, we investigate how the measurement procedures are technically realized. 




Most proposals consider NFPs, such as elapsed time in seconds. 
In essence, NFPs 
 consist of a \textit{name}, a \textit{domain}, a \textit{value}, and a \textit{unit}~\cite{benavides2005}.
The domain type of a NFP can be either quantitative (\textit{e.g.}, real and integer) or qualitative (\textit{e.g.}, string and Boolean).
\textit{Quantitative (QT)} properties are typically represented as a numeric value, thus they can be measured on a metric scale, \textit{e.g.}, the configuration is executed in 13.63 seconds. 
\textit{Qualitative (QL)} properties are represented using an ordinal scale, such as low $(-)$ and high $(+)$; \textit{e.g.}, the configuration produces a high \texttt{video quality}. 

\begin{figure}
	\centering
	\includegraphics[width=0.55\textwidth]{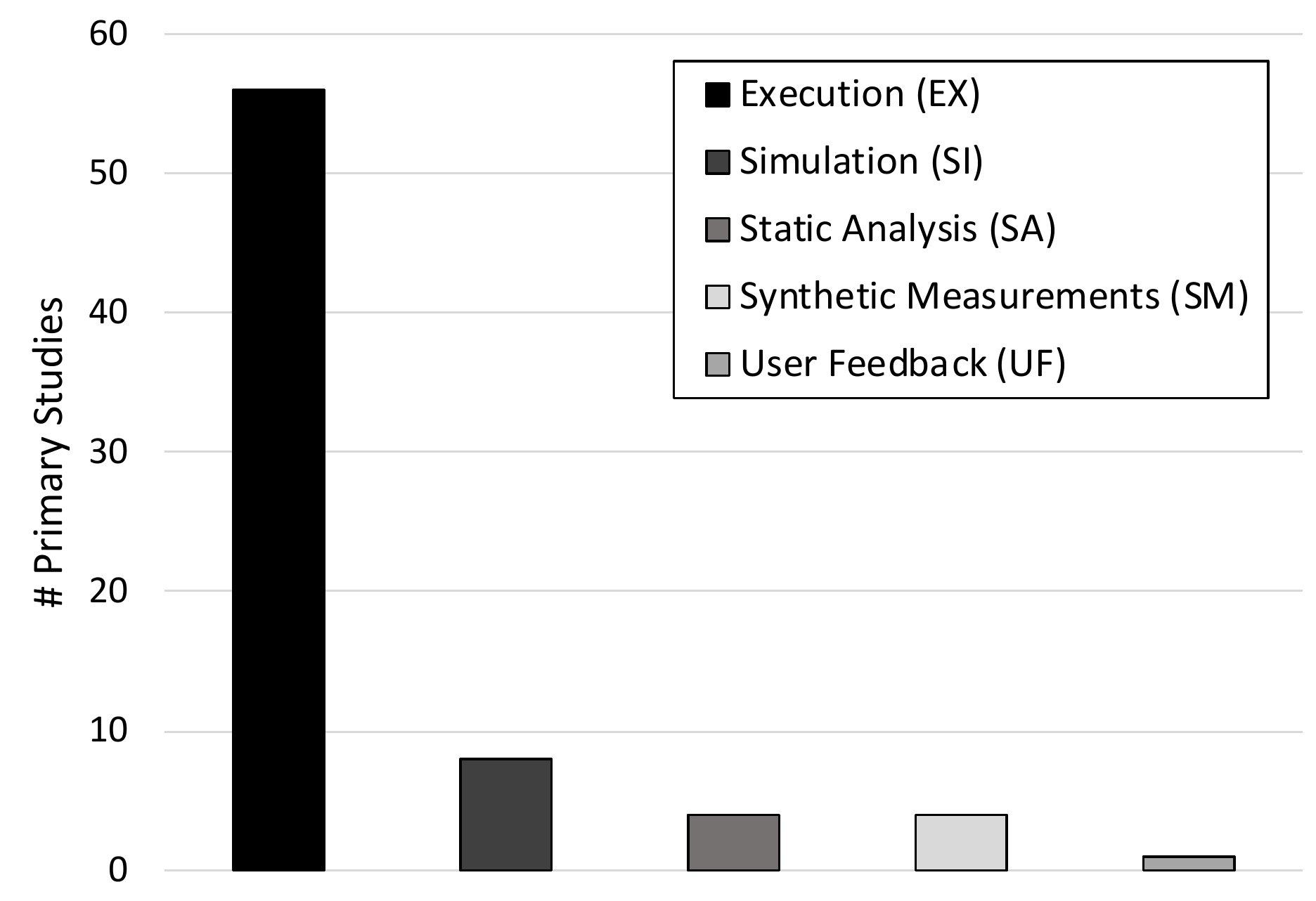}
	\caption{Strategies applied to measure the sample of configurations.}
    \label{fig:RQ3_strategy}
\end{figure}

As described in Section~\ref{preliminaries}, measurements can be obtained through five different strategies: \textit{execution} (EX), \textit{static analysis} (SA), \textit{simulation} (SI), \textit{user feedback} (UF), and \textit{synthetic measurements} (SM). Fig.~\ref{fig:RQ3_strategy} shows that automated execution is by far the most used technique to measure configuration properties.  
 
In Table~\ref{tab:RQ3_systems}, we present the subject systems used in the literature together with NFPs. 
The first column identifies the references.
The second and third columns describe the name and domain of the system, respectively.
The fourth column points out the measured NFP(s).
There are 71 real-world SPLs documented in the literature.
They are of different sizes, complexities, implemented in different programming languages (C, C++, and Java), varying implementation techniques (conditional compilation and feature-oriented programming), and from several different application domains (\textit{e.g.}, operating systems, video encoder, database system) and developers (both academic and industrial).
Therefore, they cover a broad spectrum of scenarios.
It is important to mention that the same subject systems may differ in the number of features, feature interactions, and in the number of valid configurations -- the experimental setup is simply different.


\begin{small}
\begin{longtable}[t]{p{0.18\textwidth} p{0.17\textwidth} p{0.22\textwidth} p{0.32\textwidth}} 
	\caption{Subject Systems supported in the literature.} \\
	
    	\hline
		    \textbf{References} & \textbf{Name} & \textbf{Domain} & \textbf{Non-Functional Properties} \\ 
        \hline
        \endhead
        
		    \cite{amand2019}	&	Thingiverse’s	&	3D printer	&	defects	\\
            \cite{xi2004}	&	IBM WebSphere	&	Application server	&	throughput	\\
            \cite{guo2017,kolesnikov2018, siegmund2015}	&	Clasp	&	ASP solver	&	response time	\\
            \cite{zuluaga2016}	&	SNW	&	Asset management	&	area and throughput	\\
            \cite{ding2015}	&	Binpacking	&	Binpacking algorithm	&	execution
            time and accuracy	\\
            \cite{jamshidi2018}	&	XGBoost	&	Boosting algorithms	&	training time	\\
            \cite{sharifloo2016}	&	SaaS system	&	Cloud computing	&	response time	\\
            \cite{ding2015}	&	Clustering	&	Clustering algorithm	&	execution
            time and accuracy	\\
            \cite{guo2017,kolesnikov2018, siegmund2015, siegmund2013b}	&	AJStats	&	Code analyzer	&	response time	\\
            \cite{nair2018a, jamshidi2017b, siegmund2015, jamshidi2018}	&	SaC	&	Code analyzer	&	I/O time, response time	\\
            \cite{kaltenecker2019}	&	POLLY	&	Code optimizer	&	runtime	\\
            \cite{gargantini2017}	&	Libssh	&	Combinatorial model	&	defects	\\
            \cite{gargantini2017}	&	Telecom	&	Communication system	&	defects	\\
            \cite{guo2013,guo2017,jehooh2017,kolesnikov2018,nair2018a,sarkar2015,siegmund2012a,siegmund2013,temple2017a, siegmund2015, zhang2015, nair2018c, kaltenecker2019, zuluaga2016, zhang2016}	&	LLVM	&	Compiler	&	memory footprint, performance, response time, code complexity, compilation time	\\
            \cite{lillacka2013}	&	Compressor SPL	&	Compression library	&	compression time, memory usage and compression ratio	\\
            \cite{kaltenecker2019}	&	7Z	&	Compression library	&	Compression time	\\
            \cite{siegmund2015, guo2017,kolesnikov2018,nair2017, kaltenecker2019}	&	LRZIP	&	Compression library	&	compressed size, compression time, compilation time	\\
            \cite{siegmund2013}	&	RAR	&	Compression library	&	code complexity	\\
            \cite{valov2017}	&	XZ	&	Compression library	&	compression time	\\
            \cite{siegmund2011,siegmund2012b,siegmund2013, siegmund2013b}	&	ZipMe	&	Compression library	&	memory footprint, performance, code complexity, time	\\
            \cite{murwantara2014}	&	WordPress	&	Content management	&	CPU power consumption	\\
            \cite{siegmund2011,siegmund2012b,siegmund2013, siegmund2008}	&	LinkedList	&	Data structures	&	memory footprint, performance, maintainability, binary size	\\
            \cite{siegmund2013}	&	Curl	&	Data transfer	&	code complexity	\\
            \cite{siegmund2013, nair2017}	&	Wget	&	Data transfer	&	memory footprint, code complexity	\\
            \cite{aken2017}	&	Actian Vector	&	Database system	&	runtime	\\
            \cite{jamshidi2017a}	&	Apache Cassandra	&	Database system	&	latency	\\
            \cite{guo2013,guo2017,jehooh2017,kolesnikov2018,nair2017,sarkar2015,siegmund2012a,siegmund2013,temple2017a, siegmund2011,siegmund2012b, siegmund2015, zhang2015, nair2018c, kaltenecker2019, siegmund2008, zhang2016}	&	Berkeley DB	&	Database system	&	I/O time, memory footprint, performance, response time, code complexity, maintainability, binary size	\\
            \cite{siegmund2008}	&	FAME-DBMS	&	Database system	&	maintainability, binary size, performance	\\
            \cite{yilmaz2014, aken2017, zheng2007, song2013}	&	MySQL	&	Database system	&	defects, throughput, latency	\\
            \cite{aken2017}	&	Postgres	&	Database system	&	throughput, latency	\\
            \cite{siegmund2011,siegmund2012b,siegmund2013}	&	Prevayler	&	Database system	&	memory footprint, performance	\\
            \cite{guo2013,guo2017,nair2017,sarkar2015,siegmund2012a,siegmund2013,temple2017a, siegmund2011,siegmund2012b,siegmund2013, jamshidi2017b, valov2017, nair2018c, kolesnikov2017}	&	SQLite	&	Database system	&	memory footprint, performance, response time, code complexity, runtime	\\
            \cite{chen2005}	&	StockOnline	&	Database system	&	response time	\\
            \cite{bao2018}	&	Kafka	&	Distributed systems	&	throughput	\\
            \cite{ghamizi2019}	&	DNN	&	DNNs algorithms	&	accuracy of predictions	\\
            \cite{acher2018}	&	Curriculum vitae	&	Document	&	number of pages	\\
            \cite{acher2018}	&	Paper	&	Document	&	number of pages	\\
            \cite{duarte2018}	&	RUBiS	&	E-commerce application	&	response time	\\
            \cite{siegmund2013b}	&	EMAIL	&	E-mail client	&	time	\\
            \cite{kolesnikov2017}	&	MBED TLS	&	Encryption library	&	response time	\\
            \cite{westermann2012}	&	SAP
             ERP	&	Enterprise Application	&	response time	\\
            \cite{nair2018a}	&	noc-CM-log	&	FPGA	&	CPU power consumption, runtime	\\
            \cite{nair2018a}	&	sort-256	&	FPGA	&	area, throughput	\\
            \cite{etxeberria2014}	&	E-Health System	&	Health	&	response time	\\
            \cite{guo2017, siegmund2015,temple2017a, kaltenecker2019}	&	$HIPA^{cc}$	&	Image processing	&	response time	\\
            \cite{couto2017}	&	Disparity SPL	&	Image processing	&	energy consumption	\\
            \cite{siegmund2011,siegmund2012b,siegmund2013}	&	PKJab	&	Instant messenger	&	memory footprint, performance	\\
            \cite{hutter2011}	&	IBM ILOG CPLEX	&	Integer solver	&	runtime	\\
            \cite{westermann2012}	&	SPECjjbb2005	&	Java Server	&	response time, throughput	\\
            \cite{thornton2013}	&	WEKA	&	Learning algorithm	&	accuracy of predictions	\\
            \cite{ding2015}	&	SVD	&	Linear algebra	&	execution
            time and accuracy	\\
            \cite{nair2018a}	&	Trimesh	&	Mesh solver	&	iterations, response time	\\
            \cite{siegmund2013b}	&	MBENCH	&	Micro benchmark	&	time	\\
            \cite{yilmaz2006, porter2007}	&	ACE+TAO system	&	Middleware software	&	defects	\\
            \cite{siegmund2011,siegmund2012b,siegmund2013}	&	SensorNetwork	&	Network simulator	&	memory footprint, performance	\\
            \cite{weckesser2018}	&	Simonstrator	&	Network simulator	&	latency	\\
            \cite{zuluaga2016}	&	NoC	&	Network-based system	&	energy and runtime	\\
            \cite{ding2015}	&	Helmholtz 3D	&	Numerical analysis 	&	execution
            time and accuracy	\\
            \cite{ding2015}	&	Poisson 2D	&	Numerical analysis 	&	execution
            time and accuracy	\\
            \cite{sincero2010, siegmund2011,siegmund2012b,siegmund2013}	&	Linux kernel	&	Operating system	&	memory footprint, performance	\\
            \cite{jamshidi2018}	&	DNN	&	Optimization algorithm	&	response time	\\
            \cite{martinez2018}	&	Art system	&	Paint	&	user likeability	\\
            \cite{grebhahn2017}	&	Multigrid system 	&	Equations solving	&	time of each interaction	\\
            \cite{jamshidi2017a}	&	CoBot System	&	Robotic system	&	CPU usage	\\
            \cite{siegmund2015,temple2017a, kaltenecker2019}	&	JavaGC	&	Runtime environment	&	response time	\\
            \cite{svogor2019}	&	Robot	&	Runtime environment	&	energy consumption and execution time	\\
            \cite{elafia2018, xu2008}	&	Hand	&	SAT solver	&	runtime	\\
            \cite{elafia2018, xu2008}	&	Indu	&	SAT solver	&	runtime	\\
            \cite{elafia2018, xu2008}	&	Rand	&	SAT solver	&	runtime	\\
            \cite{hutter2011}	&	SAPS	&	SAT solver	&	runtime	\\
            \cite{jamshidi2017b, hutter2011}	&	SPEAR	&	SAT solver	&	runtime, response time	\\
            \cite{saleem2015}	&	QWS dataset	&	Services	&	availability, throughput, successability, reliability, compliance, best practice, documentation	\\
            \cite{queiroz2016}	&	BusyBox	&	Software suite	&	defect, process metrics 	\\
            \cite{krismayer2017}	&	Plant automation	&	Software-intensive SoS	&	defects	\\
            \cite{ding2015}	&	Sort 	&	Sort algorithm	&	execution time 	\\
            \cite{kolesnikov2018, nair2017,temple2017a, siegmund2015, kaltenecker2019}	&	DUNE	&	Stencil code	&	response time	\\
            \cite{kolesnikov2018, nair2017,siegmund2015,temple2017a}	&	HSMGP	&	Stencil code	&	response time, runtime	\\
            \cite{nair2017, jamshidi2017a,nair2018a, zhang2015, jamshidi2016, jamshidi2018, zhang2016}	&	Apache	&	Stream processing	&	latency, throughput, performance	\\
            \cite{gargantini2017}	&	Concurrency	&	Testing problem	&	defects	\\
            \cite{samreen2016}	&	VARD on EC2	&	Text manager	&	response time	\\
            \cite{gargantini2017}	&	Aircraft	&	Toy example	&	defects	\\
            \cite{siegmund2013b}	&	ELEVATOR	&	Toy example	&	time	\\
            \cite{gargantini2017}	&	WashingMachine	&	Toy example	&	defects	\\
            \cite{siegmund2011,siegmund2012b,siegmund2013}	&	Violet	&	UML editor	&	memory footprint, performance	\\
            \cite{temple2016, temple2017a, temple2018}	&	MOTIV	&	Video encoder	&	video quality	\\
            \cite{kaltenecker2019}	&	VP9	&	Video encoder	&	encoding time	\\
            \cite{guo2013,guo2017,jehooh2017,kolesnikov2018,nair2017,sarkar2015,siegmund2012a,siegmund2013, nair2018a, jamshidi2017b, temple2017a, siegmund2015, zhang2015, valov2017, nair2018c, kaltenecker2019, zhang2016}	&	x264	&	Video encoder	&	CPU power consumption, encoding time, Peak Signal to Noise Ratio, response time, code complexity, video quality, performance	\\
            \cite{temple2017a}	&	OpenCV	&	Video tracking	&	performance	\\
            \cite{safdar2017}	&	C60 and MX300	&	Virtual environment	&	defects	\\
            \cite{alipourfard2017}	&	Amazon EC2	&	Web coud service	&	performance	\\
            \cite{guo2013,guo2017,jehooh2017,kolesnikov2018,nair2017,sarkar2015,siegmund2012a,temple2017a, siegmund2015, nair2018c, yilmaz2014, zheng2007}	&	Apache	&	Web server	&	response rate, response time, workload, defects, throughput	\\
            \cite{osogami2007}	&	Stock Brokerage 	&	Web system	&	throughput, response time	\\
            \cite{song2013}	&	vsftpd	&	FTP daemon	&	defects	\\
            \cite{song2013}	&	ngIRCd	&	IRC daemon	&	defects	\\
        \hline
	\label{tab:RQ3_systems}
\end{longtable}
\end{small}

Next, we detail the particularities of NFPs.
Specifically, we describe how the measurement is performed, what process and strategies are adopted to avoid biases in the results, and also discuss the cost of measuring. 

\begin{figure}
	\centering
	\includegraphics[width=1\textwidth]{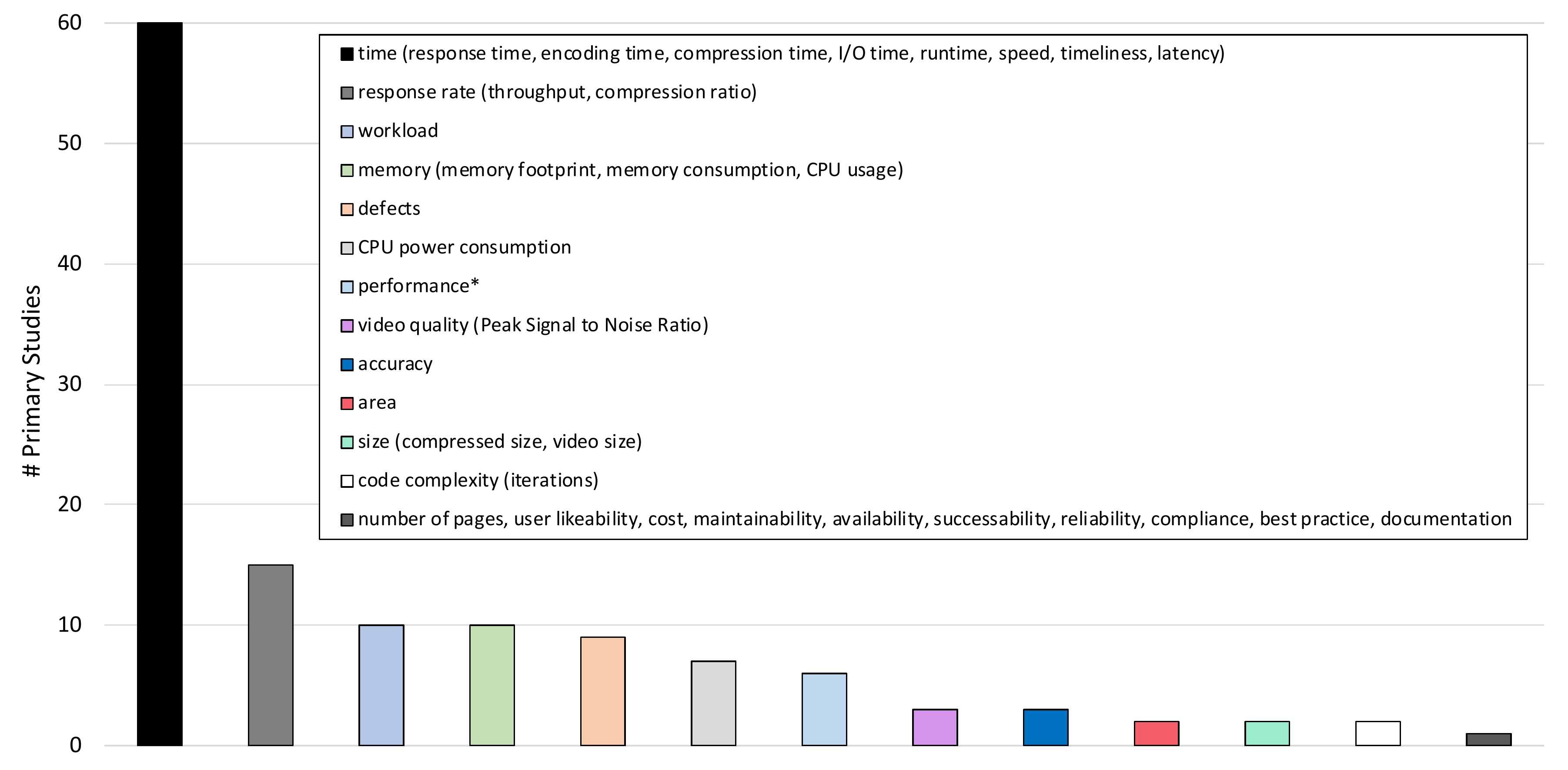}
	\caption{Non-functional properties measured in the literature.\\
	\footnotesize{*Each application domain or even each system has special definition of performance. These works did not specify the NFP(s) that refer to performance.}}
    \label{fig:RQ3_NFPs}
\end{figure}

\textbf{Time.} The time spent by a software configuration to realize a task is an important concern and has been intensively considered under different flavors and terminologies (see Fig.~\ref{fig:RQ3_NFPs}). 
Siegmund et al.\cite{siegmund2015} invested more than two months (24/7) for measuring \emph{response time} of all configurations of different subject systems (Dune MGS, HIPAcc, HSMGP, JavaGC, SaC, x264).
 For each system they use a different configuration of hardware.
The configurations' measurements are reused in many papers, mainly for evaluating the proposed learning process~\cite{guo2017, nair2018c, temple2017}. 
Chen et al.~\cite{chen2005} used a simulator to measure the response time of a sequence of 10 service requests by a client to the server. They use two implementation technologies, CORBA and EJB.

Time is a general notion and can be refined according to the applicaton domain. 
For instance, Kolesnikov et al.~\cite{kolesnikov2018} considered \emph{compression time} of a file on Lrzip (Long Range ZIP), a file archiver written. The measurements were conducted on a dedicated server and repeated multiple times to control measurements noise. 
Kaltenecker et al.~\cite{kaltenecker2019} considered 7-ZIP, a file archiver written in C++, and measured the compression time of the Canterbury corpus\footnote{https://corpus.canterbury.ac.nz/}. 

In another engineering context, several works \cite{guo2013, kolesnikov2018, nair2017, zhang2015, nair2018c, valov2015,  kaltenecker2019} measured the \emph{encoding time} of a input video over 1,152 valid configurations of x264. 
x264 is a configurable system for encoding video streams into the H.264/MPEG-4 AVC format.
As benchmark, the Sintel trailer (735 MB) is used and an encoding from AVI is considered. 
Kaltenecker et al.~\cite{kaltenecker2019} measured the encoding time of a short piece of the Big Buck Bunny trailer over 216,000 valid configurations of VPXENC (VP9), a video encoder that uses the VP9 format. 

\emph{Latency} has caught several attention. Jamshidi et al.~\cite{jamshidi2016} and \cite{jamshidi2017a} measure the average latency (\textit{i.e.}, how fast it can respond to a request) of three stream processing applications on Apache Storm (WordCount, RollingSort, SOL) over a window of 8 minutes and 2 hours, respectively.
Jamshidi et al.~\cite{jamshidi2017a} also measure the average latency of the NoSQL database system on Apache Cassandra over a window of 10 minutes.
Jamshidi et al.~\cite{jamshidi2016} performed the measurements on a multi-node cluster on the EC2 cloud.
Jamshidi et al.~\cite{jamshidi2017a} performed the same measurement procedure reported for CPU usage.
The measurements used by Nair et al.~\cite{nair2017, nair2018a} were derived from Jamshidi et al.~\cite{jamshidi2016}.
Weckesser et al.~\cite{weckesser2018} measure the latency of transferred messages over an adaptive Wireless Sensor Networks (WSNs) via simulation.
100 fully-charged nodes were distributed randomly onto a square region for each simulation run.
Aken et al.~\cite{aken2017} measure the latency for two OLTP DBMSs (MySQL v5.6, Postgres v9.3) over three workloads (The Yahoo Cloud Serving Benchmark (YCSB), TPC-C, and Wikipedia) during five minutes observation periods.
The OLTP DBMS are deployed on m4.large instances with 4 vCPUs and 16 GB RAM on Amazon EC2.
They also consider the total execution time of the OLAP DBMS (Actian Vector v4.2) over the TPC-H workload on m3.xlarge instances with 4 vCPUs and 15 GB RAM on Amazon EC2.
All of the training data was collected using the DBMSs’ default isolation level.

 



Sharifloo et al.~\cite{sharifloo2016} measures time throughout the evolution of a configurable system (an SPL). In an SPL evolution scenario, the set of measured configurations may include a removed feature or violate changed constraints. In this case, configurations are removed from the sample. Moreover, a modified feature implies to recompute the configuration measurements.


\textbf{Other NFPs (CPU power consumption, CPU usage, size, etc.)} Beyond time, other quantitative properties have been considered. 

Nair et al.~\cite{nair2017} measured the compressed size of a specific input over 432 valid configurations of Lrzip, a compression program optimized for large files.
 Siegmund et al.~\cite{siegmund2011, siegmund2012b, siegmund2013} measured the \textit{memory footprint} of nine configurable system: LinkedList, Prevayler, ZipMe, PKJab, SensorNetwork, Violet, Berkeley DB, SQLite, and Linux kernel.
Nair et al.~\cite{nair2018a} and Zuluaga et al.~\cite{zuluaga2016} used LLVM, a configurable modular compiler infrastructure. 
Footprint is measured as the binary size of a compiled configuration.
%

Several works~\cite{murwantara2014, nair2018a, temple2017a, couto2017, zuluaga2016, jamshidi2019} measured \textit{CPU power consumption}. Murwantara et al.~\cite{murwantara2014} used a simulator to compute CPU power consumption over a web service under different loads. 
They used a kernel-based virtual machine to conduct the measurements based on several combinations of HTTP servers and variant PHP technologies connected to a MySQL database system. They divided the experiment into blocks of 10 seconds for 100 increasing stages. In the first 10 seconds, they produce loads of one user per second, and in the next 10 seconds, they produce the load of two users per seconds and so on. This results in 1-100 users per period of 10 seconds.
Couto et al.~\cite{couto2017} propose a static learning approach of CPU power consumption. 
Their approach assures that the source code of every feature from a configuration is analyzed only once.
To prove its accuracy, they measured at runtime the worst-case CPU power consumption to execute a given instruction on 7 valid configurations of the disparity SPL~\cite{venkata2009}.
They repeated the measurements 200 times for each configuration using the same input.
Since their goal was to determine the worst case CPU power consumption, they removed the outliers (5 highest and lowest values) and retrieved the highest value from every configuration for validation proposes.

 
Jamshidi et al.~\cite{jamshidi2017a} use simulation measurements.
Jamshidi et al.~\cite{jamshidi2017a} have repeatedly executed a specific robot mission to navigate along a corridor offline in a simulator and measured performance in terms of \textit{CPU usage} on the CoBot system. 
To understand the power of transfer learning techniques, they consider several simple hardware changes (\textit{e.g.}, processor capacity) as well as severe changes 
(\textit{e.g.}, local desktop computer to virtual machines in the cloud)\footnote{For a complete list of hardware/software variability, see https://github.com/pooyanjamshidi/transferlearning}.
Each measurement took about 30 seconds.
 They measured each configuration of each system and environment 3 times.

To overcome the cost of measuring realistic (non-)functional properties, Siegmund et al.~\cite{siegmund2017} proposed a tool, called Thor, to generate artificial and realistic synthetic measurements, based on different distribution patterns of property values for features, interactions, and configurations from a real-world system.
Jamshidi et al.~\cite{jamshidi2019} adopted Thor to synthetically measure the property energy consumption for a robot to complete a mission consisting of randomly chosen tasks within a map.

\textbf{Qualitative properties.} Instead of measuring a numerical value, several papers assign a qualitative value to configurations. In Acher et al.~\cite{acher2018}, a variant is considered as acceptable or not thanks to an automated procedure. In Temple et al.~\cite{temple2016, temple2018}, a quantitative value is originally measured and then transformed into a qualitative one through the definition of a threshold. 
In \cite{queiroz2016}, a variant is considered as acceptable or not based on the static evolution historical analysis of commit messages (\textit{i.e.}, they identify a defect by searching for the following keywords: \textit{bug}, \textit{fix}, \textit{error}, and \textit{fail}).
The learning process then aims to predict the class of the configuration -- whether configurations are acceptable or not, as defined by the threshold. 

Software \emph{defects} are considered in~\cite{yilmaz2006, krismayer2017, gargantini2017, yilmaz2014, porter2007, amand2019}. 
Yilmaz et al.~\cite{yilmaz2006} and Porter et al.~\cite{porter2007} characterize defects of the ACE+TAO system. 
Yilmaz et al.~\cite{yilmaz2006} test each supposed valid configuration on the Red Hat Linux 2.4.9-3 platform and on Windows XP Professional using 96 developer-supplied regression tests. 
In~\cite{porter2007}, developers currently run the tests continuously on more than 100 largely uncoordinated workstations and servers at a dozen sites around the world.
The platforms vary in versions of UNIX, Windows, Mac OS, as well as to real-time operating systems. 
To examine the impact of masking effects on Apache v2.3.11-beta and MySQL v5.1, Yilmaz et al.~\cite{yilmaz2014} grouped the test outcomes into three classes: passed, failed, and skipped.
They call this approach a \textit{ternary-class fault characterization}.
Gargantini et al.~\cite{gargantini2017} interest is in comparing the defect detection capability 
on different sample heuristics (see Section~\ref{RQ2}), while Amand et al.~\cite{amand2019} interest is in comparing the accuracy of several learning algorithms to predict whether a configuration will lead to a defect. 
In the dynamic configuration scenario, Krismayer et al.~\cite{krismayer2017} use event logs (via simulation) from a real-world automation SoS to mine different types of constraints according to real-time defects.

Finally, Martinez et al.~\cite{martinez2018} consider a 5-point \textit{user likeability} scale with values ranging from 1 (strong dislike) to 5 (strong like). In this work, humans have reviewed and labeled configurations. 

\textbf{Accuracy of measurements.} In general, measuring NFPs (\textit{e.g.}, time) is a difficult process since several confounding factors should be controlled. The need to gather measures over numerous configurations exacerbates the problem. 

Weckesser et al.~\cite{weckesser2018} mitigated the construct thread of the inherent randomness and repeated all runs five times with different random seeds. Measurements started after a warm-up time of 5 minutes.
Kaltenecker et al.~\cite{kaltenecker2019} measured each configuration between 5 to 10 times until reaching a standard deviation of less than 10\%.
Zhang et al.~\cite{zhang2015} repeated the measurements 10 times. 

To investigate the influence of measurement errors on the resulting model, Kolesnikov et al.~\cite{kolesnikov2018} and Duarte et al.~\cite{duarte2018} conducted a separate experiment. 
They injected measurement errors to the original measurements and repeated the learning process with \textit{polluted} datasets.
Then, they compared the prediction error of the noisy models to the prediction error of the original models to see the potential influence of measurement errors.
For each subject system, Kolesnikov et al.~\cite{kolesnikov2018} repeated the learning process five times for different increasing measurement errors.

Dynamic properties are susceptible to measurement errors (due non-controlled external influences) which may bias the results of the measuring process. 
To account for measurement noise and be subject to external influences, these properties need to be measured multiple times on dedicated systems.
Thus, the total measurement cost to obtain the whole data used in the experiments is overly expensive and time-consuming
(\textit{e.g.}, Kaltenecker et al.~\cite{kaltenecker2019} spent multiple years of CPU time).
According to Lillacka et al.~\cite{lillacka2013}, there are a warm-up phase followed by multiple times runs; and the memory must be set up large enough to prevent disturbing effects from the Garbage Collector, as well as all operations, must be executed in memory so that disk or network I/O will also produce no disturbing effects. 
Most of the works only consider the variability of the subject system, while they use static inputs and hardware/software environments.
Therefore, the resulting model may not properly characterize the performance of a different input or environment, 
since most of the properties (\textit{e.g.}, CPU power consumption and compression time) are dependent of the input task and the used hardware/software.
Consequently, hardware/software must also be taken into account as dependent variables as considered by Jamshidi et al.~\cite{jamshidi2017a, jamshidi2017b}.

There are some properties that are much accurate, because \textit{e.g.} they are not influenced by the used hardware, such as footprint.
Siegmund et al.~\cite{siegmund2013} parallelized the measurements of footprint on three systems and used the same compiler.
Moreover, footprint can be measured quickly only once, without measurement bias.

\textbf{Cost of measurements.} The cost of observing and measuring software can be important, especially when multiple configurations should be considered. 
The cost can be related to computational resources needed (in time and space). It can also be related to human resources involved in labelling some configurations~\cite{martinez2018}. 

Zuluaga et al.~\cite{zuluaga2016} and Nair et al.~\cite{nair2018a} measure the quantitative NFP \textit{area} of a field-programmable gate array (FPGA) platform consisting of 206 different hardware implementations of a sorting network for 256 inputs.
They report that the measurement of each configuration is very costly and can take up to many hours. 
Porter et al.~\cite{porter2007} characterization of defects for each configuration ranges from 3 hours on quad-CPU machines to 12-18 hours on less powerful machines. Yilmaz et al.~\cite{yilmaz2006} took over two machine years to run the total of 18,792 valid configurations.
In Siegmund et al.~\cite{siegmund2012b}, a single measurement of memory footprint took approximately 5 minutes. 
Temple et al.~\cite{temple2016} report that the execution of a configuration to measure video quality took 30 minutes on average. They used a grid computing to distribute the computation and scale the measurement for handling 4,000+ configurations. 
The average time reported by Murwantara et al.~\cite{murwantara2014} to measure CPU power consumption of all sampled configurations was 1,000 seconds. The authors set up a predefined threshold to speed up the process.
 
There are fifteen main NFPs supported in the literature. 
Some of them are less costly, such as \textit{code complexity}, which can be measured statically by analyzing the number of code lines~\cite{siegmund2012b}.
As a result, the measurements can be parallelized and quickly done only once.
Otherwise, dynamic properties, such as \textit{CPU power consumption} and \textit{response time} are directly related to hardware and external influences. 
While \textit{CPU power consumption} might be measured under different loads over a predefined threshold time, the property \textit{response time} is much costly as a threshold can not be defined and the user does not know how long it will take.
To easy the labeling of features, some works use synthetic NFPs and statistical analysis strategies.
As the same NFP may be measured in different ways (\textit{e.g.}, video quality can be measured either by \textit{user feedback} or \textit{execution} of a program to automatically attribute labels to features).
Practitioners need to find a sweet spot to have accurate measurements with a small sample.
 
Depending on the subject system and application domain (see the dataset of~\cite{siegmund2015}), there are more favorable cases with only a few seconds per configuration. 
However, even in this case, the overall cost can quickly become prohibitive when the number of configurations to measure is too high. In ~\cite{weckesser2018}, all training simulation runs took approximately 412 hours of CPU time.\\

\begin{mdframed}[backgroundcolor=gray!10] 
    Numerous qualitative and quantitative properties of configurations are measured mainly through the use of automated software procedures. 
    For a given subject system and its application domain, there may be more than one measure (\textit{e.g.}, CPU power consumption and video quality for x264).
    Time is the most considered performance measure and is obtained in the literature through either execution, simulation, static analysis, or synthetic measurement. 
    The general problem is to find a good tradeoff between the cost and the accuracy of measuring numerous configurations (\textit{e.g.}, simulation can speed up the measurement process at the price of approximating the real observations).
\end{mdframed}

\subsection{RQ4: Which learning techniques are used?}
\label{RQ4}
In this section, we aim at understanding which learning techniques were applied and in which context. 
Table~\ref{tab:RQ4} sketches which learning techniques are supported in the literature and what application objective they address. 
The first column identifies the study reference(s). 
The second and third columns identify the name of the learning technique and its application objective, respectively.
(Notice that the application objective is related to the scenarios in which each learning technique has been already used in the literature; it means some learning techniques could well be applied for other scenarios in the future). 

\begin{table} \footnotesize{}
	\centering
    \begin{tabular}{p{0.28\textwidth} p{0.46\textwidth} p{0.18\textwidth}}
    	\hline
		    \textbf{Reference} & \textbf{Learning Technique} & \textbf{Applicability} \\
        \hline		
            \cite{yilmaz2014}	&	Adaptive ELAs, Multi-Class FDA-CIT, Static Error Locating Arrays (ELAs), Ternary-Class FDA-CIT, Test Case-Aware CIT, Traditional CIT	&	\textit{A1}	\\
            \cite{valov2015}	&	Bagging	&	\textit{A1}	\\
            \cite{zhang2015}	&	Fourier Learning of Boolean Functions	&	\textit{A1}	\\
            \cite{kolesnikov2017}	&	Frequent Item Set Mining	&	\textit{A1}	\\
            \cite{siegmund2013b}	&	Graph Family-Based Variant Simulator	&	\textit{A1}	\\
            \cite{couto2017}	&	Implicit Path Enumeration Technique (IPET)	&	\textit{A1}	\\
            \cite{queiroz2016}	&	Naive Bayes	&	\textit{A1}	\\
            \cite{jamshidi2017b,kolesnikov2018,siegmund2015,weckesser2018,kolesnikov2017,kaltenecker2019,jamshidi2019, siegmund2008}	&	Step-Wise Linear Regression	&	\textit{A1}, \textit{A2}, \textit{A3}, \textit{A4}	\\
            \cite{guo2013,guo2017,jamshidi2017b,murwantara2014,nair2017,nair2018a,sarkar2015,temple2016,temple2017a,valov2015,valov2017,westermann2012,acher2018,nair2018c,yilmaz2006,krismayer2017,porter2007,zheng2007, zhang2016, song2013}	&	Classification and Regression Trees (CART)	&	\textit{A1}, \textit{A2}, \textit{A3}, \textit{A4}, \textit{A5}	\\
            \cite{siegmund2017}	&	Kernel Density Estimation and NSGA-II	&	\textit{A1}, \textit{A2}, \textit{A3}, \textit{A4}, \textit{A5}, \textit{A6}	\\
            \cite{siegmund2011,siegmund2012a,siegmund2012b,siegmund2013}	&	Feature's Influence Delta	&	\textit{A1}, \textit{A3}	\\
            \cite{alipourfard2017,westermann2012,jamshidi2017a,aken2017,jamshidi2018,jamshidi2016,zuluaga2016}	&	Gaussian Process Model	&	\textit{A1}, \textit{A3}, \textit{A4}	\\
            \cite{murwantara2014,samreen2016, chen2005, lillacka2013}	&	Linear Regression	&	\textit{A1}, \textit{A3}, \textit{A4}	\\
            \cite{valov2015, amand2019, queiroz2016, bao2018, thornton2013}	&	Random Forest	&	\textit{A1}, \textit{A3}, \textit{A5}	\\
            \cite{amand2019,temple2018,valov2015, elafia2018}	&	Support Vector Machine	&	\textit{A1}, \textit{A3}, \textit{A5}	\\
            \cite{amand2019, queiroz2016, song2013}	&	C4.5 (J48)	&	\textit{A1}, \textit{A5}	\\
            \cite{sincero2010, etxeberria2014}	&	Covariance Analysis	&	\textit{A2}	\\
            \cite{jamshidi2017b}	&	Multinomial Logistic Regression	&	\textit{A2}	\\
            \cite{duarte2018}	&	K-Plane Algorithm	&	\textit{A2}, \textit{A4}	\\
            \cite{saleem2015}	&	AdaRank	&	\textit{A3}	\\
            \cite{murwantara2014}	&	Bagging Ensembles of CART, Bagging Ensembles of MLPs	&	\textit{A3}	\\
            \cite{martinez2018}	&	Data Mining Interpolation Technique	&	\textit{A3}	\\
            \cite{aken2017}	&	Factor Analysis, k-means, Ordinary Least Squares	&	\textit{A3}	\\
            \cite{westermann2012, svogor2019}	&	Genetic Programming (GP)	&	\textit{A3}	\\
            \cite{westermann2012}	&	Kriging	&	\textit{A3}	\\
            \cite{ding2015}	&	Max-Apriori Classifier, Exhaustive Feature Subsets Classifiers, All Features Classifier, Incremental Feature Examination classifier	&	\textit{A3}	\\
            \cite{osogami2007}	&	Quick Optimization via Guessing	&	\textit{A3}	\\
            \cite{hutter2011}	&	Random Online Adaptive Racing (ROAR), Sequential Model-based Algorithm Configuration (SMAC) 	&	\textit{A3}	\\
            \cite{svogor2019}	&	Simulated Annealing	&	\textit{A3}	\\
            \cite{xi2004}	&	Smart Hill-Climbing	&	\textit{A3}	\\
            \cite{jehooh2017}	&	Statistical Recursive Searching	&	\textit{A3}	\\
            \cite{ghamizi2019}	&	Tensorflow and Keras	&	\textit{A3}	\\
            \cite{thornton2013}	&	Tree-structured Parzen Estimator (TPE)	&	\textit{A3}	\\
            \cite{samreen2016, westermann2012, grebhahn2017}	&	Multivariate Adaptive Regression Splines (MARS), Multivariate Polynomial Regression	&	\textit{A3}, \textit{A4}	\\
            \cite{samreen2016, xu2008}	&	Ridge Regression 	&	\textit{A3}, \textit{A4}	\\
            \cite{murwantara2014, amand2019}	&	Multilayer Perceptrons (MLPs)	&	\textit{A3}, \textit{A5}	\\
            \cite{chen2009}	&	Actor-Critic Learning	&	\textit{A4}	\\
            \cite{samreen2016}	&	Lasso	&	\textit{A4}	\\
            \cite{sharifloo2016}	&	Reinforcement Learning	&	\textit{A4}, \textit{A6}	\\
            \cite{gargantini2017}	&	CitLab Model	&	\textit{A5}	\\
            \cite{temple2018}	&	Evasion Attack	&	\textit{A5}	\\
            \cite{amand2019}	&	Hoeffding Tree, K*, kNN, Logistic Model Tree, Logistic Regression, Naive Bayes, PART Decision List, Random Committee, REP Tree, RIPPER	&	\textit{A5}	\\
            \cite{safdar2017}	&	Pruning Rule-Based Classification (PART)	&	\textit{A5}	\\
        \hline
	\end{tabular}
	\caption{Learning techniques reported in the literature. \textit{A1}: Pure Prediction; \textit{A2}: Interpretability of Configurable Systems; \textit{A3}: Optimization; \textit{A4}: Dynamic Configuration; \textit{A5}: Mining Constraints; \textit{A6}: SPL Evolution.}
	\label{tab:RQ4}
\end{table}



Supervised learning problems can be grouped into regression and classification problems. In both cases, the goal is to construct machine learning model that can predict the value of the measurement from the features. The difference between the two problems is the fact that the value to predict is numerical for regression and categorical for classification. 
In the survey, we found a similar dichotomy, depending on the targeted use-case and the NFP of interest. 

A \emph{regression problem} is when the output is a real or continuous value, such as time or CPU power consumption. Most of the learning techniques tackle a supervised regression problem. 



\textbf{CART for regression.} Several authors~\cite{guo2013, guo2017, nair2017, nair2018a, jamshidi2017b, murwantara2014, sarkar2015, temple2016, temple2017a} use the \textit{Classification And Regression Trees} (CART) technique, to model the correlation between feature selections and performance. 
The sample is used to build the prediction model.
CART recursively partitions the sample 
into smaller clusters 
until the performance of the configurations in the clusters are similar.
These recursive partitions are represented as a binary decision tree. 
For each cluster, these approaches use the sample mean of the performance measurements (or even the majority vote) as the local prediction model of the cluster.
So, when they need to predict the performance of a new configuration not measured so far, they use the decision tree to find the cluster which is most similar to the new configuration.
Each split of the set of configurations is driven by the (de)selection of a feature 
that would minimize a prediction error. 

CART use two parameters to automatically control the recursive partitioning process: \textit{minbucket} and \textit{minsplit}. 
Minbucket is the minimum sample size 
for any leaf of the tree structure; and minsplit is the minimum sample size of a cluster 
before it is considered for partitioning.
Guo et al.~\cite{guo2013} compute minbucket and minsplit based on the size of the input sample, \textit{i.e.}, if $|S_C| \leq 100$, then minbucket = $|\frac{|S_C|}{10} + \frac{1}{2}|$ and minsplit = $2 \times$ minbucket; if $|S_C| > 100$, then minsplit = $|\frac{|S_C|}{10} + \frac{1}{2}|$ and minbucket = $|\frac{minsplit}{2}|$; the minimum of minbucket is 2; and the minimum of minsplit is 4.
It should be noted that CART can also be used for classification problems (see hereafter). 

Instead of using a set of empirically-determined rules, Guo et al.~\cite{guo2017} combine the previous CART approach~\cite{guo2013} with automated resampling and parameter tuning, which they call a data-efficient learning approach (DECART). 
Using resampling, DECART learns a prediction model by using different sample designs (see Section~\ref{tab:RQ5_1}).
Using parameter tuning, DECART ensures that the prediction model has been learned using optimal parameter settings of CART based on the currently available sample. 
They compare three parameter-tuning techniques: random search, grid search, and Bayesian optimization.
Westermann et al.~\cite{westermann2012} also used grid search for tuning CART parameters.

Nair et al.~\cite{nair2017, nair2018a, sarkar2015} approaches build a prediction model in a progressive way by using CART.
They start with a small training sample and subsequently add samples to improve performance predictions based on the model accuracy (see Section~\ref{RQ5}). 
In each step, while training the prediction model, Nair et al.~\cite{nair2017} compare the current accuracy of the model with the previous accuracy from the prior iteration (before adding the new set of configurations to the training set).
If the current accuracy (with more data) 
does not improve the previous accuracy (with lesser data), then the learning reaches a termination criterion 
(\textit{i.e.}, adding more sample will not result in significant accuracy improvements). 

\textbf{Performance-influence models.} Siegmund et al.~\cite{siegmund2015} combines machine learning and sampling heuristics to build so-called performance-influence models. A step-wise linear regression algorithm is used to select relevant features as relevant terms of a linear function and learn their coefficient to explain the observations.
In each iterative step, the algorithm selects the sample configuration with strongest influence regarding prediction accuracy (\textit{i.e.}, yields the model’s lowest prediction error) 
until improvements of model accuracy become marginal or a threshold for expected accuracy is reached (below 19\%).
The algorithm concludes with a backward learning step, in which every relevant feature is tested for whether its removal would decrease model accuracy. This can happen if initially a single feature is selected because it better explains the measurements, but it becomes obsolete by other features (\textit{e.g.}, because of feature interactions) later in the learning process.
Linear regression allows them to learn a formula that can be understood by humans.
It also makes it easy to incorporate domain knowledge about an option’s influence on the formula. 
However, the complete learning of a model using this technique required from 1 to 5 hours, depending on the size of the learning set and the size of the models.
Kolesnikov et al.~\cite{kolesnikov2018} 
investigate how significant are the trade-offs among prediction error, model size, and computation time. 

\textbf{Other learning algorithms for regression.} Westermann et al.~\cite{westermann2012}  used Multivariate Adaptive Regression Splines (MARS), Genetic Programming (GP), and Kriging.
Zhang et al.~\cite{zhang2015} used Fourier learning algorithm.
In all these works, for a set of sample, it verifies if the resulting accuracy is acceptable for stakeholders (\textit{e.g.}, prediction error rate below 10\%). 
While the accuracy is not satisfactory, the process continues by obtaining an additional sample of measured configurations and iterates again to produce an improved prediction model. 
Sarkar et al.~\cite{sarkar2015} propose a sampling cost metric as a stopping criterion, where the objective is to ensure the most optimal trade-off between measurement effort and prediction accuracy. 
Sampling stops when the counts of features selected and deselected is, at least, at a predefined threshold.

Chen et al.~\cite{chen2005} propose a linear regression approach
to describe the generic performance behavior of application server components running on component-based middleware technologies.
The model focuses on two performance factors: workload and degree of concurrency.
Sincero et al.~\cite{sincero2010} employ analysis of covariance for identifying factors with significant effects on the response or interactions among features. 
They aim at proposing a configuration process where the user is informed about the impact of their feature selection on the NFPs of interest.

Murwantara et al.~\cite{murwantara2014} use a set of five ML techniques (\textit{i.e.}, Linear regression, CART, Multilayer Perceptrons (MLPs), Bagging Ensembles of CART, and Bagging Ensembles of MLPs) to learn how to predict the energy consumption of web service systems. 
They use the WEKA's implementation of these techniques with its default parameters~\cite{weka2009}.



\textbf{Learning to rank.} 
 Instead of predicting the raw performance value, it can be of interest to predict the rank of a configuration (typically to identify \emph{optimal} configurations, see RQ1). 

Jehooh et al.~\cite{jehooh2017} adopt statistical learning techniques to progressively shrink a configuration space and search for near-optimal configurations.
First, the approach samples and measures a set of configurations.
For each pair of sampled configurations, this approach identifies features that are common (de)selected.
Then, it computes the performance influence of each common decision to find the best regions for future sampling.
The performance influence measures the average performance over the samples that have the feature selected against the samples that have the feature deselected, \textit{i.e.}, the sample is partitioned by whether a configuration includes a particular feature or not.
In addition, Welch’s t-test evaluates whether the performance mean of one sample group is higher than the other group with 95\% confidence.
The most influential decisions are added to the sample. 
This process continues recursively until they identify all decisions that are statistically certain to improve program performance. 
They call this a \textit{Statistical Recursive Searching} technique.

Nair et al.~\cite{nair2017} compute accuracy by using the mean rank difference measurement (the predicted rank order is compared to the optimal rank order). 
They demonstrate that their approach can find optimal configurations of a software system using fewer measurements than the approach proposed by~\cite{sarkar2015}.
One drawback with this approach is that it requires a holdout set, against which the current model (built interactively) is compared.
To overcome this issue, instead of making comparisons, Nair et al.~\cite{nair2018a} consider a predefined stopping criterion (budget associated with the optimization process). 
While the criterion is not met, the approach finds the configuration with the best accuracy and add the configuration to the training set. 
The model adds the next most promising configuration to evaluate.
Consequently, in each step, the approach discards less satisfactory configurations which have a high probability of being dominated, \textit{a.k.a.} active learning. 
This process terminates when the predefined stopping condition is reached. 
Nair et al.~\cite{nair2018a} demonstrate that their approach is much more effective for multi-objective configuration optimization than state-of-the-art approaches~\cite{zuluaga2016, sarkar2015, nair2017}.

Martinez et al.~\cite{martinez2018} propose the use of data mining interpolation techniques (\textit{i.e.}, similarity distance, similarity radius, weighted mean) for ranking configurations through user feedback on a configuration sample.
They estimate the user perceptions of each feature by computing the value of the chi-squared statistic with respect to the correlation 
score given by users on configurations. 

\textbf{Transfer learning.} The previous approaches assume a static environment (\textit{e.g.}, hardware, workload) and NFP such that learning has to be repeated once the environment and NFP changes. 
In this scenario, some approaches adopt transfer learning techniques to reuse (already available) knowledge from other relevant sources to learn a performance for a target system instead of relearn a model from scratch~\cite{valov2017, jamshidi2017a, jamshidi2017b, jamshidi2018}.

Valov et al.~\cite{valov2017} investigate the use of transfer learning across different hardware platforms.
They used 25 different hardware platforms to understand the similarity on performance prediction.
They created a prediction model using CART.
With CART, the resulting prediction models can be easily understood by end users. 
To transfer knowledge from a related source to a target source, they used a simple linear regression model.

Jamshidi et al.~\cite{jamshidi2017a} use \textit{Gaussian Process Models} (GPM) to model the correlation between source and target sources using the measure of \textit{similarity}.
GPM offers a framework in which predictions can be done using mean estimates with a confidence interval for each estimation.
Moreover, GPM computations are based on linear algebra which is cheap to compute.
This is especially useful in the domain of dynamic SPL configuration where learning a prediction model at runtime in a feedback loop under time and resource constraints is typically time constrained. 

Jamshidi et al.~\cite{jamshidi2017b} combine many statistical and ML techniques (\textit{i.e.}, \textit{Pearson linear correlation}, \textit{Kullback-Leibler divergence}, \textit{Spearman correlation coefficient}, \textit{paired t-test}, \textit{CART}, \textit{step-wise linear regression}, and \textit{multinomial logistic regression}) to identify when transfer learning can be applied. 
They use CART for estimating 
the relative importance of configuration options by examining how the prediction error will change for the trained trees on the source and target.
To investigate whether interactions across environments will be preserved, they use step-wise linear regression models (\textit{a.k.a.}, performance-influence models). 
This model learns all pairwise interactions, then it compares the coefficients of the pairwise interaction terms independently in the source and target environments. 
To avoid exploration of invalid configurations and reduce measurement effort, they use a multinomial logistic regression model to predict the probability of a configuration being invalid, then they compute the correlation between the probabilities from both environments.

Jamshidi et al.~\cite{jamshidi2018} propose a sampling strategy, called L2S, that exploits common similarities across environments from \cite{jamshidi2017b}.
L2S extracts transferable knowledge from the source to drive the selection of more informative samples in the target environment. Based on identifying interesting regions from the performance model of the source environment, it generates and selects configurations in the target environment iteratively.

\textbf{Classification problem.} Temple et al.~\cite{temple2016, temple2017a} and Acher et al.~\cite{acher2018} use CART to infer constraints to avoid the derivation of invalid (non-acceptable) configurations.
CART considers a path in a tree as a set of decisions, where each decision corresponds to the value of a single feature.
The approach creates new constraints in the variability model by building the negation of the conjunction of a path to reach a faulty leaf.
They learn constraints among Boolean and numerical options.
As an academic example, Acher et al.~\cite{acher2018} introduce Vary\LaTeX~to guide researchers to meet paper constraints by using annotated \LaTeX~sources.
To improve the learning process, Temple et al.~\cite{temple2018} specifically target low confidence areas for sampling.
The authors apply the idea of using an adversarial learning technique, called evasion attack, after a classifier is trained with a Support Vector Machine (SVM). 
In addition, Temple et al.~\cite{temple2017a} support the specialization of configurable systems for a deployment at runtime. 
In a similar context, Safdar et al.~\cite{safdar2017} infer constraints over multi-SPLs (\textit{i.e.}, they take into account cross-SPLs rules).

\textbf{Input sensitivity.}
Several works consider the influence of the input data on the resulting prediction model~\cite{lillacka2013, ding2015, thornton2013, xu2008, hutter2011, grebhahn2017, elafia2018, ghamizi2019}.
Many authors~\cite{ding2015, thornton2013, xu2008, hutter2011, grebhahn2017, elafia2018, ghamizi2019} address input sensitivity in algorithm auto-tuning.
They use learning techniques to search for the best algorithmic variants and parameter settings to achieve optimal performance for a given input instance. 
It is well known that the performance of SAT solver and learning methods strongly depends on making the right algorithmic and parameter choices, therefore SATzilla~\cite{xu2008} and Auto-WEKA~\cite{thornton2013} search for the best SAT solver and learning technique for a given input instance.
Lillacka et al.~\cite{lillacka2013} treat the variability caused by the input data as the variability of the SPL.
As it is not feasible to model every possible input data, they cluster the data based on its relevant properties (\textit{e.g.}, 10kB and 20MB in an or-group for file inputs of a compression library). 

\textbf{Reinforcement learning.} Sharifloo et al.~\cite{sharifloo2016} use a reinforcement learning technique to automatically reconfigure dynamic SPLs to deal with context changes. 
In their approach, learning continuously observes measured configurations and evaluates their ability to meet the contextual requirements.
Unmet requirements are addressed by learning new adaptation rules dynamically, or by modifying and improving the set of existing adaptation rules.
This stage takes software evolution into account to address new contexts faced at run-time.\\

\begin{mdframed}[backgroundcolor=gray!10] 
    Numerous statistical learning algorithms are used in the literature to learn software configuration spaces. The most used are standard machine learning techniques, such as polynomial linear regressions, decision trees, or Gaussian process models.
    The targeted use-case and the engineering context explain the diversity of solutions: either a supervised classification or regression problem is tackled; the requirements in terms of interpretability and accuracy may differ; there is some innovation in the sampling phase to progressively improve the learning. 
    Still the use of others (more powerful) ML techniques such as deep learning, adversarial learning, and even the idea of learning different models (\textit{e.g.}, one for each application objective) that could co-evolve can be further explored in future works.
\end{mdframed}

\subsection{RQ5: How are learning-based techniques validated?}
\label{RQ5}
In this section, we aim at understanding how the validation process is conducted in the literature. 


\begin{table} 
	\centering
    \begin{tabular}{p{0.3\textwidth} p{0.6\textwidth}}
    	\hline
		    \textbf{Sample Design} & \textbf{Reference}  \\ 
		\hline
            Merge	&	\cite{chen2005,jehooh2017,kolesnikov2018,nair2018a,siegmund2011,siegmund2012a,siegmund2012b,siegmund2013,siegmund2015,sincero2010,kaltenecker2019}	\\
            Hold-Out	&	\cite{jamshidi2016,guo2013,jamshidi2017a,sarkar2015,temple2016,temple2017a,valov2015,valov2017,weckesser2018,acher2018,temple2018,nair2018c,yilmaz2006,krismayer2017,jamshidi2018,zuluaga2016,amand2019,alipourfard2017,nair2017,guo2017,westermann2012,queiroz2016,zhang2016,ghamizi2019,lillacka2013,grebhahn2017,saleem2015,bao2018,svogor2019,elafia2018, xu2008, duarte2018, chen2009, song2013}	\\
            Cross-Validation	&	\cite{murwantara2014,martinez2018,samreen2016,yilmaz2014,guo2017,safdar2017,ding2015, duarte2018}	\\
            Bootstrapping	&	\cite{aken2017,jamshidi2016,nair2017,guo2017,thornton2013}	\\
            Dynamic Sector Validation	&	\cite{westermann2012}	\\
        \hline
	\end{tabular}
	\caption{Sample designs reported in the literature.} 
	\label{tab:RQ5_1}
\end{table}

There are five design strategies documented in the literature to explore the sample data for learning and validation (see Table~\ref{tab:RQ5_1}).

\paragraph{Merge.}
Training, testing, and validation sets are merged.
Kolesnikov et al.~\cite{kolesnikov2018} studied and analyzed the trade-offs among prediction error, model size, and computation time of performance-prediction models.
For the purpose of their study, they were specifically interested to explore the evolution of the model properties to see the maximum possible extent of the corresponding trade-offs after each iteration of the learning algorithm. 
So, they performed a whole-population exploration of the largest possible learning set (\textit{i.e.}, all valid configurations).
Therefore, in their validation process, training, testing and validation sets are merged.
In a similar scenario, other authors~\cite{chen2005, jehooh2017, nair2018a} also used the whole sample for both, learning and validation, 
even they considered a small set of valid configurations as the sample.
Some studies justify the use of a merge pool due the need to compare different sampling methods.
However, training the algorithm on the validation pool may introduce bias to the results of the accurateness of the approach.
To overcome this issue, studies have used other well-established ML design strategies. 

In particular, Guo et al.~\cite{guo2017} study the trade-offs among hold-out, cross-validation, and bootstrapping.
For most of the case studies, 10-fold cross-validation outperformed hold-out and bootstrapping.
In terms of the running time, although these three stategies usually takes seconds to run, Guo et al.~\cite{guo2017} shows that hold-out is the fastest one, and 10-fold cross-validation tends to be faster than bootstrapping.

\paragraph{Hold-Out (HO)}
Most of the works~\cite{jamshidi2016,guo2013,jamshidi2017a,sarkar2015,temple2016,temple2017a,valov2015, valov2017,weckesser2018,acher2018,temple2018,nair2018c,yilmaz2006,krismayer2017,jamshidi2018,zuluaga2016,amand2019,alipourfard2017,nair2017,guo2017,westermann2012} used a hold-out design.
Hold-out splits an input sample $S_{C}$ into two disjoined sets $S_t$ and $S_v$, one for training and the other for validation, \textit{i.e.} $S_{C} = S_t \cup S_v$ and $S_t \cap S_v = \emptyset$.
To avoid bias in the splitting procedure, some works repeated it multiple times.
For example, in Valov et al.~\cite{valov2015}, the training set for each sample size were selected randomly 10 times. 
For transfer-learning applications~\cite{jamshidi2017a, valov2017, jamshidi2018}, the training set comes from samples of the target and related sources, while the validation set comes from samples only from the target source. 
    
\paragraph{Cross-Validation (CV)}
Cross-validation splits an input sample $S_{C}$ into $k$ disjoined subsets of the same size, \textit{i.e.}, $S = S_1 \cup ... \cup S_k$, where $S_i \cap S_j = \emptyset$ ($i \neq j$); each subset $S_i$ is selected as the validation set $S_v$, and all the remaining $k-1$ subsets are selected as the training set $S_t$. 
Yilmaz et al.~\cite{yilmaz2014} used a 5-fold cross-validation to create multiple models from different subsets of the input data.
In a 5-fold cross-validation, the sample $S$ is partitioned into 5 subsets (\textit{i.e.}, $S = S_1 \cup S_2 \cup S_3 \cup S_4 \cup S_5$) of the same size.
Each subset is selected as the validation set and all of the remaining subsets form the training set.
Most authors~\cite{guo2017, samreen2016, martinez2018, murwantara2014, safdar2017} relayed on a 10-fold cross-validation.
10-fold cross-validation follows the same idea of a 5-fold cross-validation, producing $k=10$ groups of training and validation sets.
Guo et al.~\cite{guo2017} show that a 10-fold cross-validation design does not work well for very small samples, \textit{e.g.}, it did not work for Apache system when the sample size was 9 configurations.

\paragraph{Bootstrapping (BT)}
Four studies~\cite{aken2017,jamshidi2016,nair2017,guo2017} relayed on the bootstrapping design.
Bootstrapping relies on random sampling with replacement.
Given an input sample $S_{C}$ with $k$ configurations, bootstrapping randomly selects a configuration $C_b$, with $1 \leq b \leq k$ and copies it to 
the training set $S_t$.
However, it keeps $C_b$ in $S_{C}$ for the next selection.
This process is repeated $k$ times.
Given the training sample, bootstrapping uses $S_{C} \backslash S_t$ as the validation set $S_v$.
Nair et al.~\cite{nair2017} used a non-parametric bootstrap test with 95\% confidence for evaluating the statistical significance of their approach.
In non-parametric bootstrap, the input sample $S_{C}$ is drawn from a discrete set, \textit{i.e.}, 95\% of the resulting sample $S_t$ should fall within the 95\% confidence limits about $S_{C}$. 

\paragraph{Dynamic Sector Validation.} Westermann et al.~\cite{westermann2012} relayed on two types of Dynamic Sector Validation designs: with local prediction error scope (DSL) and with global prediction error scope (DSG).
Dynamic Sector Validation decides if a sample configuration $C_i \in S_{C}$ is part either of the training or testing set based on the sector's prediction error of the adopted learning techniques (see Section~\ref{RQ4}).
In DSL, all sectors have a prediction error that is less than the predefined threshold, while in DSG the average prediction error of all sectors is less than the predefined threshold.\\

There are studies where a sample design is \textit{not applicable}~\cite{jamshidi2017b,gargantini2017,kolesnikov2017,couto2017,porter2007,siegmund2017,xi2004,zheng2007,jamshidi2019, siegmund2008, etxeberria2014},
as well as studies~\cite{sharifloo2016,zhang2015, siegmund2013b,  hutter2011, osogami2007} without further details about the sample design. 
Given the sampling design, evaluation metrics are used to (learn and) validate the resulting prediction model.
Table~\ref{tab:RQ5} sketches which evaluation metrics are used in the literature.
The first column identifies the study reference(s). 
The second and third columns identify the name of the evaluation metric and its application objective, respectively.
Similar to Table~\ref{tab:RQ4}, here the application objective is related to the scenarios in which each evaluation metric has been already used in the SPL literature.
Therefore, in future researches, some metrics may be explored in other scenarios as well.
Notice that some studies~\cite{sharifloo2016, siegmund2012b, sincero2010} did not report any detail about the validation process. 

\begin{table} 
	\centering \small{}
    \begin{tabular}{p{0.34\textwidth} p{0.45\textwidth} p{0.13\textwidth}} 
    	\hline
		    \textbf{Reference} & \textbf{Evaluation Metric} & \textbf{Applicability} \\ 
        \hline
		    \cite{valov2015}	&	Closeness Range, Winner Probability	&	\textit{A1}	\\
            \cite{zhang2016, song2013}	&	Coverage	&	\textit{A1}	\\
            \cite{kolesnikov2017}	&	Jaccard Similarity 	&	\textit{A1}	\\
            \cite{zhang2016}	&	Performance-Relevant Feature Interactions Detection Accuracy	&	\textit{A1}	\\
            \cite{yilmaz2014}	&	t-masked metric	&	\textit{A1}	\\
            \cite{queiroz2016}	&	True Positive (TP) Rate, False Positive (FP) Rate, Receiver Operating Characteristic (ROC)	&	\textit{A1}	\\
            \cite{chen2005,guo2013,guo2017,kolesnikov2018,siegmund2011,siegmund2012a,siegmund2013,siegmund2015,valov2015,valov2017,westermann2012,zhang2015,nair2018c,couto2017,kaltenecker2019,jamshidi2019, jamshidi2018, grebhahn2017, thornton2013, duarte2018}	&	Mean Relative Error (MRE)	&	\textit{A1}, \textit{A2}, \textit{A3}, \textit{A4}	\\
            \cite{alipourfard2017,sarkar2015,weckesser2018, nair2018a, nair2018c, grebhahn2017, duarte2018, song2013}	&	Sampling Cost	&	\textit{A1}, \textit{A2}, \textit{A3}, \textit{A4}	\\
            \cite{westermann2012, zhang2015}	&	Highest Error (HE)	&	\textit{A1}, \textit{A3}	\\
            \cite{jamshidi2017a, jehooh2017, murwantara2014, martinez2018, jamshidi2018, jamshidi2016, siegmund2013b, ghamizi2019, lillacka2013, elafia2018, bao2018}	&	Mean Absolute Error (MAE)	&	\textit{A1}, \textit{A3}, \textit{A4}	\\
            \cite{kaltenecker2019, safdar2017, hutter2011}	&	Mann-Whitney U-test	&	\textit{A1}, \textit{A3}, \textit{A5}	\\
            \cite{samreen2016, kaltenecker2019}	&	F-test	&	\textit{A1}, \textit{A4}	\\
            \cite{temple2016,temple2017a,acher2018,kolesnikov2017,amand2019}	&	Precision, Recall	&	\textit{A1}, \textit{A4}, \textit{A5}	\\
            \cite{yilmaz2006,yilmaz2014, amand2019, safdar2017, queiroz2016}	&	Precision, Recall, F-measure	&	\textit{A1}, \textit{A5}	\\
            \cite{etxeberria2014}	&	GAP	&	\textit{A2}	\\
            \cite{jamshidi2017b}	&	Kullback-Leibler (KL), Pearson Linear Correlation, Spearman Correlation Coefficient	&	\textit{A2}	\\
            \cite{valov2017, kolesnikov2018}	&	Structure of Prediction Models	&	\textit{A2}	\\
            \cite{jamshidi2017b, jamshidi2019}	&	Rank Correlation	&	\textit{A2}, \textit{A4}	\\
            \cite{aken2017, grebhahn2017}	&	Domain Experts Feedback	&	\textit{A3}	\\
            \cite{osogami2007}	&	Error Probability	&	\textit{A3}	\\
            \cite{martinez2018}	&	Global Confidence, Neighbors Density Confidence, Neighbors Similarity Confidence	&	\textit{A3}	\\
            \cite{westermann2012}	&	LT15, LT30	&	\textit{A3}	\\
            \cite{nair2017, nair2018a}	&	Mean Rank Difference	&	\textit{A3}	\\
            \cite{murwantara2014}	&	Median Magnitude of the Relative Error (MdMRE)	&	\textit{A3}	\\
            \cite{zuluaga2016}	&	Pareto Prediction Error	&	\textit{A3}	\\
            \cite{bao2018, svogor2019, saleem2015}	&	Rank Accuracy (RA)	&	\textit{A3}	\\
            \cite{aken2017, ding2015, xu2008, osogami2007}	&	Tuning Time	&	\textit{A3}	\\
            \cite{murwantara2014, elafia2018, samreen2016}	&	Mean Square Error (MSE)	&	\textit{A3}, \textit{A4}	\\
            \cite{samreen2016}	&	p-value, R2, Residual Standard Error (RSE)	&	\textit{A4}	\\
            \cite{chen2009}	&	Reward	&	\textit{A4}	\\
            \cite{porter2007}	&	Statistical Significance	&	\textit{A4}	\\
            \cite{temple2018,porter2007}	&	Qualitative Analysis	&	\textit{A4}, \textit{A5}	\\
            \cite{krismayer2017}	&	Ranking Constraints	&	\textit{A4}, \textit{A5}	\\
            \cite{safdar2017}	&	Delaney’s Statistics	&	\textit{A5}	\\
            \cite{safdar2017}	&	Distance Function, Hyper-volume (HV)	&	\textit{A5}	\\
            \cite{gargantini2017}	&	Equivalence among Combinatorial Models 	&	\textit{A5}	\\
            \cite{gargantini2017}	&	Failure Index Delta (FID), Totally Repaired Models (TRM)	&	\textit{A5}	\\
        \hline
	\end{tabular}
	\caption{Validation procedure reported in the literature. \textit{A1}: Pure Prediction; \textit{A2}: Interpretability of Configurable Systems; \textit{A3}: Optimization; \textit{A4}: Dynamic Configuration; \textit{A5}: Mining Constraints; \textit{A6}: SPL Evolution.}
	\label{tab:RQ5}
\end{table}

\textbf{Evaluation metrics.} State-of-the-art techniques rely on 50 evaluation metrics from which it is possible to evaluate the accuracy of the resulting models in different application scenarios. 
There are metrics dedicated to supervised classification problems (\textit{e.g.}, precision, recall, F-measure). 
In such settings, the goal is to quantify the ratio of correct classifications to the total number of input samples. For instance, Temple et al.~\cite{temple2016} used precision and recall to control whether acceptable and non-acceptable configurations are correctly classified according to the ground truth. 
Others~\cite{temple2018,porter2007} use qualitative analysis to identify features with significant effects on defects, and understand feature interactions and decide whether further investigation of features is justified. 

There are also well known metrics for regression problems, such as, \textit{Mean Relative Error} (MRE - Equation~\ref{eq:mre}) and \textit{Mean Absolute Error} (MAE - Equation~\ref{eq:mae}). 
These metrics aim at estimating the accuracy between the exact measurements and the predicted one.

\begin{equation}
    MRE = \frac{1}{|S_v|} \sum_{C \in S_v} \frac{|C_{(p_i)} - C_{(p'_i)}|}{C_{(p_i)}}
    \label{eq:mre}
\end{equation}

\begin{equation}
    MAE = \frac{|C_{(p'_i)} - C_{(p_i)}|}{C_{(p_i)}}
    \label{eq:mae}
\end{equation}

Where $S_v$ is the validation set, and $C_{(p_i)}$ and $C_{(p'_i)}$ indicate the exact and predicted value of $p_i$ for $C \in S_v$, respectively.

Contributions addressing learning-to-rank problems develop specific metrics capable of assessing the ability of a learning method to correctly rank configurations. 
For example, Nair et al.~\cite{nair2017, nair2018a} use the error difference between the ranks of the predicted configurations and the true measured configurations.
To get insights about when stop sampling, they also discuss the trade-off between the size of the training set and rank difference.



\paragraph{Interpretability metrics.}
Some metrics are also used to better \emph{understand} configuration spaces and their distributions, for example, \emph{when} to use a transfer learning technique.  
In this context, Jamshidi et al.~\cite{jamshidi2017b} use a set of similarity metrics (Kullback-Leibler, Pearson Linear Correlation, Spearman Correlation Coefficient, and Rank Correlation) to investigate the relatedness of the source and target environments.
These metrics compute the correlation between predicted and exact values from source and target systems.
Moreover, Kolesnikov et al.~\cite{kolesnikov2018} and Valov et al.~\cite{valov2017} use size metrics as insights of interpretability.
Valov et al.~\cite{valov2017} compare the structure (size) of prediction models built for the same configurable system and trained on different hardware platforms.
They measured the size of a performance model by counting the features used in the nodes of a regression tree, while Kolesnikov et al.~\cite{kolesnikov2018} define the model size metric as the number of configuration options in every term of the linear regression model.
Temple et al.~\cite{temple2016} reported that constraints extracted from decision trees were consistent with the domain knowledge of a video generator and could help developers preventing non-acceptable configurations. In their context, interpretability is important both for validating the insights of the learning and for encoding the knowledge into a variability model.  

\paragraph{Sampling cost.}
Several works~\cite{alipourfard2017, sarkar2015, weckesser2018, nair2018a, nair2018c} use a sampling cost measurement to evaluate their prediction approach. 
In order to reduce measurement effort, these approaches aims at sampling configurations in a smart way by using as few configurations as possible. 
In \cite{alipourfard2017, weckesser2018, nair2018a, nair2018c}, sampling cost is considered as the total number of measurements required to train the model.
These authors show the trade-off between the size of the training set and the accuracy of the model.
Nair et al.~\cite{nair2018a, nair2018c} compare different learning algorithms, along with different sampling techniques to determine exactly those configurations for measurement that reveal key performance characteristics and consequently reach the minimum possible sampling cost.
Sarkar et al.~\cite{sarkar2015} introduce a cost model, which they consider the measurement cost involved in building the training and testing sets, as well as the cost of building the prediction model and computing the prediction error.
Sampling cost can be seen as an additional objective of the problem of learning configuration spaces, \textit{i.e.} not only the usual learning accuracy should be considered. 


\begin{mdframed}[backgroundcolor=gray!10] 
    The evaluation of the learning process requires to confront a trained model with new, unmeasured configurations. Hold-out is the most considered strategy for splitting configuration data into training set and testing set. Depending on the targeted task (regression, ranking, classification, understanding), several metrics have been reused or defined. In addition to learning accuracy, sampling cost and interpretability of the learning model can be considered as part of the problem and thus assessed.  
\end{mdframed}



\subsection{RQ6: What are the limitations faced by the current techniques and open challenges that need attention in the future?}
\label{RQ6}
In this section, we aim at identifying new areas of research that can lead to further enrichment of this field. Our recommendations are build upon previous insights and the analysis of papers' survey. 

%


We start with the revisit of RQ1, RQ2, RQ3, RQ4, and RQ5.

\textbf{Application objective.} 
Despite a wide applicability being in terms of application domains, subject systems, or measurement properties, there is few concrete evidence that the proposed learning techniques are adopted in practice and daily used. To the best of our knowledge there is still a \emph{gap}: we observe that end-users or developers of many configurable systems (\textit{e.g.}, x264) are not yet using learning-based solutions. 
The reasons that prevent practical adoption are a first important issue to explore and understand.
There are also research directions to further investigate to improve state-of-the-art approaches (see hereafter). 

\emph{More applications.} We report the use of learning techniques to target six main scenarios.
However, there is still room to target other applications (\textit{e.g.}, learning of multiple and composed configurable systems) or to combine scenarios (\textit{e.g.}, mining constraint to support dynamic adaptation of multiple SPLs).


\emph{Integrated tooling support.} We observe that several studies provide either open source or publicly available implementations. 
Tools could assist end-users during the configuration process and the choice of options. 
Variability management tools could be extended to encompass the learning stages reported in this SLR. It could also benefit to software developers in charge of maintaining configurable systems. 
However, none of the reported studies provide a tooling support fully integrated to existing SPL platforms or mainstream software tools. An exception is SPLConqueror~\cite{siegmund2012b} that supports an interactive configuration process.

\textbf{Sampling.}
We reported 23 high-level sampling techniques used by state-of-the-art approaches.
Most of the reported techniques only use as input a model of the configuration space (\textit{e.g.}, a feature model).
The incorporation of knowledge and other types of input data in the sampling phase needs more investigation in future research (\textit{e.g.}, system documentation, implementation artifacts, code smells, and system evolution). 

\emph{Constrained sampling.} A difficult challenge when sampling is to generate configurations conform to constraints. It boils down to solve a satisfiability (SAT) problem. Recent results show that uniform, random sampling is still challenging for configurable systems~\cite{plazar:hal-01991857}. 
Hence true random sampling is hard to instrument. 
In the context of testing SPL, there are numerous available heuristics, algorithms, and empirical studies~\cite{medeiros2016, lopez2015, machado2014, lee2012, thum2014classification}. In this case, sampling is used to efficiently test a subset of configurations. Though the sample has not the same purpose, we believe learning-based process could benefit from works done in the software testing domain. An interesting question is whether heuristics or sampling criterion devised for testing are as efficient for learning configuration spaces. 

 
\emph{Sampling numeric options.} The majority of sampling techniques operates over Boolean options. However, many configurable systems provide numerical options. 
In the literature, three numeric-sampling strategies have been reported: \textit{random}~\cite{temple2016, temple2017a, temple2018}, \textit{Cartesian product of numerical sampling}~\cite{amand2019}, and \textit{Plackett-Burman design}~\cite{siegmund2015}.
Numeric sampling have substantially different value ranges in comparison with binary sampling. The number of options' values to select can be huge while constraints should still be respected. Sampling with numeric options is still an open issue -- not a pure SAT problem.
    
\textbf{Measuring.}
There are a large number of NFPs reported in the literature (see Fig.~\ref{fig:RQ3_NFPs} and Table~\ref{tab:RQ3_systems}), most of them being quantitative and performance properties. 
The measurement of performance is subject to intensive research: there can be bias, noise or uncertainty~\cite{DBLP:journals/jss/AletiTHJ18, DBLP:journals/jss/ColmantRKSFS18, apel2019PLUS}.
Quantifying the influence of such factors on the system behaviour is important to be able to avoid unexpected issues.
Trubiani and Apel introduce the concept of uncertainty-influence models and present PLUS, a preliminary approach to learn and quantify the influence of uncertain parameters on system performance.
Only a few works deal with uncertainty when learning configuration spaces (\textit{e.g.}, ~\cite{jamshidi2017a}).
There are questions we could ask, such as how many times do we need to repeat measurements? Can we trust?
Furthermore, there is a tension between the accuracy of measurements and the need to scale up the process to thousands of configurations. For example, the use of a cloud has the merit of offering numerous computing resources at the expense of networking issues and heterogenity~\cite{DBLP:journals/toit/LeitnerC16, bao2018}. 
Though some countermeasures can be considered, it remains an important practical problem. 
At the opposite end of the spectrum, simulation and static analysis are less costly but may approximate the real measures: few studies explore them which demand further investigation. 
Another issue is how performance measurements transfer to computational environment (\textit{e.g.}, hardware). The recent rise of transfer learning techniques is an interesting direction, but much more research is needed. 
 

\textbf{Learning.}
Numerous learning techniques have been proposed and can be applied: which one practitioners should use and under which conditions? It is still missing an actionable framework that would automatically choose or recommend a suited learning based on an engineering context (\textit{e.g.}, targeted NFPs, need of interpretability). 
Our SLR is a first step in this direction, since we provide a repository of learning techniques together with their application objective. However more research is needed.


\emph{Parameter tuning and resampling.} Most of the used learning techniques have parameters that guide their execution and have a strong influence on their prediction accuracy.
However, most of the proposed approaches set these parameters manually based on the knowledge of a domain expert or even use default parameters.
Also, the use of different resampling techniques is hardly explored by current works (except by \cite{guo2017, jamshidi2016, nair2017}).

\emph{Unsupervised learning.} 
Although all studies focused only on supervised learning, unsupervised learning is another potential candidate to support the exploration of large software configuration spaces.
Still, analyzing the feasibility of unsupervised learning techniques remains unexplored. 

\emph{Transfer learning.} We notice that several recent works aim to transfer the learning-based knowledge of a configuration space to another setting (e.g., the use of another hardware). It is an important research direction, since many factors can influence software configuration spaces. 

\textbf{Validation.} There is a wide range of validation metrics reported in the literature. We observe that some metrics can be used for the same task and there is not necessarily a consensus. On the other hand, most of the studies use an unique metric for validation, which may provide incomplete quantification of the accuracy.

\emph{Interpretability.} There are few studies reported in the literature whose application objective is the comprehension of configurable systems.
However, none of these studies use an empirical user studies to validate the proposed approach level of comprehension by domain experts. 
The existing approaches use only a model size metric to report the level of comprehension~\cite{valov2017, kolesnikov2018}. 

\emph{Comprehensive evaluation.}
The reported studies analyse prediction error, model size, and computation time.
As future work, we aim at analyzing to what extent the trade-offs among ``prediction error, model size, and computation time" affect the application objective of prediction models: interactions, code smells, variability model characteristics (feature type, model height, etc), system evolution, and others. 
In addition, sampling cost could also be considered as part of the evaluation. 
Overall, we call for more comprehensive evaluation that aims to trade different aspects of concrete engineering context. 

\emph{Unified Benchmark.}
Although we noticed the use of the same system across some studies, we also noticed the discrepancy of measurements among them.
As future work, we aim at building a single trust benchmark repository with a portfolio of all data from the state-of-the-art systems in Table~\ref{tab:RQ3_systems} and we advise researchers to use these benchmarks. 
Providing a repository of the measured subject systems from various domains will facilitate the comparison of different approaches under the same scenario. 






\textbf{Synergies with other community.} 
As a final note, we notice that the problem of learning configuration spaces is at the intersection of: 
\begin{itemize}
\item artificial intelligence (mainly statistical machine learning and constraint reasoning). Though many works in this field do not specifically address software systems or configuration problems~\cite{hutter2011, putri2017}, there is an important potential to connect the dots and to reuse state-of-the-art methods for learning highly dimensional spaces like software configuration spaces; 
\item software engineering in general (including performance engineering and software testing). There are many challenges related to pure software: how to deploy, execute, and observe software configurations? How to synthesize knowledge actionable by software developers? Hence software frames the overall problem and calls to investigate specific issues (\textit{e.g.}, interpretability or transferability). 
\end{itemize}




\section{Threats to Validity} \label{threats_validity}

This section discusses potential threats to validity that might have affected the results of the SLR. 
We faced similar threats to validity as any other SLR.
The findings of this SLR may have been affected by bias in the selection of the primary studies, inaccuracy in the data extraction and in the classification of the primary studies, and incompleteness in defining the 
open challenges.
Next, we summarize the main threats to the validity of our work and the strategies we have followed to minimize their impact.
We discussed the SLR validity with respect to the two groups of common threats to validity: \textit{internal} and \textit{external} validity \citep{wohlin2000}.

\paragraph{Internal validity} 
An internal validity threat concerns the reliability of the selection and data extraction process. 
To further increase the internal validity of the review results, the search for relevant studies was conducted in several relevant scientific databases, and it was focused not only on journals but also on conferences, symposiums, and workshops.
Moreover, we conducted the inclusion and exclusion processes in parallel by involving three researchers and we cross-checked the outcome after each phase. 
In the case of disagreements, we discussed until a final decision was achieved.
Furthermore, we documented potentially relevant studies that were excluded. 
Therefore, we believe that we have not omitted any relevant study.
However, the quality of the search engines could have influenced the completeness of the identified primary studies (\textit{i.e.}, our search may have missed those studies whose authors did not use the terms we used in our search string to specify keywords).

For the selected papers, a potential threat to validity is the reliability and accuracy of the data extraction process, since not all information was obvious to extract (\textit{e.g.}, many papers lacked details about the measurement procedure and the validation design of the reported study).
Consequently, some data had to be interpreted which involved subjective decisions by the researchers.
Therefore, to ensure the validity, multiple sources of data were analyzed, \textit{i.e.}, papers, websites, technical reports, manuals, and executable. 
Moreover, whenever there was a doubt about some extracted data in a particular paper, we discussed the reported data from different perspectives in order to resolve all discrepancies.
However, we are aware that the data extraction process is a subjective activity and likely to yield different results when executed by different researchers.

\paragraph{External validity} 
A major external validity to this study was during the identified primary studies.
Key terms are directly related to the \textit{scope}
of the paper and they can suffer a high variation.
We limited the search for studies mainly targeting software systems (\textit{e.g.}, software product lines) and thus mainly focus on software engineering conferences. 
This may affect the completeness of our search results since we are aware of some studies outside the software engineering community that also address the learning of software configurable systems. 
To minimize this limitation and avoid missing relevant papers, we also analyzed the references of the primary studies to identify other relevant studies. 
In addition, this SLR was based on a strict protocol described in Section~\ref{methodology} which was discussed before the start of the review to increase the reliability of the selection and data extraction processes of the primary studies and allow other researchers to replicate this review.

Another external validity concerns the description of open challenges.
We are aware that the completeness of open challenges is another limitation that should be considered while interpreting the results of this review.
It is also important to explore general contributions from other fields outside the software domain to fully gather the spread knowledge, which may extend the list of findings in this field.
Therefore, in a future work, the list of findings highlighted in Section~\ref{RQ6} may be extended by conducting an additional review, making use of other keywords to be able to find additional relevant studies outside the software community.

\section{Related Work} \label{related_work}

After the introduction of SLR in software engineering in 2004, the number of published reviews in this field has grown significantly~\cite{kitchenham2009}.
A broad SLR has been conducted by Heradio et al.~\cite{heradio2016} to identify the most influential researched topics in SPL, and how the interest in those topics has evolved over the years.
Although these reviews are not directly related to ours, the high level of detail of their research methodology supported to structure and define our own methodology.

Benavides et al.~\cite{benavides2010} presented the results of a literature review to identify a set of operations and techniques that provide support to the automatic analysis of variability models. 
In a similar scenario, Lisboa et al.~\cite{lisboa2010} and Pereira et al.~\cite{pereira2015} conducted an SLR on variability management of SPLs. 
They reported several dozens of approaches and tools to support stakeholders in an interactive and automatic configuration process.
Strict to the application engineering phase, Ochoa et al.~\cite{ochoa2018} and Harman et al.~\cite{harman2014} conducted a literature review on the use of search-based software engineering techniques for optimization of SPL configurations.
In contrast to these previous reviews, our SLR provides further details on the use of automatic learning techniques to explore large configuration spaces. 
We contribute with a catalogue of sampling approaches, measurement procedures, learning techniques, and validation steps that serves as a summarization of the results in this specific field.

In~\cite{sayagh2018}, 14 software engineering experts and 229 Java software engineers were interviewed to identify major activities and challenges related to configuration engineering. In complement, a systematic literature review was conducted. In order to select papers, authors focused on those in which the title and abstract contain the keyword "config*", "misconfig*", or "mis-config*". On the one hand, the scope is much broader: It spans all engineering activities of configurable systems whereas our survey specifically targets learning-based approaches and applications. On the other hand, we learn that configuration spaces are studied in many different engineering contexts and we take care of considering a broader terminology (see Table~\ref{tab:keywords}, page~\pageref{tab:keywords}). 
Despite different scope and reviewing methodology, the two surveys are complementary. A research direction is to further explore how learning-based approaches classified in this survey could support configuration engineering activities identified in~\cite{sayagh2018}.

Finally, complementary to our review, several works~\cite{varshosaz2018, medeiros2016, lopez2015, machado2014, lee2012, thum2014classification} discussed product sampling techniques in the context of testing and verifying a software product line.
 They classify the proposed techniques into different categories and discuss the required input and criteria used to evaluate these techniques, such as the sample size, the rate of fault detection, and the tool support.
Our SLR could benefit from their results through the use of sampling techniques still not explored by learning techniques (\textit{e.g.} code and requirements coverage), as well as the SPL testing community could benefit from the sampling techniques reported in this paper still not used for testing. 
In~\cite{ebrahim2013}, 111 real-world Web configurators are analyzed but do not consider learning mechanisms.

Overall, the main topics of previous SLRs include variability model analysis, variability management, configuration engineering, and SPL testing. None of the aforementioned surveys directly address the use of learning techniques to explore the behaviour of large configuration spaces.



\section{Conclusion} \label{conclusion}


We presented a systematic literature review related to the use of learning techniques to analyze large configuration software spaces. 
 We analysed the literature in terms of a four-stage process: sampling, measuring, learning, and validation (see Section~\ref{preliminaries}).
Our contributions are fourfold.
 First, we identified the application of each approach which can guide researchers and industrial practitioners when searching for an appropriate technique that fits their current needs. 
Second, we classified the literature with respect to each learning stage.
Mainly, we give an in-depth view of \textit{(i)} sampling techniques and employed design; \textit{(ii)} measurement properties and effort for measurement; \textit{(iii)} employed learning techniques; and \textit{(iv)} how these techniques are empirically validated. 
Third, we provide a repository of all available subject systems used in the literature together with their application domains and qualitative or quantitative properties of interest.
We welcome any contribution by the community: The list of selected studies and their classification can be found in our Web supplementary material~\cite{pereira2019}.
Fourth, we identify the main shortcomings from existing approaches and non-addressed research areas to be explored by future work.

Our results reveal that the research in this field is application-dependent: Though the overall scheme remains the same ("sampling, measuring, learning"), the concrete choice of techniques should trade various criterion like safety, cost, accuracy, and interpretability. 
The proposed techniques have typically been validated with respect to different metrics depending on their tasks (\textit{e.g.}, performance prediction). 
Although the results are quite accurate, there is still need to decrease learning errors or to generalize predictions to multiple computing environments. Given the increasing interest and importance of this field, there are many exciting opportunities of research at the interplay of artificial intelligence and software engineering. 

\begin{acks}
 This research was partially funded by the ANR-17-CE25-0010-01 VaryVary project. We would like to thank Paul Temple for his early comments on a draft of this article. 
\end{acks}

\bibliographystyle{ACM-Reference-Format}
\bibliography{IEEEfull,MYfull,ml}


\begin{thebibliography}{121}


\ifx \showCODEN    \undefined \def \showCODEN     #1{\unskip}     \fi
\ifx \showDOI      \undefined \def \showDOI       #1{#1}\fi
\ifx \showISBNx    \undefined \def \showISBNx     #1{\unskip}     \fi
\ifx \showISBNxiii \undefined \def \showISBNxiii  #1{\unskip}     \fi
\ifx \showISSN     \undefined \def \showISSN      #1{\unskip}     \fi
\ifx \showLCCN     \undefined \def \showLCCN      #1{\unskip}     \fi
\ifx \shownote     \undefined \def \shownote      #1{#1}          \fi
\ifx \showarticletitle \undefined \def \showarticletitle #1{#1}   \fi
\ifx \showURL      \undefined \def \showURL       {\relax}        \fi
\providecommand\bibfield[2]{#2}
\providecommand\bibinfo[2]{#2}
\providecommand\natexlab[1]{#1}
\providecommand\showeprint[2][]{arXiv:#2}

\bibitem[\protect\citeauthoryear{??}{per}{2019}]%
        {pereira2019}
 \bibinfo{year}{2019}\natexlab{}.
\newblock \bibinfo{title}{Learning Software Configuration Spaces: A Systematic
  Literature Review}.
\newblock
\newblock
\urldef\tempurl%
\url{https://github.com/VaryVary/ML-configurable-SLR}
\showURL{%
\tempurl}
\newblock
\shownote{Accessed: 2019-06-04.}


\bibitem[\protect\citeauthoryear{Acher, Collet, Lahire, and France}{Acher
  et~al\mbox{.}}{2013}]%
        {acher2013}
\bibfield{author}{\bibinfo{person}{Mathieu Acher}, \bibinfo{person}{Philippe
  Collet}, \bibinfo{person}{Philippe Lahire}, {and} \bibinfo{person}{Robert~B
  France}.} \bibinfo{year}{2013}\natexlab{}.
\newblock \showarticletitle{Familiar: A domain-specific language for large
  scale management of feature models}.
\newblock \bibinfo{journal}{\emph{Science of Computer Programming (SCP)}}
  \bibinfo{volume}{78}, \bibinfo{number}{6} (\bibinfo{year}{2013}),
  \bibinfo{pages}{657--681}.
\newblock


\bibitem[\protect\citeauthoryear{Acher, Temple, Jezequel, Galindo, Martinez,
  and Ziadi}{Acher et~al\mbox{.}}{2018}]%
        {acher2018}
\bibfield{author}{\bibinfo{person}{Mathieu Acher}, \bibinfo{person}{Paul
  Temple}, \bibinfo{person}{Jean-Marc Jezequel}, \bibinfo{person}{Jos{\'e}~A
  Galindo}, \bibinfo{person}{Jabier Martinez}, {and} \bibinfo{person}{Tewfik
  Ziadi}.} \bibinfo{year}{2018}\natexlab{}.
\newblock \showarticletitle{VaryLaTeX: Learning Paper Variants That Meet
  Constraints}. In \bibinfo{booktitle}{\emph{Proceedings of the 12th
  International Workshop on Variability Modelling of Software-Intensive
  Systems}}. ACM, \bibinfo{pages}{83--88}.
\newblock


\bibitem[\protect\citeauthoryear{Akers}{Akers}{1978}]%
        {akers1978}
\bibfield{author}{\bibinfo{person}{Sheldon~B. Akers}.}
  \bibinfo{year}{1978}\natexlab{}.
\newblock \showarticletitle{Binary decision diagrams}.
\newblock \bibinfo{journal}{\emph{IEEE Transactions on computers}}
  \bibinfo{number}{6} (\bibinfo{year}{1978}), \bibinfo{pages}{509--516}.
\newblock


\bibitem[\protect\citeauthoryear{Aleti, Trubiani, van Hoorn, and
  Jamshidi}{Aleti et~al\mbox{.}}{2018}]%
        {DBLP:journals/jss/AletiTHJ18}
\bibfield{author}{\bibinfo{person}{Aldeida Aleti}, \bibinfo{person}{Catia
  Trubiani}, \bibinfo{person}{Andr{\'{e}} van Hoorn}, {and}
  \bibinfo{person}{Pooyan Jamshidi}.} \bibinfo{year}{2018}\natexlab{}.
\newblock \showarticletitle{An efficient method for uncertainty propagation in
  robust software performance estimation}.
\newblock \bibinfo{journal}{\emph{Journal of Systems and Software}}
  \bibinfo{volume}{138} (\bibinfo{year}{2018}), \bibinfo{pages}{222--235}.
\newblock
\urldef\tempurl%
\url{https://doi.org/10.1016/j.jss.2018.01.010}
\showDOI{\tempurl}


\bibitem[\protect\citeauthoryear{Alipourfard, Liu, Chen, Venkataraman, Yu, and
  Zhang}{Alipourfard et~al\mbox{.}}{2017}]%
        {alipourfard2017}
\bibfield{author}{\bibinfo{person}{Omid Alipourfard},
  \bibinfo{person}{Hongqiang~Harry Liu}, \bibinfo{person}{Jianshu Chen},
  \bibinfo{person}{Shivaram Venkataraman}, \bibinfo{person}{Minlan Yu}, {and}
  \bibinfo{person}{Ming Zhang}.} \bibinfo{year}{2017}\natexlab{}.
\newblock \showarticletitle{Cherrypick: Adaptively unearthing the best cloud
  configurations for big data analytics}. In \bibinfo{booktitle}{\emph{14th
  $\{$USENIX$\}$ Symposium on Networked Systems Design and Implementation
  ($\{$NSDI$\}$ 17)}}. \bibinfo{pages}{469--482}.
\newblock


\bibitem[\protect\citeauthoryear{Amand, Cordy, Heymans, Acher, Temple, and
  J{\'e}z{\'e}quel}{Amand et~al\mbox{.}}{2019}]%
        {amand2019}
\bibfield{author}{\bibinfo{person}{Benoit Amand}, \bibinfo{person}{Maxime
  Cordy}, \bibinfo{person}{Patrick Heymans}, \bibinfo{person}{Mathieu Acher},
  \bibinfo{person}{Paul Temple}, {and} \bibinfo{person}{Jean-Marc
  J{\'e}z{\'e}quel}.} \bibinfo{year}{2019}\natexlab{}.
\newblock \showarticletitle{Towards Learning-Aided Configuration in 3D
  Printing: Feasibility Study and Application to Defect Prediction}. In
  \bibinfo{booktitle}{\emph{Proceedings of the 13th International Workshop on
  Variability Modelling of Software-Intensive Systems}}. ACM,
  \bibinfo{pages}{7}.
\newblock


\bibitem[\protect\citeauthoryear{Apel, Batory, K{\"a}stner, and Saake}{Apel
  et~al\mbox{.}}{2013}]%
        {apel2013book}
\bibfield{author}{\bibinfo{person}{Sven Apel}, \bibinfo{person}{Don Batory},
  \bibinfo{person}{Christian K{\"a}stner}, {and} \bibinfo{person}{Gunter
  Saake}.} \bibinfo{year}{2013}\natexlab{}.
\newblock \bibinfo{booktitle}{\emph{Feature-Oriented Software Product Lines:
  Concepts and Implementation}}.
\newblock \bibinfo{publisher}{Springer-Verlag}.
\newblock


\bibitem[\protect\citeauthoryear{B{\k{a}}k, Diskin, Antkiewicz, Czarnecki, and
  W{\k{a}}sowski}{B{\k{a}}k et~al\mbox{.}}{2016}]%
        {bkak2016}
\bibfield{author}{\bibinfo{person}{Kacper B{\k{a}}k}, \bibinfo{person}{Zinovy
  Diskin}, \bibinfo{person}{Micha{\l} Antkiewicz}, \bibinfo{person}{Krzysztof
  Czarnecki}, {and} \bibinfo{person}{Andrzej W{\k{a}}sowski}.}
  \bibinfo{year}{2016}\natexlab{}.
\newblock \showarticletitle{Clafer: unifying class and feature modeling}.
\newblock \bibinfo{journal}{\emph{Software \& Systems Modeling}}
  \bibinfo{volume}{15}, \bibinfo{number}{3} (\bibinfo{year}{2016}),
  \bibinfo{pages}{811--845}.
\newblock


\bibitem[\protect\citeauthoryear{Bao, Liu, Xu, and Fang}{Bao
  et~al\mbox{.}}{2018}]%
        {bao2018}
\bibfield{author}{\bibinfo{person}{Liang Bao}, \bibinfo{person}{Xin Liu},
  \bibinfo{person}{Ziheng Xu}, {and} \bibinfo{person}{Baoyin Fang}.}
  \bibinfo{year}{2018}\natexlab{}.
\newblock \showarticletitle{AutoConfig: automatic configuration tuning for
  distributed message systems}. In \bibinfo{booktitle}{\emph{IEEE/ACM
  International Conference on Automated Software Engineering (ASE)}}. ACM,
  \bibinfo{pages}{29--40}.
\newblock


\bibitem[\protect\citeauthoryear{Benavides, Mart{\'\i}n-Arroyo, and
  Cort{\'e}s}{Benavides et~al\mbox{.}}{2005}]%
        {benavides2005}
\bibfield{author}{\bibinfo{person}{David Benavides},
  \bibinfo{person}{Pablo~Trinidad Mart{\'\i}n-Arroyo}, {and}
  \bibinfo{person}{Antonio~Ruiz Cort{\'e}s}.} \bibinfo{year}{2005}\natexlab{}.
\newblock \showarticletitle{Automated reasoning on feature models.}. In
  \bibinfo{booktitle}{\emph{International Conference on Advanced Information
  Systems Engineering (CAiSE)}}, Vol.~\bibinfo{volume}{5}. Springer,
  \bibinfo{pages}{491--503}.
\newblock


\bibitem[\protect\citeauthoryear{Benavides, Segura, and
  Ruiz-Cort{\'e}s}{Benavides et~al\mbox{.}}{2010}]%
        {benavides2010}
\bibfield{author}{\bibinfo{person}{David Benavides}, \bibinfo{person}{Sergio
  Segura}, {and} \bibinfo{person}{Antonio Ruiz-Cort{\'e}s}.}
  \bibinfo{year}{2010}\natexlab{}.
\newblock \showarticletitle{{Automated analysis of feature models 20 years
  later: a literature review}}.
\newblock \bibinfo{journal}{\emph{Information Systems}} \bibinfo{volume}{35},
  \bibinfo{number}{6} (\bibinfo{year}{2010}), \bibinfo{pages}{615--708}.
\newblock


\bibitem[\protect\citeauthoryear{Cashman, Cohen, Ranjan, and
  Cottingham}{Cashman et~al\mbox{.}}{2018}]%
        {cohenASE2018}
\bibfield{author}{\bibinfo{person}{Mikaela Cashman}, \bibinfo{person}{Myra~B.
  Cohen}, \bibinfo{person}{Priya Ranjan}, {and} \bibinfo{person}{Robert~W.
  Cottingham}.} \bibinfo{year}{2018}\natexlab{}.
\newblock \showarticletitle{Navigating the maze: the impact of configurability
  in bioinformatics software}. In \bibinfo{booktitle}{\emph{IEEE/ACM
  International Conference on Automated Software Engineering (ASE)}}.
  \bibinfo{pages}{757--767}.
\newblock
\urldef\tempurl%
\url{https://doi.org/10.1145/3238147.3240466}
\showDOI{\tempurl}


\bibitem[\protect\citeauthoryear{Chen, Jiang, Zhang, and Yoshihira}{Chen
  et~al\mbox{.}}{2009}]%
        {chen2009}
\bibfield{author}{\bibinfo{person}{Haifeng Chen}, \bibinfo{person}{Guofei
  Jiang}, \bibinfo{person}{Hui Zhang}, {and} \bibinfo{person}{Kenji
  Yoshihira}.} \bibinfo{year}{2009}\natexlab{}.
\newblock \showarticletitle{Boosting the performance of computing systems
  through adaptive configuration tuning}. In \bibinfo{booktitle}{\emph{ACM
  Symposium on Applied Computing (SAC)}}. ACM, \bibinfo{pages}{1045--1049}.
\newblock


\bibitem[\protect\citeauthoryear{Chen, Liu, Gorton, and Liu}{Chen
  et~al\mbox{.}}{2005}]%
        {chen2005}
\bibfield{author}{\bibinfo{person}{Shiping Chen}, \bibinfo{person}{Yan Liu},
  \bibinfo{person}{Ian Gorton}, {and} \bibinfo{person}{Anna Liu}.}
  \bibinfo{year}{2005}\natexlab{}.
\newblock \showarticletitle{Performance prediction of component-based
  applications}.
\newblock \bibinfo{journal}{\emph{Journal of Systems and Software}}
  \bibinfo{volume}{74}, \bibinfo{number}{1} (\bibinfo{year}{2005}),
  \bibinfo{pages}{35--43}.
\newblock


\bibitem[\protect\citeauthoryear{Colmant, Rouvoy, Kurpicz, Sobe, Felber, and
  Seinturier}{Colmant et~al\mbox{.}}{2018}]%
        {DBLP:journals/jss/ColmantRKSFS18}
\bibfield{author}{\bibinfo{person}{Maxime Colmant}, \bibinfo{person}{Romain
  Rouvoy}, \bibinfo{person}{Mascha Kurpicz}, \bibinfo{person}{Anita Sobe},
  \bibinfo{person}{Pascal Felber}, {and} \bibinfo{person}{Lionel Seinturier}.}
  \bibinfo{year}{2018}\natexlab{}.
\newblock \showarticletitle{The next 700 {CPU} power models}.
\newblock \bibinfo{journal}{\emph{Journal of Systems and Software}}
  \bibinfo{volume}{144} (\bibinfo{year}{2018}), \bibinfo{pages}{382--396}.
\newblock
\urldef\tempurl%
\url{https://doi.org/10.1016/j.jss.2018.07.001}
\showDOI{\tempurl}


\bibitem[\protect\citeauthoryear{Couto, Borba, Cunha, Fernandes, Pereira, and
  Saraiva}{Couto et~al\mbox{.}}{2017}]%
        {couto2017}
\bibfield{author}{\bibinfo{person}{Marco Couto}, \bibinfo{person}{Paulo Borba},
  \bibinfo{person}{J{\'a}come Cunha}, \bibinfo{person}{Jo{\~a}o~Paulo
  Fernandes}, \bibinfo{person}{Rui Pereira}, {and} \bibinfo{person}{Jo{\~a}o
  Saraiva}.} \bibinfo{year}{2017}\natexlab{}.
\newblock \showarticletitle{Products go green: Worst-case energy consumption in
  software product lines}. In \bibinfo{booktitle}{\emph{Proceedings of the 21st
  International Systems and Software Product Line Conference-Volume A}}. ACM,
  \bibinfo{pages}{84--93}.
\newblock


\bibitem[\protect\citeauthoryear{Ding, Ansel, Veeramachaneni, Shen, O’Reilly,
  and Amarasinghe}{Ding et~al\mbox{.}}{2015}]%
        {ding2015}
\bibfield{author}{\bibinfo{person}{Yufei Ding}, \bibinfo{person}{Jason Ansel},
  \bibinfo{person}{Kalyan Veeramachaneni}, \bibinfo{person}{Xipeng Shen},
  \bibinfo{person}{Una-May O’Reilly}, {and} \bibinfo{person}{Saman
  Amarasinghe}.} \bibinfo{year}{2015}\natexlab{}.
\newblock \showarticletitle{Autotuning algorithmic choice for input
  sensitivity}. In \bibinfo{booktitle}{\emph{ACM SIGPLAN Notices}},
  Vol.~\bibinfo{volume}{50}. ACM, \bibinfo{pages}{379--390}.
\newblock


\bibitem[\protect\citeauthoryear{do~Carmo~Machado, Mcgregor, Cavalcanti, and
  De~Almeida}{do~Carmo~Machado et~al\mbox{.}}{2014}]%
        {machado2014}
\bibfield{author}{\bibinfo{person}{Ivan do Carmo~Machado},
  \bibinfo{person}{John~D Mcgregor}, \bibinfo{person}{Yguarat{\~a}~Cerqueira
  Cavalcanti}, {and} \bibinfo{person}{Eduardo~Santana De~Almeida}.}
  \bibinfo{year}{2014}\natexlab{}.
\newblock \showarticletitle{On strategies for testing software product lines: A
  systematic literature review}.
\newblock \bibinfo{journal}{\emph{Information and Software Technology}}
  \bibinfo{volume}{56}, \bibinfo{number}{10} (\bibinfo{year}{2014}),
  \bibinfo{pages}{1183--1199}.
\newblock


\bibitem[\protect\citeauthoryear{Duarte, Gil, Romano, Lopes, and
  Rodrigues}{Duarte et~al\mbox{.}}{2018}]%
        {duarte2018}
\bibfield{author}{\bibinfo{person}{Francisco Duarte}, \bibinfo{person}{Richard
  Gil}, \bibinfo{person}{Paolo Romano}, \bibinfo{person}{Ant{\'o}nia Lopes},
  {and} \bibinfo{person}{Lu{\'\i}s Rodrigues}.}
  \bibinfo{year}{2018}\natexlab{}.
\newblock \showarticletitle{Learning non-deterministic impact models for
  adaptation}. In \bibinfo{booktitle}{\emph{Proceedings of the 13th
  International Conference on Software Engineering for Adaptive and
  Self-Managing Systems}}. ACM, \bibinfo{pages}{196--205}.
\newblock


\bibitem[\protect\citeauthoryear{Eichelberger, Qin, Sizonenko, and
  Schmid}{Eichelberger et~al\mbox{.}}{2016}]%
        {eichelberger2016}
\bibfield{author}{\bibinfo{person}{Holger Eichelberger}, \bibinfo{person}{Cui
  Qin}, \bibinfo{person}{Roman Sizonenko}, {and} \bibinfo{person}{Klaus
  Schmid}.} \bibinfo{year}{2016}\natexlab{}.
\newblock \showarticletitle{Using IVML to model the topology of big data
  processing pipelines}. In \bibinfo{booktitle}{\emph{International Systems and
  Software Product Line Conference (SPLC)}}. ACM, \bibinfo{pages}{204--208}.
\newblock


\bibitem[\protect\citeauthoryear{El~Afia and Sarhani}{El~Afia and
  Sarhani}{2017}]%
        {elafia2018}
\bibfield{author}{\bibinfo{person}{Abdellatif El~Afia} {and}
  \bibinfo{person}{Malek Sarhani}.} \bibinfo{year}{2017}\natexlab{}.
\newblock \showarticletitle{Performance prediction using support vector machine
  for the configuration of optimization algorithms}. In
  \bibinfo{booktitle}{\emph{2017 3rd International Conference of Cloud
  Computing Technologies and Applications (CloudTech)}}. IEEE,
  \bibinfo{pages}{1--7}.
\newblock


\bibitem[\protect\citeauthoryear{Etxeberria, Trubiani, Cortellessa, and
  Sagardui}{Etxeberria et~al\mbox{.}}{2014}]%
        {etxeberria2014}
\bibfield{author}{\bibinfo{person}{Leire Etxeberria}, \bibinfo{person}{Catia
  Trubiani}, \bibinfo{person}{Vittorio Cortellessa}, {and}
  \bibinfo{person}{Goiuria Sagardui}.} \bibinfo{year}{2014}\natexlab{}.
\newblock \showarticletitle{Performance-based selection of software and
  hardware features under parameter uncertainty}. In
  \bibinfo{booktitle}{\emph{Proceedings of the 10th international ACM Sigsoft
  conference on Quality of software architectures}}. ACM,
  \bibinfo{pages}{23--32}.
\newblock


\bibitem[\protect\citeauthoryear{Gargantini, Petke, and Radavelli}{Gargantini
  et~al\mbox{.}}{2017}]%
        {gargantini2017}
\bibfield{author}{\bibinfo{person}{Angelo Gargantini}, \bibinfo{person}{Justyna
  Petke}, {and} \bibinfo{person}{Marco Radavelli}.}
  \bibinfo{year}{2017}\natexlab{}.
\newblock \showarticletitle{Combinatorial interaction testing for automated
  constraint repair}. In \bibinfo{booktitle}{\emph{2017 IEEE International
  Conference on Software Testing, Verification and Validation Workshops
  (ICSTW)}}. IEEE, \bibinfo{pages}{239--248}.
\newblock


\bibitem[\protect\citeauthoryear{Ghamizi, Cordy, Papadakis, and Traon}{Ghamizi
  et~al\mbox{.}}{2019}]%
        {ghamizi2019}
\bibfield{author}{\bibinfo{person}{Salah Ghamizi}, \bibinfo{person}{Maxime
  Cordy}, \bibinfo{person}{Mike Papadakis}, {and} \bibinfo{person}{Yves~Le
  Traon}.} \bibinfo{year}{2019}\natexlab{}.
\newblock \showarticletitle{Automated Search for Configurations of Deep Neural
  Network Architectures}.
\newblock \bibinfo{journal}{\emph{arXiv preprint arXiv:1904.04612}}
  (\bibinfo{year}{2019}).
\newblock


\bibitem[\protect\citeauthoryear{Grebhahn, Rodrigo, Siegmund, Gaspar, and
  Apel}{Grebhahn et~al\mbox{.}}{2017}]%
        {grebhahn2017}
\bibfield{author}{\bibinfo{person}{Alexander Grebhahn}, \bibinfo{person}{Carmen
  Rodrigo}, \bibinfo{person}{Norbert Siegmund}, \bibinfo{person}{Francisco~J
  Gaspar}, {and} \bibinfo{person}{Sven Apel}.} \bibinfo{year}{2017}\natexlab{}.
\newblock \showarticletitle{Performance-influence models of multigrid methods:
  A case study on triangular grids}.
\newblock \bibinfo{journal}{\emph{Concurrency and Computation: Practice and
  Experience}} \bibinfo{volume}{29}, \bibinfo{number}{17}
  (\bibinfo{year}{2017}), \bibinfo{pages}{e4057}.
\newblock


\bibitem[\protect\citeauthoryear{Guo, Czarnecki, Apel, Siegmund, and
  Wasowski}{Guo et~al\mbox{.}}{2013}]%
        {guo2013}
\bibfield{author}{\bibinfo{person}{Jianmei Guo}, \bibinfo{person}{Krzysztof
  Czarnecki}, \bibinfo{person}{Sven Apel}, \bibinfo{person}{Norbert Siegmund},
  {and} \bibinfo{person}{Andrzej Wasowski}.} \bibinfo{year}{2013}\natexlab{}.
\newblock \showarticletitle{Variability-aware performance prediction: A
  statistical learning approach}. In \bibinfo{booktitle}{\emph{ASE}}.
\newblock


\bibitem[\protect\citeauthoryear{Guo, White, Wang, Li, and Wang}{Guo
  et~al\mbox{.}}{2011}]%
        {guo2011}
\bibfield{author}{\bibinfo{person}{Jianmei Guo}, \bibinfo{person}{Jules White},
  \bibinfo{person}{Guangxin Wang}, \bibinfo{person}{Jian Li}, {and}
  \bibinfo{person}{Yinglin Wang}.} \bibinfo{year}{2011}\natexlab{}.
\newblock \showarticletitle{A genetic algorithm for optimized feature selection
  with resource constraints in software product lines}.
\newblock \bibinfo{journal}{\emph{Journal of Systems and Software}}
  \bibinfo{volume}{84}, \bibinfo{number}{12} (\bibinfo{year}{2011}),
  \bibinfo{pages}{2208--2221}.
\newblock


\bibitem[\protect\citeauthoryear{Guo, Yang, Siegmund, Apel, Sarkar, Valov,
  Czarnecki, Wasowski, and Yu}{Guo et~al\mbox{.}}{2017}]%
        {guo2017}
\bibfield{author}{\bibinfo{person}{Jianmei Guo}, \bibinfo{person}{Dingyu Yang},
  \bibinfo{person}{Norbert Siegmund}, \bibinfo{person}{Sven Apel},
  \bibinfo{person}{Atrisha Sarkar}, \bibinfo{person}{Pavel Valov},
  \bibinfo{person}{Krzysztof Czarnecki}, \bibinfo{person}{Andrzej Wasowski},
  {and} \bibinfo{person}{Huiqun Yu}.} \bibinfo{year}{2017}\natexlab{}.
\newblock \showarticletitle{Data-efficient performance learning for
  configurable systems}.
\newblock \bibinfo{journal}{\emph{Empirical Software Engineering}}
  (\bibinfo{year}{2017}), \bibinfo{pages}{1--42}.
\newblock


\bibitem[\protect\citeauthoryear{Halin, Nuttinck, Acher, Devroey, Perrouin, and
  Baudry}{Halin et~al\mbox{.}}{2018}]%
        {halin:hal-01829928}
\bibfield{author}{\bibinfo{person}{Axel Halin}, \bibinfo{person}{Alexandre
  Nuttinck}, \bibinfo{person}{Mathieu Acher}, \bibinfo{person}{Xavier Devroey},
  \bibinfo{person}{Gilles Perrouin}, {and} \bibinfo{person}{Benoit Baudry}.}
  \bibinfo{year}{2018}\natexlab{}.
\newblock \showarticletitle{Test them all, is it worth it? Assessing
  configuration sampling on the JHipster Web development stack}.
\newblock \bibinfo{journal}{\emph{{Empirical Software Engineering}}}
  (\bibinfo{date}{July} \bibinfo{year}{2018}).
\newblock
\urldef\tempurl%
\url{https://doi.org/10.07980}
\showDOI{\tempurl}
\newblock
\shownote{Empirical Software Engineering journal.}


\bibitem[\protect\citeauthoryear{Hall, Frank, Holmes, Pfahringer, Reutemann,
  and Witten}{Hall et~al\mbox{.}}{2009}]%
        {weka2009}
\bibfield{author}{\bibinfo{person}{Mark Hall}, \bibinfo{person}{Eibe Frank},
  \bibinfo{person}{Geoffrey Holmes}, \bibinfo{person}{Bernhard Pfahringer},
  \bibinfo{person}{Peter Reutemann}, {and} \bibinfo{person}{Ian~H Witten}.}
  \bibinfo{year}{2009}\natexlab{}.
\newblock \showarticletitle{The WEKA data mining software: an update}.
\newblock \bibinfo{journal}{\emph{ACM SIGKDD explorations newsletter}}
  \bibinfo{volume}{11}, \bibinfo{number}{1} (\bibinfo{year}{2009}),
  \bibinfo{pages}{10--18}.
\newblock


\bibitem[\protect\citeauthoryear{Hallsteinsen, Hinchey, Park, and
  Schmid}{Hallsteinsen et~al\mbox{.}}{2008}]%
        {DBLP:journals/computer/HallsteinsenHPS08}
\bibfield{author}{\bibinfo{person}{Svein~O. Hallsteinsen},
  \bibinfo{person}{Mike Hinchey}, \bibinfo{person}{Sooyong Park}, {and}
  \bibinfo{person}{Klaus Schmid}.} \bibinfo{year}{2008}\natexlab{}.
\newblock \showarticletitle{Dynamic Software Product Lines}.
\newblock \bibinfo{journal}{\emph{{IEEE} Computer}} \bibinfo{volume}{41},
  \bibinfo{number}{4} (\bibinfo{year}{2008}), \bibinfo{pages}{93--95}.
\newblock
\urldef\tempurl%
\url{https://doi.org/10.1109/MC.2008.123}
\showDOI{\tempurl}


\bibitem[\protect\citeauthoryear{Harman, Jia, Krinke, Langdon, Petke, and
  Zhang}{Harman et~al\mbox{.}}{2014}]%
        {harman2014}
\bibfield{author}{\bibinfo{person}{Mark Harman}, \bibinfo{person}{Yue Jia},
  \bibinfo{person}{Jens Krinke}, \bibinfo{person}{William~B Langdon},
  \bibinfo{person}{Justyna Petke}, {and} \bibinfo{person}{Yuanyuan Zhang}.}
  \bibinfo{year}{2014}\natexlab{}.
\newblock \showarticletitle{Search based software engineering for software
  product line engineering: a survey and directions for future work}. In
  \bibinfo{booktitle}{\emph{Proceedings of the 18th International Software
  Product Line Conference-Volume 1}}. ACM, \bibinfo{pages}{5--18}.
\newblock


\bibitem[\protect\citeauthoryear{Heradio, Perez-Morago, Fernandez-Amoros,
  Cabrerizo, and Herrera-Viedma}{Heradio et~al\mbox{.}}{2016}]%
        {heradio2016}
\bibfield{author}{\bibinfo{person}{Ruben Heradio}, \bibinfo{person}{Hector
  Perez-Morago}, \bibinfo{person}{David Fernandez-Amoros},
  \bibinfo{person}{Francisco~Javier Cabrerizo}, {and} \bibinfo{person}{Enrique
  Herrera-Viedma}.} \bibinfo{year}{2016}\natexlab{}.
\newblock \showarticletitle{A bibliometric analysis of 20 years of research on
  software product lines}.
\newblock \bibinfo{journal}{\emph{Information and Software Technology}}
  \bibinfo{volume}{72} (\bibinfo{year}{2016}), \bibinfo{pages}{1--15}.
\newblock


\bibitem[\protect\citeauthoryear{Hoda, Salleh, Grundy, and Tee}{Hoda
  et~al\mbox{.}}{2017}]%
        {hoda2017}
\bibfield{author}{\bibinfo{person}{Rashina Hoda}, \bibinfo{person}{Norsaremah
  Salleh}, \bibinfo{person}{John Grundy}, {and} \bibinfo{person}{Hui~Mien
  Tee}.} \bibinfo{year}{2017}\natexlab{}.
\newblock \showarticletitle{Systematic literature reviews in agile software
  development: A tertiary study}.
\newblock \bibinfo{journal}{\emph{Information and software technology}}
  \bibinfo{volume}{85} (\bibinfo{year}{2017}), \bibinfo{pages}{60--70}.
\newblock


\bibitem[\protect\citeauthoryear{Hutter, Hoos, and Leyton-Brown}{Hutter
  et~al\mbox{.}}{2011}]%
        {hutter2011}
\bibfield{author}{\bibinfo{person}{Frank Hutter}, \bibinfo{person}{Holger~H
  Hoos}, {and} \bibinfo{person}{Kevin Leyton-Brown}.}
  \bibinfo{year}{2011}\natexlab{}.
\newblock \showarticletitle{Sequential model-based optimization for general
  algorithm configuration}. In \bibinfo{booktitle}{\emph{International
  Conference on Learning and Intelligent Optimization}}. Springer,
  \bibinfo{pages}{507--523}.
\newblock


\bibitem[\protect\citeauthoryear{Jamshidi, C{\'a}mara, Schmerl, K{\"a}stner,
  and Garlan}{Jamshidi et~al\mbox{.}}{2019}]%
        {jamshidi2019}
\bibfield{author}{\bibinfo{person}{Pooyan Jamshidi}, \bibinfo{person}{Javier
  C{\'a}mara}, \bibinfo{person}{Bradley Schmerl}, \bibinfo{person}{Christian
  K{\"a}stner}, {and} \bibinfo{person}{David Garlan}.}
  \bibinfo{year}{2019}\natexlab{}.
\newblock \showarticletitle{Machine Learning Meets Quantitative Planning:
  Enabling Self-Adaptation in Autonomous Robots}.
\newblock \bibinfo{journal}{\emph{arXiv preprint arXiv:1903.03920}}
  (\bibinfo{year}{2019}).
\newblock


\bibitem[\protect\citeauthoryear{Jamshidi and Casale}{Jamshidi and
  Casale}{2016}]%
        {jamshidi2016}
\bibfield{author}{\bibinfo{person}{Pooyan Jamshidi} {and}
  \bibinfo{person}{Giuliano Casale}.} \bibinfo{year}{2016}\natexlab{}.
\newblock \showarticletitle{An uncertainty-aware approach to optimal
  configuration of stream processing systems}. In
  \bibinfo{booktitle}{\emph{2016 IEEE 24th International Symposium on Modeling,
  Analysis and Simulation of Computer and Telecommunication Systems
  (MASCOTS)}}. IEEE, \bibinfo{pages}{39--48}.
\newblock


\bibitem[\protect\citeauthoryear{Jamshidi, Siegmund, Velez, K{\"{a}}stner,
  Patel, and Agarwal}{Jamshidi et~al\mbox{.}}{2017a}]%
        {jamshidi2017b}
\bibfield{author}{\bibinfo{person}{Pooyan Jamshidi}, \bibinfo{person}{Norbert
  Siegmund}, \bibinfo{person}{Miguel Velez}, \bibinfo{person}{Christian
  K{\"{a}}stner}, \bibinfo{person}{Akshay Patel}, {and} \bibinfo{person}{Yuvraj
  Agarwal}.} \bibinfo{year}{2017}\natexlab{a}.
\newblock \showarticletitle{Transfer learning for performance modeling of
  configurable systems: an exploratory analysis}. In
  \bibinfo{booktitle}{\emph{IEEE/ACM International Conference on Automated
  Software Engineering (ASE)}}. IEEE Press, \bibinfo{pages}{497--508}.
\newblock
\urldef\tempurl%
\url{http://dl.acm.org/citation.cfm?id=3155625}
\showURL{%
\tempurl}


\bibitem[\protect\citeauthoryear{Jamshidi, Velez, K{\"a}stner, and
  Siegmund}{Jamshidi et~al\mbox{.}}{2018}]%
        {jamshidi2018}
\bibfield{author}{\bibinfo{person}{Pooyan Jamshidi}, \bibinfo{person}{Miguel
  Velez}, \bibinfo{person}{Christian K{\"a}stner}, {and}
  \bibinfo{person}{Norbert Siegmund}.} \bibinfo{year}{2018}\natexlab{}.
\newblock \showarticletitle{Learning to sample: exploiting similarities across
  environments to learn performance models for configurable systems}. In
  \bibinfo{booktitle}{\emph{Proceedings of the 2018 26th ACM Joint Meeting on
  European Software Engineering Conference and Symposium on the Foundations of
  Software Engineering}}. ACM, \bibinfo{pages}{71--82}.
\newblock


\bibitem[\protect\citeauthoryear{Jamshidi, Velez, K{\"{a}}stner, Siegmund, and
  Kawthekar}{Jamshidi et~al\mbox{.}}{2017b}]%
        {jamshidi2017a}
\bibfield{author}{\bibinfo{person}{Pooyan Jamshidi}, \bibinfo{person}{Miguel
  Velez}, \bibinfo{person}{Christian K{\"{a}}stner}, \bibinfo{person}{Norbert
  Siegmund}, {and} \bibinfo{person}{Prasad Kawthekar}.}
  \bibinfo{year}{2017}\natexlab{b}.
\newblock \showarticletitle{Transfer Learning for Improving Model Predictions
  in Highly Configurable Software}. In \bibinfo{booktitle}{\emph{12th
  {IEEE/ACM} International Symposium on Software Engineering for Adaptive and
  Self-Managing Systems, SEAMS@ICSE 2017, Buenos Aires, Argentina, May 22-23,
  2017}}. \bibinfo{pages}{31--41}.
\newblock
\urldef\tempurl%
\url{https://doi.org/10.1109/SEAMS.2017.11}
\showDOI{\tempurl}


\bibitem[\protect\citeauthoryear{Kaltenecker, Grebhahn, Siegmund, Guo, and
  Apel}{Kaltenecker et~al\mbox{.}}{2019}]%
        {kaltenecker2019}
\bibfield{author}{\bibinfo{person}{Christian Kaltenecker},
  \bibinfo{person}{Alexander Grebhahn}, \bibinfo{person}{Norbert Siegmund},
  \bibinfo{person}{Jianmei Guo}, {and} \bibinfo{person}{Sven Apel}.}
  \bibinfo{year}{2019}\natexlab{}.
\newblock \showarticletitle{Distance-Based Sampling of Software Configuration
  Spaces}. In \bibinfo{booktitle}{\emph{Proceedings of the IEEE/ACM
  International Conference on Software Engineering (ICSE). ACM}}.
\newblock


\bibitem[\protect\citeauthoryear{Kang, Cohen, Hess, Novak, and Peterson}{Kang
  et~al\mbox{.}}{1990}]%
        {kang1990}
\bibfield{author}{\bibinfo{person}{K. Kang}, \bibinfo{person}{S. Cohen},
  \bibinfo{person}{J. Hess}, \bibinfo{person}{W. Novak}, {and}
  \bibinfo{person}{S. Peterson}.} \bibinfo{year}{1990}\natexlab{}.
\newblock \bibinfo{booktitle}{\emph{{Feature-Oriented Domain Analysis
  (FODA)}}}.
\newblock \bibinfo{type}{{T}echnical {R}eport} CMU/SEI-90-TR-21.
  \bibinfo{institution}{SEI}.
\newblock


\bibitem[\protect\citeauthoryear{Khalil~Abbasi, Hubaux, Acher, Boucher, and
  Heymans}{Khalil~Abbasi et~al\mbox{.}}{2013}]%
        {ebrahim2013}
\bibfield{author}{\bibinfo{person}{Ebrahim Khalil~Abbasi},
  \bibinfo{person}{Arnaud Hubaux}, \bibinfo{person}{Mathieu Acher},
  \bibinfo{person}{Quentin Boucher}, {and} \bibinfo{person}{Patrick Heymans}.}
  \bibinfo{year}{2013}\natexlab{}.
\newblock \showarticletitle{The Anatomy of a Sales Configurator: An Empirical
  Study of 111 Cases}. In \bibinfo{booktitle}{\emph{CAiSE'13}}.
\newblock


\bibitem[\protect\citeauthoryear{Kitchenham, Brereton, Budgen, Turner, Bailey,
  and Linkman}{Kitchenham et~al\mbox{.}}{2009}]%
        {kitchenham2009}
\bibfield{author}{\bibinfo{person}{Barbara Kitchenham},
  \bibinfo{person}{O~Pearl Brereton}, \bibinfo{person}{David Budgen},
  \bibinfo{person}{Mark Turner}, \bibinfo{person}{John Bailey}, {and}
  \bibinfo{person}{Stephen Linkman}.} \bibinfo{year}{2009}\natexlab{}.
\newblock \showarticletitle{Systematic literature reviews in software
  engineering--a systematic literature review}.
\newblock \bibinfo{journal}{\emph{Information and Software Technology (IST)}}
  \bibinfo{volume}{51}, \bibinfo{number}{1} (\bibinfo{year}{2009}),
  \bibinfo{pages}{7--15}.
\newblock


\bibitem[\protect\citeauthoryear{Kitchenham and Charters}{Kitchenham and
  Charters}{2007}]%
        {kitchenham2007}
\bibfield{author}{\bibinfo{person}{Barbara Kitchenham} {and}
  \bibinfo{person}{Stuart Charters}.} \bibinfo{year}{2007}\natexlab{}.
\newblock \showarticletitle{Guidelines for performing Systematic Literature
  Reviews in Software Engineering}.
\newblock \bibinfo{publisher}{Citeseer}.
\newblock


\bibitem[\protect\citeauthoryear{Kolesnikov, Siegmund, K{\"a}stner, and
  Apel}{Kolesnikov et~al\mbox{.}}{2017}]%
        {kolesnikov2017}
\bibfield{author}{\bibinfo{person}{Sergiy Kolesnikov}, \bibinfo{person}{Norbert
  Siegmund}, \bibinfo{person}{Christian K{\"a}stner}, {and}
  \bibinfo{person}{Sven Apel}.} \bibinfo{year}{2017}\natexlab{}.
\newblock \showarticletitle{On the relation of external and internal feature
  interactions: A case study}.
\newblock \bibinfo{journal}{\emph{arXiv preprint arXiv:1712.07440}}
  (\bibinfo{year}{2017}).
\newblock


\bibitem[\protect\citeauthoryear{Kolesnikov, Siegmund, K{\"a}stner, Grebhahn,
  and Apel}{Kolesnikov et~al\mbox{.}}{2018}]%
        {kolesnikov2018}
\bibfield{author}{\bibinfo{person}{Sergiy Kolesnikov}, \bibinfo{person}{Norbert
  Siegmund}, \bibinfo{person}{Christian K{\"a}stner},
  \bibinfo{person}{Alexander Grebhahn}, {and} \bibinfo{person}{Sven Apel}.}
  \bibinfo{year}{2018}\natexlab{}.
\newblock \showarticletitle{Tradeoffs in modeling performance of highly
  configurable software systems}.
\newblock \bibinfo{journal}{\emph{Software \& Systems Modeling}}
  (\bibinfo{year}{2018}), \bibinfo{pages}{1--19}.
\newblock


\bibitem[\protect\citeauthoryear{Krismayer, Rabiser, and
  Gr{\"u}nbacher}{Krismayer et~al\mbox{.}}{2017}]%
        {krismayer2017}
\bibfield{author}{\bibinfo{person}{Thomas Krismayer}, \bibinfo{person}{Rick
  Rabiser}, {and} \bibinfo{person}{Paul Gr{\"u}nbacher}.}
  \bibinfo{year}{2017}\natexlab{}.
\newblock \showarticletitle{Mining constraints for event-based monitoring in
  systems of systems}. In \bibinfo{booktitle}{\emph{IEEE/ACM International
  Conference on Automated Software Engineering (ASE)}}. IEEE Press,
  \bibinfo{pages}{826--831}.
\newblock


\bibitem[\protect\citeauthoryear{Lee, Kang, and Lee}{Lee et~al\mbox{.}}{2012}]%
        {lee2012}
\bibfield{author}{\bibinfo{person}{Jihyun Lee}, \bibinfo{person}{Sungwon Kang},
  {and} \bibinfo{person}{Danhyung Lee}.} \bibinfo{year}{2012}\natexlab{}.
\newblock \showarticletitle{A survey on software product line testing}. In
  \bibinfo{booktitle}{\emph{Proceedings of the 16th International Software
  Product Line Conference-Volume 1}}. ACM, \bibinfo{pages}{31--40}.
\newblock


\bibitem[\protect\citeauthoryear{Leitner and Cito}{Leitner and Cito}{2016}]%
        {DBLP:journals/toit/LeitnerC16}
\bibfield{author}{\bibinfo{person}{Philipp Leitner} {and}
  \bibinfo{person}{J{\"{u}}rgen Cito}.} \bibinfo{year}{2016}\natexlab{}.
\newblock \showarticletitle{Patterns in the Chaos - {A} Study of Performance
  Variation and Predictability in Public IaaS Clouds}.
\newblock \bibinfo{journal}{\emph{{ACM} Trans. Internet Techn.}}
  \bibinfo{volume}{16}, \bibinfo{number}{3} (\bibinfo{year}{2016}),
  \bibinfo{pages}{15:1--15:23}.
\newblock
\urldef\tempurl%
\url{https://doi.org/10.1145/2885497}
\showDOI{\tempurl}


\bibitem[\protect\citeauthoryear{Lillack, M{\"u}ller, and Eisenecker}{Lillack
  et~al\mbox{.}}{2013}]%
        {lillacka2013}
\bibfield{author}{\bibinfo{person}{Max Lillack}, \bibinfo{person}{Johannes
  M{\"u}ller}, {and} \bibinfo{person}{Ulrich~W Eisenecker}.}
  \bibinfo{year}{2013}\natexlab{}.
\newblock \showarticletitle{Improved prediction of non-functional properties in
  software product lines with domain context}.
\newblock \bibinfo{journal}{\emph{Software Engineering 2013}}
  (\bibinfo{year}{2013}).
\newblock


\bibitem[\protect\citeauthoryear{Lisboa, Garcia, Lucr{\'e}dio, de~Almeida,
  de~Lemos~Meira, and de~Mattos~Fortes}{Lisboa et~al\mbox{.}}{2010}]%
        {lisboa2010}
\bibfield{author}{\bibinfo{person}{Liana~Barachisio Lisboa},
  \bibinfo{person}{Vinicius~Cardoso Garcia}, \bibinfo{person}{Daniel
  Lucr{\'e}dio}, \bibinfo{person}{Eduardo~Santana de Almeida},
  \bibinfo{person}{Silvio~Romero de Lemos~Meira}, {and}
  \bibinfo{person}{Renata~Pontin de Mattos~Fortes}.}
  \bibinfo{year}{2010}\natexlab{}.
\newblock \showarticletitle{A systematic review of domain analysis tools}.
\newblock \bibinfo{journal}{\emph{Information and Software Technology}}
  \bibinfo{volume}{52}, \bibinfo{number}{1} (\bibinfo{year}{2010}),
  \bibinfo{pages}{1--13}.
\newblock


\bibitem[\protect\citeauthoryear{Lopez-Herrejon, Fischer, Ramler, and
  Egyed}{Lopez-Herrejon et~al\mbox{.}}{2015}]%
        {lopez2015}
\bibfield{author}{\bibinfo{person}{Roberto~E Lopez-Herrejon},
  \bibinfo{person}{Stefan Fischer}, \bibinfo{person}{Rudolf Ramler}, {and}
  \bibinfo{person}{Alexander Egyed}.} \bibinfo{year}{2015}\natexlab{}.
\newblock \showarticletitle{A first systematic mapping study on combinatorial
  interaction testing for software product lines}. In
  \bibinfo{booktitle}{\emph{2015 IEEE Eighth International Conference on
  Software Testing, Verification and Validation Workshops (ICSTW)}}. IEEE,
  \bibinfo{pages}{1--10}.
\newblock


\bibitem[\protect\citeauthoryear{Martinez, Sottet, Frey, Bissyand{\'e}, Ziadi,
  Klein, Temple, Acher, and Le~Traon}{Martinez et~al\mbox{.}}{2018}]%
        {martinez2018}
\bibfield{author}{\bibinfo{person}{Jabier Martinez},
  \bibinfo{person}{Jean-S{\'e}bastien Sottet},
  \bibinfo{person}{Alfonso~Garc{\'\i}a Frey}, \bibinfo{person}{Tegawend{\'e}~F
  Bissyand{\'e}}, \bibinfo{person}{Tewfik Ziadi}, \bibinfo{person}{Jacques
  Klein}, \bibinfo{person}{Paul Temple}, \bibinfo{person}{Mathieu Acher}, {and}
  \bibinfo{person}{Yves Le~Traon}.} \bibinfo{year}{2018}\natexlab{}.
\newblock \showarticletitle{Towards Estimating and Predicting User Perception
  on Software Product Variants}. In \bibinfo{booktitle}{\emph{International
  Conference on Software Reuse}}. Springer, \bibinfo{pages}{23--40}.
\newblock


\bibitem[\protect\citeauthoryear{Medeiros, K{\"a}stner, Ribeiro, Gheyi, and
  Apel}{Medeiros et~al\mbox{.}}{2016}]%
        {medeiros2016}
\bibfield{author}{\bibinfo{person}{Fl{\'a}vio Medeiros},
  \bibinfo{person}{Christian K{\"a}stner}, \bibinfo{person}{M{\'a}rcio
  Ribeiro}, \bibinfo{person}{Rohit Gheyi}, {and} \bibinfo{person}{Sven Apel}.}
  \bibinfo{year}{2016}\natexlab{}.
\newblock \showarticletitle{A comparison of 10 sampling algorithms for
  configurable systems}. In \bibinfo{booktitle}{\emph{Proceedings of the 38th
  International Conference on Software Engineering}}. ACM,
  \bibinfo{pages}{643--654}.
\newblock


\bibitem[\protect\citeauthoryear{Mendonca, Wasowski, Czarnecki, and
  Cowan}{Mendonca et~al\mbox{.}}{2008}]%
        {mendonca2008}
\bibfield{author}{\bibinfo{person}{Marcilio Mendonca}, \bibinfo{person}{Andrzej
  Wasowski}, \bibinfo{person}{Krzysztof Czarnecki}, {and}
  \bibinfo{person}{Donald Cowan}.} \bibinfo{year}{2008}\natexlab{}.
\newblock \showarticletitle{Efficient compilation techniques for large scale
  feature models}. In \bibinfo{booktitle}{\emph{Proceedings of the 7th
  international conference on Generative programming and component
  engineering}}. ACM, \bibinfo{pages}{13--22}.
\newblock


\bibitem[\protect\citeauthoryear{Morin, Barais, J{\'{e}}z{\'{e}}quel, Fleurey,
  and Solberg}{Morin et~al\mbox{.}}{2009}]%
        {DBLP:journals/computer/MorinBJFS09}
\bibfield{author}{\bibinfo{person}{Brice Morin}, \bibinfo{person}{Olivier
  Barais}, \bibinfo{person}{Jean{-}Marc J{\'{e}}z{\'{e}}quel},
  \bibinfo{person}{Franck Fleurey}, {and} \bibinfo{person}{Arnor Solberg}.}
  \bibinfo{year}{2009}\natexlab{}.
\newblock \showarticletitle{Models@ Run.time to Support Dynamic Adaptation}.
\newblock \bibinfo{journal}{\emph{{IEEE} Computer}} \bibinfo{volume}{42},
  \bibinfo{number}{10} (\bibinfo{year}{2009}), \bibinfo{pages}{44--51}.
\newblock
\urldef\tempurl%
\url{https://doi.org/10.1109/MC.2009.327}
\showDOI{\tempurl}


\bibitem[\protect\citeauthoryear{Murashkin, Antkiewicz, Rayside, and
  Czarnecki}{Murashkin et~al\mbox{.}}{2013}]%
        {murashkin2013}
\bibfield{author}{\bibinfo{person}{Alexandr Murashkin},
  \bibinfo{person}{Micha{\l} Antkiewicz}, \bibinfo{person}{Derek Rayside},
  {and} \bibinfo{person}{Krzysztof Czarnecki}.}
  \bibinfo{year}{2013}\natexlab{}.
\newblock \showarticletitle{Visualization and exploration of optimal variants
  in product line engineering}. In \bibinfo{booktitle}{\emph{Proceedings of the
  17th International Software Product Line Conference}}.
  \bibinfo{publisher}{ACM}, \bibinfo{pages}{111--115}.
\newblock


\bibitem[\protect\citeauthoryear{Murwantara, Bordbar, and Minku}{Murwantara
  et~al\mbox{.}}{2014}]%
        {murwantara2014}
\bibfield{author}{\bibinfo{person}{I~Made Murwantara}, \bibinfo{person}{Behzad
  Bordbar}, {and} \bibinfo{person}{Leandro~L. Minku}.}
  \bibinfo{year}{2014}\natexlab{}.
\newblock \showarticletitle{Measuring Energy Consumption for Web Service
  Product Configuration}. In \bibinfo{booktitle}{\emph{Proceedings of the 16th
  International Conference on Information Integration and Web-based
  Applications \&\#38; Services}} \emph{(\bibinfo{series}{iiWAS '14})}.
  \bibinfo{publisher}{ACM}, \bibinfo{address}{New York, NY, USA},
  \bibinfo{pages}{224--228}.
\newblock
\showISBNx{978-1-4503-3001-5}
\urldef\tempurl%
\url{https://doi.org/10.1145/2684200.2684314}
\showDOI{\tempurl}


\bibitem[\protect\citeauthoryear{Nair, Menzies, Siegmund, and Apel}{Nair
  et~al\mbox{.}}{2017}]%
        {nair2017}
\bibfield{author}{\bibinfo{person}{Vivek Nair}, \bibinfo{person}{Tim Menzies},
  \bibinfo{person}{Norbert Siegmund}, {and} \bibinfo{person}{Sven Apel}.}
  \bibinfo{year}{2017}\natexlab{}.
\newblock \showarticletitle{Using bad learners to find good configurations}. In
  \bibinfo{booktitle}{\emph{Proceedings of the 2017 11th Joint Meeting on
  Foundations of Software Engineering, {ESEC/FSE} 2017, Paderborn, Germany,
  September 4-8, 2017}}. \bibinfo{pages}{257--267}.
\newblock
\urldef\tempurl%
\url{https://doi.org/10.1145/3106237.3106238}
\showDOI{\tempurl}


\bibitem[\protect\citeauthoryear{Nair, Menzies, Siegmund, and Apel}{Nair
  et~al\mbox{.}}{2018a}]%
        {nair2018c}
\bibfield{author}{\bibinfo{person}{Vivek Nair}, \bibinfo{person}{Tim Menzies},
  \bibinfo{person}{Norbert Siegmund}, {and} \bibinfo{person}{Sven Apel}.}
  \bibinfo{year}{2018}\natexlab{a}.
\newblock \showarticletitle{Faster discovery of faster system configurations
  with spectral learning}.
\newblock \bibinfo{journal}{\emph{Automated Software Engineering}}
  (\bibinfo{year}{2018}), \bibinfo{pages}{1--31}.
\newblock


\bibitem[\protect\citeauthoryear{Nair, Yu, Menzies, Siegmund, and Apel}{Nair
  et~al\mbox{.}}{2018b}]%
        {nair2018a}
\bibfield{author}{\bibinfo{person}{Vivek Nair}, \bibinfo{person}{Zhe Yu},
  \bibinfo{person}{Tim Menzies}, \bibinfo{person}{Norbert Siegmund}, {and}
  \bibinfo{person}{Sven Apel}.} \bibinfo{year}{2018}\natexlab{b}.
\newblock \showarticletitle{Finding faster configurations using flash}.
\newblock \bibinfo{journal}{\emph{IEEE Transactions on Software Engineering}}
  (\bibinfo{year}{2018}).
\newblock


\bibitem[\protect\citeauthoryear{Ochoa, Gonzalez-Rojas, Juliana, Castro, and
  Saake}{Ochoa et~al\mbox{.}}{2018}]%
        {ochoa2018}
\bibfield{author}{\bibinfo{person}{Lina Ochoa}, \bibinfo{person}{Oscar
  Gonzalez-Rojas}, \bibinfo{person}{Alves~Pereira Juliana},
  \bibinfo{person}{Harold Castro}, {and} \bibinfo{person}{Gunter Saake}.}
  \bibinfo{year}{2018}\natexlab{}.
\newblock \showarticletitle{A systematic literature review on the
  semi-automatic configuration of extended product lines}.
\newblock \bibinfo{journal}{\emph{Journal of Systems and Software}}
  \bibinfo{volume}{144} (\bibinfo{year}{2018}), \bibinfo{pages}{511--532}.
\newblock


\bibitem[\protect\citeauthoryear{Ochoa, Gonz{\'a}lez-Rojas, and Th{\"u}m}{Ochoa
  et~al\mbox{.}}{2015}]%
        {ochoa2015}
\bibfield{author}{\bibinfo{person}{Lina Ochoa}, \bibinfo{person}{Oscar
  Gonz{\'a}lez-Rojas}, {and} \bibinfo{person}{Thomas Th{\"u}m}.}
  \bibinfo{year}{2015}\natexlab{}.
\newblock \showarticletitle{Using decision rules for solving conflicts in
  extended feature models}. In \bibinfo{booktitle}{\emph{International
  Conference on Software Language Engineering (SLE)}}. ACM,
  \bibinfo{pages}{149--160}.
\newblock


\bibitem[\protect\citeauthoryear{Oh, Batory, Myers, and Siegmund}{Oh
  et~al\mbox{.}}{2017}]%
        {jehooh2017}
\bibfield{author}{\bibinfo{person}{Jeho Oh}, \bibinfo{person}{Don~S. Batory},
  \bibinfo{person}{Margaret Myers}, {and} \bibinfo{person}{Norbert Siegmund}.}
  \bibinfo{year}{2017}\natexlab{}.
\newblock \showarticletitle{Finding near-optimal configurations in product
  lines by random sampling}. In \bibinfo{booktitle}{\emph{Proceedings of the
  2017 11th Joint Meeting on Foundations of Software Engineering, {ESEC/FSE}
  2017, Paderborn, Germany, September 4-8, 2017}}. \bibinfo{pages}{61--71}.
\newblock
\urldef\tempurl%
\url{https://doi.org/10.1145/3106237.3106273}
\showDOI{\tempurl}


\bibitem[\protect\citeauthoryear{Osogami and Kato}{Osogami and Kato}{2007}]%
        {osogami2007}
\bibfield{author}{\bibinfo{person}{Takayuki Osogami} {and} \bibinfo{person}{Sei
  Kato}.} \bibinfo{year}{2007}\natexlab{}.
\newblock \showarticletitle{Optimizing system configurations quickly by
  guessing at the performance}. In \bibinfo{booktitle}{\emph{ACM SIGMETRICS
  Performance Evaluation Review}}, Vol.~\bibinfo{volume}{35}. ACM,
  \bibinfo{pages}{145--156}.
\newblock


\bibitem[\protect\citeauthoryear{Pereira, Constantino, and Figueiredo}{Pereira
  et~al\mbox{.}}{2015}]%
        {pereira2015}
\bibfield{author}{\bibinfo{person}{Juliana~Alves Pereira},
  \bibinfo{person}{Kattiana Constantino}, {and} \bibinfo{person}{Eduardo
  Figueiredo}.} \bibinfo{year}{2015}\natexlab{}.
\newblock \showarticletitle{{A systematic literature review of software product
  line management tools}}. In \bibinfo{booktitle}{\emph{International
  Conference on Software Reuse (ICSR)}}. Springer, \bibinfo{pages}{73--89}.
\newblock


\bibitem[\protect\citeauthoryear{Pereira, Matuszyk, Krieter, Spiliopoulou, and
  Saake}{Pereira et~al\mbox{.}}{2018}]%
        {pereira2018a}
\bibfield{author}{\bibinfo{person}{Juliana~Alves Pereira},
  \bibinfo{person}{Pawel Matuszyk}, \bibinfo{person}{Sebastian Krieter},
  \bibinfo{person}{Myra Spiliopoulou}, {and} \bibinfo{person}{Gunter Saake}.}
  \bibinfo{year}{2018}\natexlab{}.
\newblock \showarticletitle{Personalized recommender systems for product-line
  configuration processes}.
\newblock \bibinfo{journal}{\emph{Computer Languages, Systems \& Structures}}
  (\bibinfo{year}{2018}).
\newblock


\bibitem[\protect\citeauthoryear{Plazar, Acher, Perrouin, Devroey, and
  Cordy}{Plazar et~al\mbox{.}}{2019}]%
        {plazar:hal-01991857}
\bibfield{author}{\bibinfo{person}{Quentin Plazar}, \bibinfo{person}{Mathieu
  Acher}, \bibinfo{person}{Gilles Perrouin}, \bibinfo{person}{Xavier Devroey},
  {and} \bibinfo{person}{Maxime Cordy}.} \bibinfo{year}{2019}\natexlab{}.
\newblock \showarticletitle{Uniform Sampling of SAT Solutions for Configurable
  Systems: Are We There Yet?}. In \bibinfo{booktitle}{\emph{ICST 2019 - 12th
  International Conference on Software Testing, Verification, and Validation}}.
  \bibinfo{address}{Xian, China}, \bibinfo{pages}{1--12}.
\newblock
\urldef\tempurl%
\url{https://hal.inria.fr/hal-01991857}
\showURL{%
\tempurl}


\bibitem[\protect\citeauthoryear{Pohl, B{\"o}ckle, and van~der Linden}{Pohl
  et~al\mbox{.}}{2005}]%
        {pohl2005}
\bibfield{author}{\bibinfo{person}{Klaus Pohl}, \bibinfo{person}{G{\"u}nter
  B{\"o}ckle}, {and} \bibinfo{person}{Frank~J. van~der Linden}.}
  \bibinfo{year}{2005}\natexlab{}.
\newblock \bibinfo{booktitle}{\emph{{Software product line engineering:
  foundations, principles and techniques}}}.
\newblock \bibinfo{publisher}{Springer}, \bibinfo{address}{Berlin Heidelberg}.
\newblock


\bibitem[\protect\citeauthoryear{Pohl, Lauenroth, and Pohl}{Pohl
  et~al\mbox{.}}{2011}]%
        {pohl2011}
\bibfield{author}{\bibinfo{person}{Richard Pohl}, \bibinfo{person}{Kim
  Lauenroth}, {and} \bibinfo{person}{Klaus Pohl}.}
  \bibinfo{year}{2011}\natexlab{}.
\newblock \showarticletitle{A performance comparison of contemporary
  algorithmic approaches for automated analysis operations on feature models}.
  In \bibinfo{booktitle}{\emph{IEEE/ACM International Conference on Automated
  Software Engineering (ASE)}}. IEEE Computer Society,
  \bibinfo{pages}{313--322}.
\newblock


\bibitem[\protect\citeauthoryear{Porter, Yilmaz, Memon, Schmidt, and
  Natarajan}{Porter et~al\mbox{.}}{2007}]%
        {porter2007}
\bibfield{author}{\bibinfo{person}{Adam Porter}, \bibinfo{person}{Cemal
  Yilmaz}, \bibinfo{person}{Atif~M Memon}, \bibinfo{person}{Douglas~C Schmidt},
  {and} \bibinfo{person}{Bala Natarajan}.} \bibinfo{year}{2007}\natexlab{}.
\newblock \showarticletitle{Skoll: A process and infrastructure for distributed
  continuous quality assurance}.
\newblock \bibinfo{journal}{\emph{IEEE Transactions on Software Engineering}}
  \bibinfo{volume}{33}, \bibinfo{number}{8} (\bibinfo{year}{2007}),
  \bibinfo{pages}{510--525}.
\newblock


\bibitem[\protect\citeauthoryear{Putri et~al\mbox{.}}{Putri
  et~al\mbox{.}}{2017}]%
        {putri2017}
\bibfield{author}{\bibinfo{person}{Sukmawati~Anggraeni Putri} {et~al\mbox{.}}}
  \bibinfo{year}{2017}\natexlab{}.
\newblock \showarticletitle{Combining integreted sampling technique with
  feature selection for software defect prediction}. In
  \bibinfo{booktitle}{\emph{2017 5th International Conference on Cyber and IT
  Service Management (CITSM)}}. IEEE, \bibinfo{pages}{1--6}.
\newblock


\bibitem[\protect\citeauthoryear{Queiroz, Berger, and Czarnecki}{Queiroz
  et~al\mbox{.}}{2016}]%
        {queiroz2016}
\bibfield{author}{\bibinfo{person}{Rodrigo Queiroz}, \bibinfo{person}{Thorsten
  Berger}, {and} \bibinfo{person}{Krzysztof Czarnecki}.}
  \bibinfo{year}{2016}\natexlab{}.
\newblock \showarticletitle{Towards predicting feature defects in software
  product lines}. In \bibinfo{booktitle}{\emph{Proceedings of the 7th
  International Workshop on Feature-Oriented Software Development}}. ACM,
  \bibinfo{pages}{58--62}.
\newblock


\bibitem[\protect\citeauthoryear{Roos-Frantz, Benavides, Ruiz-Cort{\'e}s,
  Heuer, and Lauenroth}{Roos-Frantz et~al\mbox{.}}{2012}]%
        {frantz2012}
\bibfield{author}{\bibinfo{person}{Fabricia Roos-Frantz},
  \bibinfo{person}{David Benavides}, \bibinfo{person}{Antonio Ruiz-Cort{\'e}s},
  \bibinfo{person}{Andr{\'e} Heuer}, {and} \bibinfo{person}{Kim Lauenroth}.}
  \bibinfo{year}{2012}\natexlab{}.
\newblock \showarticletitle{Quality-aware analysis in product line engineering
  with the orthogonal variability model}.
\newblock \bibinfo{journal}{\emph{Software Quality Journal}}
  \bibinfo{volume}{20}, \bibinfo{number}{3-4} (\bibinfo{year}{2012}),
  \bibinfo{pages}{519--565}.
\newblock


\bibitem[\protect\citeauthoryear{Safdar, Lu, Yue, and Ali}{Safdar
  et~al\mbox{.}}{2017}]%
        {safdar2017}
\bibfield{author}{\bibinfo{person}{Safdar~Aqeel Safdar}, \bibinfo{person}{Hong
  Lu}, \bibinfo{person}{Tao Yue}, {and} \bibinfo{person}{Shaukat Ali}.}
  \bibinfo{year}{2017}\natexlab{}.
\newblock \showarticletitle{Mining cross product line rules with
  multi-objective search and machine learning}. In
  \bibinfo{booktitle}{\emph{Proceedings of the Genetic and Evolutionary
  Computation Conference}}. ACM, \bibinfo{pages}{1319--1326}.
\newblock


\bibitem[\protect\citeauthoryear{Saleem, Ding, Liu, and Chi}{Saleem
  et~al\mbox{.}}{2015}]%
        {saleem2015}
\bibfield{author}{\bibinfo{person}{Muhammad~Suleman Saleem},
  \bibinfo{person}{Chen Ding}, \bibinfo{person}{Xumin Liu}, {and}
  \bibinfo{person}{Chi-Hung Chi}.} \bibinfo{year}{2015}\natexlab{}.
\newblock \showarticletitle{Personalized decision-strategy based web service
  selection using a learning-to-rank algorithm}.
\newblock \bibinfo{journal}{\emph{IEEE Transactions on Services Computing}}
  \bibinfo{volume}{8}, \bibinfo{number}{5} (\bibinfo{year}{2015}),
  \bibinfo{pages}{727--739}.
\newblock


\bibitem[\protect\citeauthoryear{Samreen, Elkhatib, Rowe, and Blair}{Samreen
  et~al\mbox{.}}{2016}]%
        {samreen2016}
\bibfield{author}{\bibinfo{person}{Faiza Samreen}, \bibinfo{person}{Yehia
  Elkhatib}, \bibinfo{person}{Matthew Rowe}, {and} \bibinfo{person}{Gordon~S
  Blair}.} \bibinfo{year}{2016}\natexlab{}.
\newblock \showarticletitle{Daleel: Simplifying cloud instance selection using
  machine learning}. In \bibinfo{booktitle}{\emph{NOMS 2016-2016 IEEE/IFIP
  Network Operations and Management Symposium}}. IEEE,
  \bibinfo{pages}{557--563}.
\newblock


\bibitem[\protect\citeauthoryear{Sarkar, Guo, Siegmund, Apel, and
  Czarnecki}{Sarkar et~al\mbox{.}}{2015}]%
        {sarkar2015}
\bibfield{author}{\bibinfo{person}{Atri Sarkar}, \bibinfo{person}{Jianmei Guo},
  \bibinfo{person}{Norbert Siegmund}, \bibinfo{person}{Sven Apel}, {and}
  \bibinfo{person}{Krzysztof Czarnecki}.} \bibinfo{year}{2015}\natexlab{}.
\newblock \showarticletitle{Cost-efficient sampling for performance prediction
  of configurable systems (t)}. In \bibinfo{booktitle}{\emph{IEEE/ACM
  International Conference on Automated Software Engineering (ASE)}}. IEEE,
  \bibinfo{pages}{342--352}.
\newblock


\bibitem[\protect\citeauthoryear{Sayagh, Kerzazi, Adams, and Petrillo}{Sayagh
  et~al\mbox{.}}{2018}]%
        {sayagh2018}
\bibfield{author}{\bibinfo{person}{Mohammed Sayagh},
  \bibinfo{person}{Noureddine Kerzazi}, \bibinfo{person}{Bram Adams}, {and}
  \bibinfo{person}{Fabio Petrillo}.} \bibinfo{year}{2018}\natexlab{}.
\newblock \showarticletitle{Software Configuration Engineering in Practice:
  Interviews, Survey, and Systematic Literature Review}.
\newblock \bibinfo{journal}{\emph{IEEE Transactions on Software Engineering}}
  (\bibinfo{year}{2018}).
\newblock


\bibitem[\protect\citeauthoryear{Sayyad, Menzies, and Ammar}{Sayyad
  et~al\mbox{.}}{2013}]%
        {sayyad2013}
\bibfield{author}{\bibinfo{person}{Abdel~Salam Sayyad}, \bibinfo{person}{Tim
  Menzies}, {and} \bibinfo{person}{Hany Ammar}.}
  \bibinfo{year}{2013}\natexlab{}.
\newblock \showarticletitle{On the value of user preferences in search-based
  software engineering: a case study in software product lines}. In
  \bibinfo{booktitle}{\emph{Proceedings of the 2013 International Conference on
  Software Engineering}}. IEEE Press, \bibinfo{pages}{492--501}.
\newblock


\bibitem[\protect\citeauthoryear{Sharifloo, Metzger, Quinton, Baresi, and
  Pohl}{Sharifloo et~al\mbox{.}}{2016}]%
        {sharifloo2016}
\bibfield{author}{\bibinfo{person}{Amir~Molzam Sharifloo},
  \bibinfo{person}{Andreas Metzger}, \bibinfo{person}{Cl{\'e}ment Quinton},
  \bibinfo{person}{Luciano Baresi}, {and} \bibinfo{person}{Klaus Pohl}.}
  \bibinfo{year}{2016}\natexlab{}.
\newblock \showarticletitle{Learning and evolution in dynamic software product
  lines}. In \bibinfo{booktitle}{\emph{2016 IEEE/ACM 11th International
  Symposium on Software Engineering for Adaptive and Self-Managing Systems
  (SEAMS)}}. IEEE, \bibinfo{pages}{158--164}.
\newblock


\bibitem[\protect\citeauthoryear{Siegmund, Grebhahn, Apel, and
  K\"{a}stner}{Siegmund et~al\mbox{.}}{2015}]%
        {siegmund2015}
\bibfield{author}{\bibinfo{person}{Norbert Siegmund},
  \bibinfo{person}{Alexander Grebhahn}, \bibinfo{person}{Sven Apel}, {and}
  \bibinfo{person}{Christian K\"{a}stner}.} \bibinfo{year}{2015}\natexlab{}.
\newblock \showarticletitle{Performance-influence Models for Highly
  Configurable Systems}. In \bibinfo{booktitle}{\emph{Proceedings of the 2015
  10th Joint Meeting on Foundations of Software Engineering}}
  \emph{(\bibinfo{series}{ESEC/FSE 2015})}. \bibinfo{pages}{284--294}.
\newblock


\bibitem[\protect\citeauthoryear{Siegmund, Kolesnikov, K{\"a}stner, Apel,
  Batory, Rosenm{\"u}ller, and Saake}{Siegmund et~al\mbox{.}}{2012a}]%
        {siegmund2012a}
\bibfield{author}{\bibinfo{person}{Norbert Siegmund},
  \bibinfo{person}{Sergiy~S. Kolesnikov}, \bibinfo{person}{Christian
  K{\"a}stner}, \bibinfo{person}{Sven Apel}, \bibinfo{person}{Don~S. Batory},
  \bibinfo{person}{Marko Rosenm{\"u}ller}, {and} \bibinfo{person}{Gunter
  Saake}.} \bibinfo{year}{2012}\natexlab{a}.
\newblock \showarticletitle{Predicting performance via automated
  feature-interaction detection}. In \bibinfo{booktitle}{\emph{ICSE}}.
  \bibinfo{pages}{167--177}.
\newblock


\bibitem[\protect\citeauthoryear{Siegmund, Rosenm{\"u}ller, K{\"a}stner,
  Giarrusso, Apel, and Kolesnikov}{Siegmund et~al\mbox{.}}{2011}]%
        {siegmund2011}
\bibfield{author}{\bibinfo{person}{Norbert Siegmund}, \bibinfo{person}{Marko
  Rosenm{\"u}ller}, \bibinfo{person}{Christian K{\"a}stner},
  \bibinfo{person}{Paolo~G Giarrusso}, \bibinfo{person}{Sven Apel}, {and}
  \bibinfo{person}{Sergiy~S Kolesnikov}.} \bibinfo{year}{2011}\natexlab{}.
\newblock \showarticletitle{Scalable prediction of non-functional properties in
  software product lines}. In \bibinfo{booktitle}{\emph{Software Product Line
  Conference (SPLC), 2011 15th International}}. \bibinfo{pages}{160--169}.
\newblock


\bibitem[\protect\citeauthoryear{Siegmund, Rosenm{\"u}ller, K{\"a}stner,
  Giarrusso, Apel, and Kolesnikov}{Siegmund et~al\mbox{.}}{2013a}]%
        {siegmund2013}
\bibfield{author}{\bibinfo{person}{Norbert Siegmund}, \bibinfo{person}{Marko
  Rosenm{\"u}ller}, \bibinfo{person}{Christian K{\"a}stner},
  \bibinfo{person}{Paolo~G Giarrusso}, \bibinfo{person}{Sven Apel}, {and}
  \bibinfo{person}{Sergiy~S Kolesnikov}.} \bibinfo{year}{2013}\natexlab{a}.
\newblock \showarticletitle{Scalable prediction of non-functional properties in
  software product lines: Footprint and memory consumption}.
\newblock \bibinfo{journal}{\emph{Information and Software Technology}}
  \bibinfo{volume}{55}, \bibinfo{number}{3} (\bibinfo{year}{2013}),
  \bibinfo{pages}{491--507}.
\newblock


\bibitem[\protect\citeauthoryear{Siegmund, Rosenm{\"u}ller, Kuhlemann,
  K{\"a}stner, Apel, and Saake}{Siegmund et~al\mbox{.}}{2012b}]%
        {siegmund2012b}
\bibfield{author}{\bibinfo{person}{Norbert Siegmund}, \bibinfo{person}{Marko
  Rosenm{\"u}ller}, \bibinfo{person}{Martin Kuhlemann},
  \bibinfo{person}{Christian K{\"a}stner}, \bibinfo{person}{Sven Apel}, {and}
  \bibinfo{person}{Gunter Saake}.} \bibinfo{year}{2012}\natexlab{b}.
\newblock \showarticletitle{SPL Conqueror: Toward optimization of
  non-functional properties in software product lines}.
\newblock \bibinfo{journal}{\emph{Software Quality Journal}}
  \bibinfo{volume}{20}, \bibinfo{number}{3-4} (\bibinfo{year}{2012}),
  \bibinfo{pages}{487--517}.
\newblock


\bibitem[\protect\citeauthoryear{Siegmund, Rosenm{\"u}ller, Kuhlemann,
  K{\"a}stner, and Saake}{Siegmund et~al\mbox{.}}{2008}]%
        {siegmund2008}
\bibfield{author}{\bibinfo{person}{Norbert Siegmund}, \bibinfo{person}{Marko
  Rosenm{\"u}ller}, \bibinfo{person}{Martin Kuhlemann},
  \bibinfo{person}{Christian K{\"a}stner}, {and} \bibinfo{person}{Gunter
  Saake}.} \bibinfo{year}{2008}\natexlab{}.
\newblock \showarticletitle{Measuring non-functional properties in software
  product line for product derivation}. In \bibinfo{booktitle}{\emph{2008 15th
  Asia-Pacific Software Engineering Conference}}. IEEE,
  \bibinfo{pages}{187--194}.
\newblock


\bibitem[\protect\citeauthoryear{Siegmund, Sobernig, and Apel}{Siegmund
  et~al\mbox{.}}{2017}]%
        {siegmund2017}
\bibfield{author}{\bibinfo{person}{Norbert Siegmund}, \bibinfo{person}{Stefan
  Sobernig}, {and} \bibinfo{person}{Sven Apel}.}
  \bibinfo{year}{2017}\natexlab{}.
\newblock \showarticletitle{Attributed variability models: outside the comfort
  zone}. In \bibinfo{booktitle}{\emph{Proceedings of the 2017 11th Joint
  Meeting on Foundations of Software Engineering}}. ACM,
  \bibinfo{pages}{268--278}.
\newblock


\bibitem[\protect\citeauthoryear{Siegmund, von Rhein, and Apel}{Siegmund
  et~al\mbox{.}}{2013b}]%
        {siegmund2013b}
\bibfield{author}{\bibinfo{person}{Norbert Siegmund},
  \bibinfo{person}{Alexander von Rhein}, {and} \bibinfo{person}{Sven Apel}.}
  \bibinfo{year}{2013}\natexlab{b}.
\newblock \showarticletitle{Family-based performance measurement}. In
  \bibinfo{booktitle}{\emph{ACM SIGPLAN Notices}}, Vol.~\bibinfo{volume}{49}.
  ACM, \bibinfo{pages}{95--104}.
\newblock


\bibitem[\protect\citeauthoryear{Sincero, Schroder-Preikschat, and
  Spinczyk}{Sincero et~al\mbox{.}}{2010}]%
        {sincero2010}
\bibfield{author}{\bibinfo{person}{Julio Sincero}, \bibinfo{person}{Wolfgang
  Schroder-Preikschat}, {and} \bibinfo{person}{Olaf Spinczyk}.}
  \bibinfo{year}{2010}\natexlab{}.
\newblock \showarticletitle{Approaching non-functional properties of software
  product lines: Learning from products}. In \bibinfo{booktitle}{\emph{Software
  Engineering Conference (APSEC), 2010 17th Asia Pacific}}.
  \bibinfo{pages}{147--155}.
\newblock


\bibitem[\protect\citeauthoryear{Song, Porter, and Foster}{Song
  et~al\mbox{.}}{2013}]%
        {song2013}
\bibfield{author}{\bibinfo{person}{Charles Song}, \bibinfo{person}{Adam
  Porter}, {and} \bibinfo{person}{Jeffrey~S Foster}.}
  \bibinfo{year}{2013}\natexlab{}.
\newblock \showarticletitle{iTree: efficiently discovering high-coverage
  configurations using interaction trees}.
\newblock \bibinfo{journal}{\emph{IEEE Transactions on Software Engineering}}
  \bibinfo{volume}{40}, \bibinfo{number}{3} (\bibinfo{year}{2013}),
  \bibinfo{pages}{251--265}.
\newblock


\bibitem[\protect\citeauthoryear{Stuckman, Walden, and Scandariato}{Stuckman
  et~al\mbox{.}}{2017}]%
        {stuckman2017}
\bibfield{author}{\bibinfo{person}{Jeffrey Stuckman}, \bibinfo{person}{James
  Walden}, {and} \bibinfo{person}{Riccardo Scandariato}.}
  \bibinfo{year}{2017}\natexlab{}.
\newblock \showarticletitle{The effect of dimensionality reduction on software
  vulnerability prediction models}.
\newblock \bibinfo{journal}{\emph{IEEE Transactions on Reliability}}
  \bibinfo{volume}{66}, \bibinfo{number}{1} (\bibinfo{year}{2017}),
  \bibinfo{pages}{17--37}.
\newblock


\bibitem[\protect\citeauthoryear{Svahnberg, van Gurp, and Bosch}{Svahnberg
  et~al\mbox{.}}{2005}]%
        {svahnberg2005}
\bibfield{author}{\bibinfo{person}{Mikael Svahnberg}, \bibinfo{person}{Jilles
  van Gurp}, {and} \bibinfo{person}{Jan Bosch}.}
  \bibinfo{year}{2005}\natexlab{}.
\newblock \showarticletitle{A taxonomy of variability realization techniques:
  Research Articles}.
\newblock \bibinfo{journal}{\emph{Softw. Pract. Exper.}} \bibinfo{volume}{35},
  \bibinfo{number}{8} (\bibinfo{year}{2005}), \bibinfo{pages}{705--754}.
\newblock
\showISSN{0038-0644}
\urldef\tempurl%
\url{https://doi.org/10.1002/spe.v35:8}
\showDOI{\tempurl}


\bibitem[\protect\citeauthoryear{{\v{S}}vogor, Crnkovi{\'c}, and
  Vr{\v{c}}ek}{{\v{S}}vogor et~al\mbox{.}}{2019}]%
        {svogor2019}
\bibfield{author}{\bibinfo{person}{Ivan {\v{S}}vogor}, \bibinfo{person}{Ivica
  Crnkovi{\'c}}, {and} \bibinfo{person}{Neven Vr{\v{c}}ek}.}
  \bibinfo{year}{2019}\natexlab{}.
\newblock \showarticletitle{An extensible framework for software configuration
  optimization on heterogeneous computing systems: Time and energy case study}.
\newblock \bibinfo{journal}{\emph{Information and software technology}}
  \bibinfo{volume}{105} (\bibinfo{year}{2019}), \bibinfo{pages}{30--42}.
\newblock


\bibitem[\protect\citeauthoryear{Temple, Acher, Biggio, J{\'e}z{\'e}quel, and
  Roli}{Temple et~al\mbox{.}}{2018}]%
        {temple2018}
\bibfield{author}{\bibinfo{person}{Paul Temple}, \bibinfo{person}{Mathieu
  Acher}, \bibinfo{person}{Battista Biggio}, \bibinfo{person}{Jean-Marc
  J{\'e}z{\'e}quel}, {and} \bibinfo{person}{Fabio Roli}.}
  \bibinfo{year}{2018}\natexlab{}.
\newblock \showarticletitle{Towards Adversarial Configurations for Software
  Product Lines}.
\newblock \bibinfo{journal}{\emph{arXiv preprint arXiv:1805.12021}}
  (\bibinfo{year}{2018}).
\newblock


\bibitem[\protect\citeauthoryear{Temple, Acher, J{\'{e}}z{\'{e}}quel, and
  Barais}{Temple et~al\mbox{.}}{2017a}]%
        {temple2017a}
\bibfield{author}{\bibinfo{person}{Paul Temple}, \bibinfo{person}{Mathieu
  Acher}, \bibinfo{person}{Jean{-}Marc J{\'{e}}z{\'{e}}quel}, {and}
  \bibinfo{person}{Olivier Barais}.} \bibinfo{year}{2017}\natexlab{a}.
\newblock \showarticletitle{Learning Contextual-Variability Models}.
\newblock \bibinfo{journal}{\emph{{IEEE} Software}} \bibinfo{volume}{34},
  \bibinfo{number}{6} (\bibinfo{year}{2017}), \bibinfo{pages}{64--70}.
\newblock
\urldef\tempurl%
\url{https://doi.org/10.1109/MS.2017.4121211}
\showDOI{\tempurl}


\bibitem[\protect\citeauthoryear{Temple, Acher, J{\'e}z{\'e}quel, Noel-Baron,
  and Galindo}{Temple et~al\mbox{.}}{2017b}]%
        {temple2017}
\bibfield{author}{\bibinfo{person}{Paul Temple}, \bibinfo{person}{Mathieu
  Acher}, \bibinfo{person}{Jean-Marc~A J{\'e}z{\'e}quel},
  \bibinfo{person}{L{\'e}o~A Noel-Baron}, {and} \bibinfo{person}{Jos{\'e}~A
  Galindo}.} \bibinfo{year}{2017}\natexlab{b}.
\newblock \bibinfo{booktitle}{\emph{Learning-Based Performance Specialization
  of Configurable Systems}}.
\newblock \bibinfo{type}{Research Report}. \bibinfo{institution}{{IRISA, Inria
  Rennes ; University of Rennes 1}}.
\newblock
\urldef\tempurl%
\url{https://hal.archives-ouvertes.fr/hal-01467299}
\showURL{%
\tempurl}


\bibitem[\protect\citeauthoryear{Temple, Galindo~Duarte, Acher, and
  J{\'e}z{\'e}quel}{Temple et~al\mbox{.}}{2016}]%
        {temple2016}
\bibfield{author}{\bibinfo{person}{Paul Temple},
  \bibinfo{person}{Jos{\'e}~Angel Galindo~Duarte}, \bibinfo{person}{Mathieu
  Acher}, {and} \bibinfo{person}{Jean-Marc J{\'e}z{\'e}quel}.}
  \bibinfo{year}{2016}\natexlab{}.
\newblock \showarticletitle{{Using Machine Learning to Infer Constraints for
  Product Lines}}. In \bibinfo{booktitle}{\emph{{Software Product Line
  Conference (SPLC)}}}. \bibinfo{address}{Beijing, China}.
\newblock
\urldef\tempurl%
\url{https://doi.org/10.1145/2934466.2934472}
\showDOI{\tempurl}


\bibitem[\protect\citeauthoryear{Thornton, Hutter, Hoos, and
  Leyton-Brown}{Thornton et~al\mbox{.}}{2013}]%
        {thornton2013}
\bibfield{author}{\bibinfo{person}{Chris Thornton}, \bibinfo{person}{Frank
  Hutter}, \bibinfo{person}{Holger~H Hoos}, {and} \bibinfo{person}{Kevin
  Leyton-Brown}.} \bibinfo{year}{2013}\natexlab{}.
\newblock \showarticletitle{Auto-WEKA: Combined selection and hyperparameter
  optimization of classification algorithms}. In
  \bibinfo{booktitle}{\emph{Proceedings of the 19th ACM SIGKDD international
  conference on Knowledge discovery and data mining}}. ACM,
  \bibinfo{pages}{847--855}.
\newblock


\bibitem[\protect\citeauthoryear{Th{\"u}m, Apel, K{\"a}stner, Schaefer, and
  Saake}{Th{\"u}m et~al\mbox{.}}{2014}]%
        {thum2014classification}
\bibfield{author}{\bibinfo{person}{Thomas Th{\"u}m}, \bibinfo{person}{Sven
  Apel}, \bibinfo{person}{Christian K{\"a}stner}, \bibinfo{person}{Ina
  Schaefer}, {and} \bibinfo{person}{Gunter Saake}.}
  \bibinfo{year}{2014}\natexlab{}.
\newblock \showarticletitle{A classification and survey of analysis strategies
  for software product lines}.
\newblock \bibinfo{journal}{\emph{ACM Computing Surveys (CSUR)}}
  \bibinfo{volume}{47}, \bibinfo{number}{1} (\bibinfo{year}{2014}),
  \bibinfo{pages}{6}.
\newblock


\bibitem[\protect\citeauthoryear{Trubiani and Apel}{Trubiani and Apel}{2019}]%
        {apel2019PLUS}
\bibfield{author}{\bibinfo{person}{Catia Trubiani} {and} \bibinfo{person}{Sven
  Apel}.} \bibinfo{year}{2019}\natexlab{}.
\newblock \showarticletitle{PLUS: Performance Learning for Uncertainty of
  Software}. In \bibinfo{booktitle}{\emph{International Conference on Software
  Engineering NIER}}. ACM.
\newblock


\bibitem[\protect\citeauthoryear{Valov, Guo, and Czarnecki}{Valov
  et~al\mbox{.}}{2015}]%
        {valov2015}
\bibfield{author}{\bibinfo{person}{Pavel Valov}, \bibinfo{person}{Jianmei Guo},
  {and} \bibinfo{person}{Krzysztof Czarnecki}.}
  \bibinfo{year}{2015}\natexlab{}.
\newblock \showarticletitle{Empirical comparison of regression methods for
  variability-aware performance prediction}. In
  \bibinfo{booktitle}{\emph{SPLC'15}}.
\newblock


\bibitem[\protect\citeauthoryear{Valov, Petkovich, Guo, Fischmeister, and
  Czarnecki}{Valov et~al\mbox{.}}{2017}]%
        {valov2017}
\bibfield{author}{\bibinfo{person}{Pavel Valov},
  \bibinfo{person}{Jean-Christophe Petkovich}, \bibinfo{person}{Jianmei Guo},
  \bibinfo{person}{Sebastian Fischmeister}, {and} \bibinfo{person}{Krzysztof
  Czarnecki}.} \bibinfo{year}{2017}\natexlab{}.
\newblock \showarticletitle{Transferring performance prediction models across
  different hardware platforms}. In \bibinfo{booktitle}{\emph{Proceedings of
  the 8th ACM/SPEC on International Conference on Performance Engineering}}.
  ACM, \bibinfo{pages}{39--50}.
\newblock


\bibitem[\protect\citeauthoryear{Van~Aken, Pavlo, Gordon, and Zhang}{Van~Aken
  et~al\mbox{.}}{2017}]%
        {aken2017}
\bibfield{author}{\bibinfo{person}{Dana Van~Aken}, \bibinfo{person}{Andrew
  Pavlo}, \bibinfo{person}{Geoffrey~J Gordon}, {and} \bibinfo{person}{Bohan
  Zhang}.} \bibinfo{year}{2017}\natexlab{}.
\newblock \showarticletitle{Automatic database management system tuning through
  large-scale machine learning}. In \bibinfo{booktitle}{\emph{Proceedings of
  the 2017 ACM International Conference on Management of Data}}. ACM,
  \bibinfo{pages}{1009--1024}.
\newblock


\bibitem[\protect\citeauthoryear{Varshosaz, Al-Hajjaji, Th{\"u}m, Runge,
  Mousavi, and Schaefer}{Varshosaz et~al\mbox{.}}{2018}]%
        {varshosaz2018}
\bibfield{author}{\bibinfo{person}{Mahsa Varshosaz}, \bibinfo{person}{Mustafa
  Al-Hajjaji}, \bibinfo{person}{Thomas Th{\"u}m}, \bibinfo{person}{Tobias
  Runge}, \bibinfo{person}{Mohammad~Reza Mousavi}, {and} \bibinfo{person}{Ina
  Schaefer}.} \bibinfo{year}{2018}\natexlab{}.
\newblock \showarticletitle{A classification of product sampling for software
  product lines}. In \bibinfo{booktitle}{\emph{Proceeedings of the 22nd
  International Conference on Systems and Software Product Line-Volume 1}}.
  ACM, \bibinfo{pages}{1--13}.
\newblock


\bibitem[\protect\citeauthoryear{Venkata, Ahn, Jeon, Gupta, Louie, Garcia,
  Belongie, and Taylor}{Venkata et~al\mbox{.}}{2009}]%
        {venkata2009}
\bibfield{author}{\bibinfo{person}{Sravanthi~Kota Venkata},
  \bibinfo{person}{Ikkjin Ahn}, \bibinfo{person}{Donghwan Jeon},
  \bibinfo{person}{Anshuman Gupta}, \bibinfo{person}{Christopher Louie},
  \bibinfo{person}{Saturnino Garcia}, \bibinfo{person}{Serge Belongie}, {and}
  \bibinfo{person}{Michael~Bedford Taylor}.} \bibinfo{year}{2009}\natexlab{}.
\newblock \showarticletitle{SD-VBS: The San Diego vision benchmark suite}. In
  \bibinfo{booktitle}{\emph{2009 IEEE International Symposium on Workload
  Characterization (IISWC)}}. IEEE, \bibinfo{pages}{55--64}.
\newblock


\bibitem[\protect\citeauthoryear{Weckesser, Kluge, Pfannem{\"u}ller,
  Matth{\'e}, Sch{\"u}rr, and Becker}{Weckesser et~al\mbox{.}}{2018}]%
        {weckesser2018}
\bibfield{author}{\bibinfo{person}{Markus Weckesser}, \bibinfo{person}{Roland
  Kluge}, \bibinfo{person}{Martin Pfannem{\"u}ller}, \bibinfo{person}{Michael
  Matth{\'e}}, \bibinfo{person}{Andy Sch{\"u}rr}, {and}
  \bibinfo{person}{Christian Becker}.} \bibinfo{year}{2018}\natexlab{}.
\newblock \showarticletitle{Optimal reconfiguration of dynamic software product
  lines based on performance-influence models}. In
  \bibinfo{booktitle}{\emph{Proceeedings of the 22nd International Conference
  on Systems and Software Product Line-Volume 1}}. ACM,
  \bibinfo{pages}{98--109}.
\newblock


\bibitem[\protect\citeauthoryear{Westermann, Happe, Krebs, and
  Farahbod}{Westermann et~al\mbox{.}}{2012}]%
        {westermann2012}
\bibfield{author}{\bibinfo{person}{Dennis Westermann}, \bibinfo{person}{Jens
  Happe}, \bibinfo{person}{Rouven Krebs}, {and} \bibinfo{person}{Roozbeh
  Farahbod}.} \bibinfo{year}{2012}\natexlab{}.
\newblock \showarticletitle{Automated inference of goal-oriented performance
  prediction functions}. In \bibinfo{booktitle}{\emph{IEEE/ACM International
  Conference on Automated Software Engineering (ASE)}}. ACM,
  \bibinfo{pages}{190--199}.
\newblock


\bibitem[\protect\citeauthoryear{Wohlin}{Wohlin}{2014}]%
        {wohlin2014}
\bibfield{author}{\bibinfo{person}{Claes Wohlin}.}
  \bibinfo{year}{2014}\natexlab{}.
\newblock \showarticletitle{Guidelines for snowballing in systematic literature
  studies and a replication in software engineering}. In
  \bibinfo{booktitle}{\emph{Proceedings of the 18th international conference on
  evaluation and assessment in software engineering}}. ACM,
  \bibinfo{pages}{38}.
\newblock


\bibitem[\protect\citeauthoryear{Wohlin, Runeson, Host, Ohlsson, Regnell, and
  Wesslen}{Wohlin et~al\mbox{.}}{2000}]%
        {wohlin2000}
\bibfield{author}{\bibinfo{person}{C Wohlin}, \bibinfo{person}{P Runeson},
  \bibinfo{person}{M Host}, \bibinfo{person}{MC Ohlsson}, \bibinfo{person}{B
  Regnell}, {and} \bibinfo{person}{A Wesslen}.}
  \bibinfo{year}{2000}\natexlab{}.
\newblock \bibinfo{title}{{Experimentation in software engineering: an
  introduction}}.
\newblock
\newblock


\bibitem[\protect\citeauthoryear{Xi, Liu, Raghavachari, Xia, and Zhang}{Xi
  et~al\mbox{.}}{2004}]%
        {xi2004}
\bibfield{author}{\bibinfo{person}{Bowei Xi}, \bibinfo{person}{Zhen Liu},
  \bibinfo{person}{Mukund Raghavachari}, \bibinfo{person}{Cathy~H Xia}, {and}
  \bibinfo{person}{Li Zhang}.} \bibinfo{year}{2004}\natexlab{}.
\newblock \showarticletitle{A smart hill-climbing algorithm for application
  server configuration}. In \bibinfo{booktitle}{\emph{Proceedings of the 13th
  international conference on World Wide Web}}. ACM, \bibinfo{pages}{287--296}.
\newblock


\bibitem[\protect\citeauthoryear{Xu, Hutter, Hoos, and Leyton-Brown}{Xu
  et~al\mbox{.}}{2008}]%
        {xu2008}
\bibfield{author}{\bibinfo{person}{Lin Xu}, \bibinfo{person}{Frank Hutter},
  \bibinfo{person}{Holger~H Hoos}, {and} \bibinfo{person}{Kevin Leyton-Brown}.}
  \bibinfo{year}{2008}\natexlab{}.
\newblock \showarticletitle{SATzilla: portfolio-based algorithm selection for
  SAT}.
\newblock \bibinfo{journal}{\emph{Journal of artificial intelligence research}}
   \bibinfo{volume}{32} (\bibinfo{year}{2008}), \bibinfo{pages}{565--606}.
\newblock


\bibitem[\protect\citeauthoryear{Xu, Jin, Fan, Zhou, Pasupathy, and
  Talwadker}{Xu et~al\mbox{.}}{2015}]%
        {DBLP:conf/sigsoft/XuJFZPT15}
\bibfield{author}{\bibinfo{person}{Tianyin Xu}, \bibinfo{person}{Long Jin},
  \bibinfo{person}{Xuepeng Fan}, \bibinfo{person}{Yuanyuan Zhou},
  \bibinfo{person}{Shankar Pasupathy}, {and} \bibinfo{person}{Rukma
  Talwadker}.} \bibinfo{year}{2015}\natexlab{}.
\newblock \showarticletitle{Hey, you have given me too many knobs!:
  understanding and dealing with over-designed configuration in system
  software}. In \bibinfo{booktitle}{\emph{Proceedings of the 2015 10th Joint
  Meeting on Foundations of Software Engineering, {ESEC/FSE} 2015, Bergamo,
  Italy, August 30 - September 4, 2015}}. \bibinfo{pages}{307--319}.
\newblock
\urldef\tempurl%
\url{https://doi.org/10.1145/2786805.2786852}
\showDOI{\tempurl}


\bibitem[\protect\citeauthoryear{Yilmaz, Cohen, and Porter}{Yilmaz
  et~al\mbox{.}}{2006}]%
        {yilmaz2006}
\bibfield{author}{\bibinfo{person}{Cemal Yilmaz}, \bibinfo{person}{Myra~B
  Cohen}, {and} \bibinfo{person}{Adam~A Porter}.}
  \bibinfo{year}{2006}\natexlab{}.
\newblock \showarticletitle{Covering arrays for efficient fault
  characterization in complex configuration spaces}.
\newblock \bibinfo{journal}{\emph{IEEE Transactions on Software Engineering}}
  \bibinfo{volume}{32}, \bibinfo{number}{1} (\bibinfo{year}{2006}),
  \bibinfo{pages}{20--34}.
\newblock


\bibitem[\protect\citeauthoryear{Yilmaz, Dumlu, Cohen, and Porter}{Yilmaz
  et~al\mbox{.}}{2014}]%
        {yilmaz2014}
\bibfield{author}{\bibinfo{person}{Cemal Yilmaz}, \bibinfo{person}{Emine
  Dumlu}, \bibinfo{person}{Myra~B Cohen}, {and} \bibinfo{person}{Adam Porter}.}
  \bibinfo{year}{2014}\natexlab{}.
\newblock \showarticletitle{Reducing masking effects in
  combinatorialinteraction testing: A feedback drivenadaptive approach}.
\newblock \bibinfo{journal}{\emph{IEEE Transactions on Software Engineering}}
  \bibinfo{volume}{40}, \bibinfo{number}{1} (\bibinfo{year}{2014}),
  \bibinfo{pages}{43--66}.
\newblock


\bibitem[\protect\citeauthoryear{Zhang, Guo, Blais, and Czarnecki}{Zhang
  et~al\mbox{.}}{2015}]%
        {zhang2015}
\bibfield{author}{\bibinfo{person}{Yi Zhang}, \bibinfo{person}{Jianmei Guo},
  \bibinfo{person}{Eric Blais}, {and} \bibinfo{person}{Krzysztof Czarnecki}.}
  \bibinfo{year}{2015}\natexlab{}.
\newblock \showarticletitle{Performance prediction of configurable software
  systems by fourier learning (t)}. In \bibinfo{booktitle}{\emph{IEEE/ACM
  International Conference on Automated Software Engineering (ASE)}}. IEEE,
  \bibinfo{pages}{365--373}.
\newblock


\bibitem[\protect\citeauthoryear{Zhang, Guo, Blais, Czarnecki, and Yu}{Zhang
  et~al\mbox{.}}{2016}]%
        {zhang2016}
\bibfield{author}{\bibinfo{person}{Yi Zhang}, \bibinfo{person}{Jianmei Guo},
  \bibinfo{person}{Eric Blais}, \bibinfo{person}{Krzysztof Czarnecki}, {and}
  \bibinfo{person}{Huiqun Yu}.} \bibinfo{year}{2016}\natexlab{}.
\newblock \showarticletitle{A mathematical model of performance-relevant
  feature interactions}. In \bibinfo{booktitle}{\emph{Proceedings of the 20th
  International Systems and Software Product Line Conference}}. ACM,
  \bibinfo{pages}{25--34}.
\newblock


\bibitem[\protect\citeauthoryear{Zheng, Bianchini, and Nguyen}{Zheng
  et~al\mbox{.}}{2007}]%
        {zheng2007}
\bibfield{author}{\bibinfo{person}{Wei Zheng}, \bibinfo{person}{Ricardo
  Bianchini}, {and} \bibinfo{person}{Thu~D Nguyen}.}
  \bibinfo{year}{2007}\natexlab{}.
\newblock \showarticletitle{Automatic configuration of internet services}.
\newblock \bibinfo{journal}{\emph{ACM SIGOPS Operating Systems Review}}
  \bibinfo{volume}{41}, \bibinfo{number}{3} (\bibinfo{year}{2007}),
  \bibinfo{pages}{219--229}.
\newblock


\bibitem[\protect\citeauthoryear{Zuluaga, Krause, and P{\"u}schel}{Zuluaga
  et~al\mbox{.}}{2016}]%
        {zuluaga2016}
\bibfield{author}{\bibinfo{person}{Marcela Zuluaga}, \bibinfo{person}{Andreas
  Krause}, {and} \bibinfo{person}{Markus P{\"u}schel}.}
  \bibinfo{year}{2016}\natexlab{}.
\newblock \showarticletitle{$\varepsilon$-pal: an active learning approach to
  the multi-objective optimization problem}.
\newblock \bibinfo{journal}{\emph{The Journal of Machine Learning Research}}
  \bibinfo{volume}{17}, \bibinfo{number}{1} (\bibinfo{year}{2016}),
  \bibinfo{pages}{3619--3650}.
\newblock


\end{thebibliography}


\end{document}